\RequirePackage{fix-cm}
\documentclass[smallcondensed,final]{svjour3}     
\journalname{Space Science Reviews}
\smartqed  
\usepackage{journal_macros}

\usepackage[T1]{fontenc}
\usepackage[utf8]{inputenc}
\usepackage{xcolor}
\usepackage[bookmarks=false, 
     pdfnewwindow=true, 
     colorlinks=true,   
     linkcolor=magenta, 
     citecolor=blue,    
     filecolor=cyan,    
     urlcolor=purple,   
	final=true]{hyperref}
\usepackage{graphicx}
\usepackage{ulem}
\usepackage{amssymb}
\usepackage{amsmath}
\usepackage{natbib}
\usepackage{multirow}
\usepackage{siunitx}
\DeclareRobustCommand{\ion}[2]{\textup{#1\,\textsc{\lowercase{#2}}}}
\newcommand*\element[1][]{%
  \def\aa@element@tr{#1}%
  \aa@element}

\usepackage{url}
\graphicspath{{FIP_FIGURES/}}
\usepackage{lineno}
\newcommand{\orcid}[1]{} 

\newcommand*{\doi}[1]{\href{http://dx.doi.org/#1}{DOI: #1}}
\begin{document}

\title{Linking the Sun to the Heliosphere Using Composition Data and Modelling}
\subtitle{A Test Case with a Coronal Jet}

\titlerunning{Linking the Sun to the Heliosphere Using Composition}

\author{{Susanna~Parenti} \orcid{0000-0002-0786-7307}  \and
        Iulia Chifu \orcid{0000-0002-5481-9228} \and
        Giulio~Del~Zanna \and
        Justin~Edmondson \and
        Alessandra~Giunta \and 
        Viggo~H.~Hansteen \and
        Aleida~Higginson \and
        J.~Martin~Laming \and
        Susan~T.~Lepri  \orcid{0000-0003-1611-227X} \and 
        Benjamin~J.~Lynch \and
        Yeimy~J.~Rivera \orcid{0000-0002-8748-2123} \and
        Rudolf~von~Steiger \orcid{0000-0002-3350-0023} \and
        {Thomas~Wiegelmann   \orcid{0000-0001-6238-0721}} \and
        Robert~F.~Wimmer-Schweingruber \and
        Natalia~Zambrana~Prado \and 
        Gabriel Pelouze \orcid{0000-0002-0397-2214}
}

\authorrunning{Parenti et al.} 

\institute{S. Parenti \at
              Institut d’Astrophysique Spatiale, CNRS-Université Paris-Saclay, 91405 Orsay, France \\
              \email{susanna.parenti@ias.u-psud.fr}       
                \orcid{0000-0003-1438-1310}   
           \and
           I. Chifu \at
           Institute for Astrophysics, University of G\"ottingen, Friedrich-Hund-Platz 1, 37077 G\"ottingen, Germany, \\
           Max-Planck-Institut f\"ur Sonnensystemforschung,
           Justus-von-Liebig-Weg 3
           37077 G\"ottingen, Germany 
           \and
            G. Del Zanna \at
              DAMTP, Centre for Mathematical Sciences,  
 University of Cambridge,
 Wilberforce Road, Cambridge CB3 0WA UK  
            \and
            Justin~Edmondson \at
            Department of Climate and Space Sciences and Engineering, University of Michigan, Ann Arbor, MI 48109, USA
            \and
            Alessandra~Giunta \at
            RAL Space, UKRI STFC Rutherford Appleton Laboratory, 
            Harwell, Didcot, OX11 0QX, United Kingdom
            \and 
            Viggo~H.~Hansteen \at
            Institute of Theoretical Astrophysics, University of Oslo, P.O. Box 1029 Blindern, 0315 Oslo, Norway
            \and 
            Aleida~Higginson \at
            Department of Climate and Space Sciences and Engineering, University of Michigan, Ann Arbor, MI 48109, USA
            \and 
           J. M. Laming \at
           Space Science Division, Code 7684,
           Naval Research Laboratory,
           Washington, DC 20375, USA
            \and
          S. T. Lepri \at
           Department of Climate and Space Sciences and Engineering,
           University of Michigan,
           Ann Arbor, MI 48109, USA
           \and
           B. J. Lynch \at
              Space Sciences Laboratory,
              University of California--Berkeley,
              Berkeley, CA 94720, USA
               \and
          Y. J. Rivera \at
          Center for Astrophysics,  
              Harvard University,
              60 Garden Street, Cambridge, MA 02138, USA
           \and     
           R. von Steiger \at 
           International Space Science Institute, Hallerstrasse 6, 3012 Bern, Switzerland\\
           Physikalisches Institut, University of Bern, Sidlerstrasse 5, 3012 Bern, Switzerland
            \and
           T. Wiegelmann \at
           Max-Planck-Institut f\"ur Sonnensystemforschung,
           Justus-von-Liebig-Weg 3,
           37077 G\"ottingen, Germany
           \and
           Robert~F.~Wimmer-Schweingruber \at
           Institute of Experimental and Applied Physics, University of Kiel, Leibnizstrasse 11, D-24118 Kiel, Germany
           \and 
           Natalia~Zambrana~Prado \at
           Institut d’Astrophysique Spatiale, CNRS-Université Paris-Saclay, 91405 Orsay, France
           \and
           Gabriel Pelouze \at
           Centre for mathematical Plasma Astrophysics, Department of Mathematics, KU Leuven, Celestijnenlaan 200B bus 2400, 3001 Leuven, Belgium
}

\date{Received: March 31, 2021 / Accepted: October, 2, 2021}

\maketitle
\begin{abstract}

Our understanding of the formation and evolution of the corona and the heliosphere is linked to our capability of properly interpret the data from remote sensing and {\it in--situ} observations. In this respect, being able to correctly connect  {\it in--situ} observations with their source regions on the Sun is the key for solving this problem. In this work we aim at testing a diagnostics method for this connectivity.

This paper makes use of a coronal jet observed on 2010 August 2nd in active region 11092  as a test for our connectivity method. This combines solar EUV and {\it in--situ} data together with magnetic field extrapolation, large scale MHD modeling and FIP (First Ionization Potential) bias modeling to provide a global picture from the source region of the jet to its possible signatures at 1AU.

Our data analysis reveals the presence of outflow areas near the jet which are within open magnetic flux regions and which present FIP bias consistent with the FIP model results. In our picture, one of these open areas is the candidate jet source. Using a back-mapping technique we identified the arrival time of this solar plasma at the ACE spacecraft. The {\it in--situ} data show signatures of changes in the plasma and magnetic field parameters, with FIP bias consistent  with the possible passage of the jet material. 

Our results highlight the importance of remote sensing and {\it in--situ} coordinated observations as a key to solve the connectivity problem. We discuss our results in view of the recent Solar Orbiter launch which is currently providing such unique data.  

\keywords{Active region \and Solar wind \and Connectivity \and Diagnostics \and Modelling \and UV emission \and {\it in--situ} data}
\end{abstract}

\section{Introduction} 
\label{intro}

It is well acknowledged that the interplanetary medium is filled by the spatial expansion of the solar magnetic field embedded to a continuous flow of particles, the solar wind; these are the main ingredients of the heliosphere. 

The first indirect evidence of the  presence of the solar wind dates back to the beginning of the twentieth century, while the first in-situ wind and magnetic field measures are from the early 1960 (see for instance \citet{hundhausen72} for an historical review). Since then, enormous progress has been made, both for the spatial and temporal monitoring of the heliosphere and the theoretical explanation of its origin and evolution \citep[see for instance][]{hansteen12, abbo16}. For instance, the initial prediction for the solar wind \citep{parker58} envisaged a wind driven by thermal forces originating in a MHD wave heated corona. The modern view is more that the MHD waves couple to the nascent solar wind directly to cause its acceleration. The role of interchange reconnection also appears to be a good candidate for releasing plasma into open regions to feed  the solar wind (see for instance \cite{abbo16}).

The solar wind is comprised of two basic regimes {\citep[e.g.][]{mccomas03}}. The fast wind, with typical speed $> 600$ km s$^{-1}$ is relatively steady, exhibits Alfv\'enic turbulence, is established to originate in coronal holes, and corresponds the closest to Parker's original idea {\citep[e.g.][]{ko18}}. The slow wind $< 500$ km s$^{-1}$ is much more variable, exhibits periods of Alfv\'enic and non-Alfv\'enic turbulence {\citep[e.g.][]{ko18}}, has composition similar to coronal loops {\citep[e.g.][]{heber21}}, and has less certain solar origins. Prior observations suggest that solar wind ions are heated in open magnetic field by microscopic (kinetic) wave-particle processes, as the plasma becomes collisionless. SOHO/UVCS (Solar and Heliosheric Observatory, Ultraviolet Coronagraph Spectrometer) showed pronounced kinetic anisotropies (with $T_{\perp} > T_{||}$) for O$^{5+}$ and Mg$^{9+}$ in polar coronal holes \citep{kohl97,kohl98,kohl06,cranmer99,cranmer08} and streamers above 1.9~R$_{\odot}$ \citep{frazin99}. Interchange reconnection may also play a role in supplying material into the slow solar wind, as well, with material previously trapped on coronal loops \citep{Fisk2001, Edmondson2010a}.

The global properties of the heliosphere are quite well understood, however there are several not well understood aspects of the smaller scales. One example is the identification at the Sun of the small scale wind source regions. To fully understand the structure and dynamics of the heliosphere requires modelling and simulation capabilities to create an evolving 3D environment validated by continuous, widely distributed observations. Ground and space based data sets are critical to constraining {our understanding} of the spatial and temporal conditions  in this 3D environment. Both {\it in--situ} and remote sensing observations provide critical vantage points to view structures and phenomena on the Sun, in the corona, as well as observe their propagation through the heliosphere. Due to the limited number of satellite and ground based observations, incomplete spatial mapping from the Sun out into the heliosphere requires new methods for connecting observations \citep[see for instance][]{Stansby2020, depablos2021}. Connecting measurements from the Sun into the heliosphere is a major challenge in solar physics and heliophysics, not only in that different quantities are observed with different sensors, at different time cadences, but also the connections between these measurements often do not have a well defined or unique solution \citep[see for instance][]{poletto96}.  In order to connect structures at the Sun with those observed {\it in--situ}, one must ensure that the same magnetic structure is sampled by both methods.  Spatial (due to the deformation of the magnetic field as it expands in the heliosphere) and the temporal changes (due to the delay between the moment of surface escape and the {\it in--situ} spacecraft crossing) of such magnetic structures have to be taken into account. Several methods can be used to back map {\it {\it in--situ}} measurements \cite[see for instance][] {neugebauer98} and their comparison made in \citet{peleikis17} or \citet{kruse-etal-2021}, but each have their limitations. An example of this kind of effort was made during the periodic  quadrature between the SOHO satellite \citep{domingo95} and the Ulysses mission \citep{wenzel1992}. In this period it was possible to derive the plasma parameters from the Sun, to the corona (e.g. density, temperature, composition, outflows) up to several solar radii, and try to link them to the same parameters  measured {\it {\it in--situ}} \citep[e.g.][]{suess01, parenti03, bemporad03}. 

One of the most promising methods for back-mapping heliospheric structures to their solar counterparts involves observing the compositional properties including the ionization states of heavy ions as well as their elemental abundances, sorted by a property called the First Ionization Potential (FIP) bias. While kinetic properties of the solar wind are affected by propagation effects such as cooling, the heavy ion properties are determined in the low corona and do not evolve as the solar wind transits through the heliosphere. The FIP bias describes the variation of the fractionation of the chemical composition of heavy elements which occurs between the photosphere and regions further out in the corona \citep[e.g.][]{sheely95, feldman92b}. In particular, elements with FIP lower than about 10 eV (e.g. Mg, Si, Fe) are observed to be enhanced compared to those with higher FIP (e.g. O, Ne, Ar), with respect to their photospheric values. This bias has different amplitudes depending on the plasma/magnetic structure observed. It was first detected in the solar wind, with an amplitude of on average 2--3 for slow wind, and around 1--1.5 for the fast wind and has served as a key discriminator between these two types of wind \citep[e.g.][]{geiss95b, vonSteiger2000, zurbuchen12}. 

Variations of the FIP bias were also inferred in different parts of the Sun and in the corona \citep{raymond97, parenti00, parenti19}, showing  temperature dependence \citep[e.g.][]{delzanna19}. In {active regions (ARs)}, which are considered one of the candidate locations for the nascent slow wind and which are the subject of our work, the FIP bias is quite variable and both coronal (FIP bias $>$ 2) or photospheric composition can be found \citep[see for instance the review in][]{delzanna18}. This variation also includes the increase of FIP bias with the ARs lifetime {\citep[e.g.][]{ko16, harra2021}} and/or with the emergence of new flux \citep[e.g.][]{baker18}. The range of variability in the corona can be higher than that observed in the wind, because at 1 AU the wind may include a mixture of different solar wind structures. It has to be noticed that the various methodologies used to derive the bias may affect the results, which may contribute to the origin of the  inferred variability of the FIP effect \citep[see for instance the discussion in][]{delzanna15}.

In addition to the FIP bias, {\it in--situ} heavy ion charge state measurements can also be used as a tool to investigate coronal heating processes that accelerate the steady-state slow and fast solar wind \citep[e.g.][]{Hundhausen1968, Owocki1983, Burgi1986,  Ko1997, vonSteiger1997, wimmer-etal-1997, wimmer-etal-1999, vonSteiger2000, Zurbuchen2002, Laming2007, vonSteiger2011, Landi2012b, zurbuchen12, Lepri2013, ZhaoL2017a, ZhaoL2017b}. The same techniques have been implemented in transients to investigate and place constraints on the thermal history of plasma as it accelerates through the corona \citep[e.g.][]{Gruesbeck2011, Rivera2019a, Rivera2019b}. The heavy ion charge states in the solar wind can reflect the plasma properties of the inner corona where the solar wind originates. As the solar wind is accelerated, the electron number density rapidly decreases with distance from the Sun, while the electron temperature increases and reaches a maximum value before decreasing again {\citep{kohl06, laming04b}}. The ionization and recombination rates for a given pair of charge states of an element are proportional to the electron density and also depend on electron temperature (or kinetic energy), so they also rapidly decrease with increasing height in the corona. As the plasma parcel accelerates through the corona it will reach a critical height where the electron density is so low that the time scale to modify the charge states exceeds the expansion time scale  and the ionization state of an element remains unaltered \citep{Owocki1983} as the solar wind continues to propagate through the corona \citep{Landi2012b}. This process is called the ``freeze-in'' process. Beyond the freeze-in point, located at $\sim 1.5 -5 R_{\odot}$ depending on which ion pair is considered {\citep{laming04b,laming07}}, the charge state distributions of heavy ions (e.g. C, N, O, Si, Mg, Fe) thus maintain a record of the thermal properties of the plasma in the inner corona and carry with them this information as the solar wind propagates throughout the heliosphere.  

{
The ESA-NASA Solar Orbiter mission \citep[launched in February 2020,][]{muller2020} has been designed to strongly contribute to solve the connectivity problem between remote and {\it in--situ} measurements below 1 AU}. The mission profile is designed to reduced the gap between {\it in--situ} and remote sensing measurements, by approaching to the Sun up to a distance of about 0.3 AU, while additionally providing periods of near-corotation. 
Some of the on board remote sensing instruments, {the Spectral Imaging of the Coronal Environment (SPICE,  \cite{spice2020}), the Extreme Ultraviolet Imager (EUI, \cite{rochus20}), the Polarimetric and Helioseismic Imager (PHI, \cite{solanki20}) and the Spectrometer Telescope for Imaging X-rays (STIX, \cite{krucker2020})} will observe the source regions at the Sun that will ultimately be sampled a few days later by the  on board {\it in--situ} instruments. 
The spacecraft will locally sample the solar wind which has just been accelerated \citep{owen-etal-2020}, and observe the heliospheric structures formed in the corona.  Close to the Sun, the plasma and magnetic field properties have only been  partially modified by propagation effects and interactions with nearby heliospheric structures, thus reducing the uncertainties in connecting these structures back to their sources. The payload instruments are designed to provide a wide complement of measurements to solve this connectivity problem below 1 AU. {The Solar Orbiter activity comes to complement the existing measurements at 1 AU, which are needed for the solar connectivity at this distance, as the dominant physical processes  may not be the same. We need to sample the environment over different physical conditions to test our heliospheric models in order to advance in our knowledge of the whole Sun-heliospheric system. Furthermore, the opportunity to have available measurements at different solar distance should be accompanied by the evolution and optimization of the diagnostics for multi-instrument observations.}

This paper presents an initial attempt to utilize both remote sensing and {\it in--situ} measurements at {1 AU}, along with a full complement of modelling techniques, to illustrate the methods required to solve the connectivity problem that will {also be addressed by Solar Orbiter at the beginning of 2022, when the nominal mission phase will start}. To this purpose we present a test case using coronal jet data to provide information on a coronal structure to connect out into the heliosphere.

AR jets are impulsive and collimated eruptions which are seen statistically observed to emerge primarily from the western periphery of the leading sunspot in a sunspot pair \citep{shimojo96}. Magnetic reconnection following magnetic field emergence or cancellation appears to be one of the main drivers of these plasma eruptions {{\citep[][and references therein]{mulay16}}}. A few studies have analyzed the possible correlation between AR jets and interplanetary radio III bursts observed co-temporally with these events. {The conclusion is that indeed, accelerated electron beams escape the jet region \citep[e.g.][]{innes11, mulay16, paraschiv19}, as well as  impulsive solar energetic particles \citep[see for instance][]{Lin1970,Reames1985, Bucik2020}. }

UV AR jets are seen to extend into the corona up to about 2 R$_\odot$ and magnetic field extrapolations suggest they are ejected in open field regions within the AR \citep{innes11, mulay16}. The  detection of {\it in--situ} signatures of an AR jet has been attempted in the past, but with no clear results \citep[see for instance][]{corti07, raouafi16}.

Based on our present knowledge of coronal jets and their expected signatures in the interplanetary region, we consider them as good candidates for testing connectivity diagnostics. In this paper we propose a methodology that will utilize FIP bias to connect the solar structure with the heliospheric counterpart, based on both data and modelling, and which may have a direct application for the Solar Orbiter Mission.

The paper is organized as follows: after a general introduction to the jet event in Section \ref{sec:1}, we describe all the building blocks for the connectivity in Sect. \ref{sec:methods:gen}. These include the use and comparison of  different models for the reconstruction the coronal magnetic field, as well as the comparison of different FIP bias estimation methods. Section \ref{sec:results} presents the results from the modeling and data analysis while we draw our conclusions in Sect. \ref{sec:discussion}.

\section{The 2nd of August 2010 Solar Jet} 
\label{sec:1}

The jet we select for our connectivity study was chosen because it had one of the most complete data coverage in time (three days around the jet) and space for the  EUV observations, as well as the interplanetary signature as radio type-III burst and {\it in--situ}. 

\begin{figure*}[!htb]
\centering
\includegraphics[width=3.2cm]{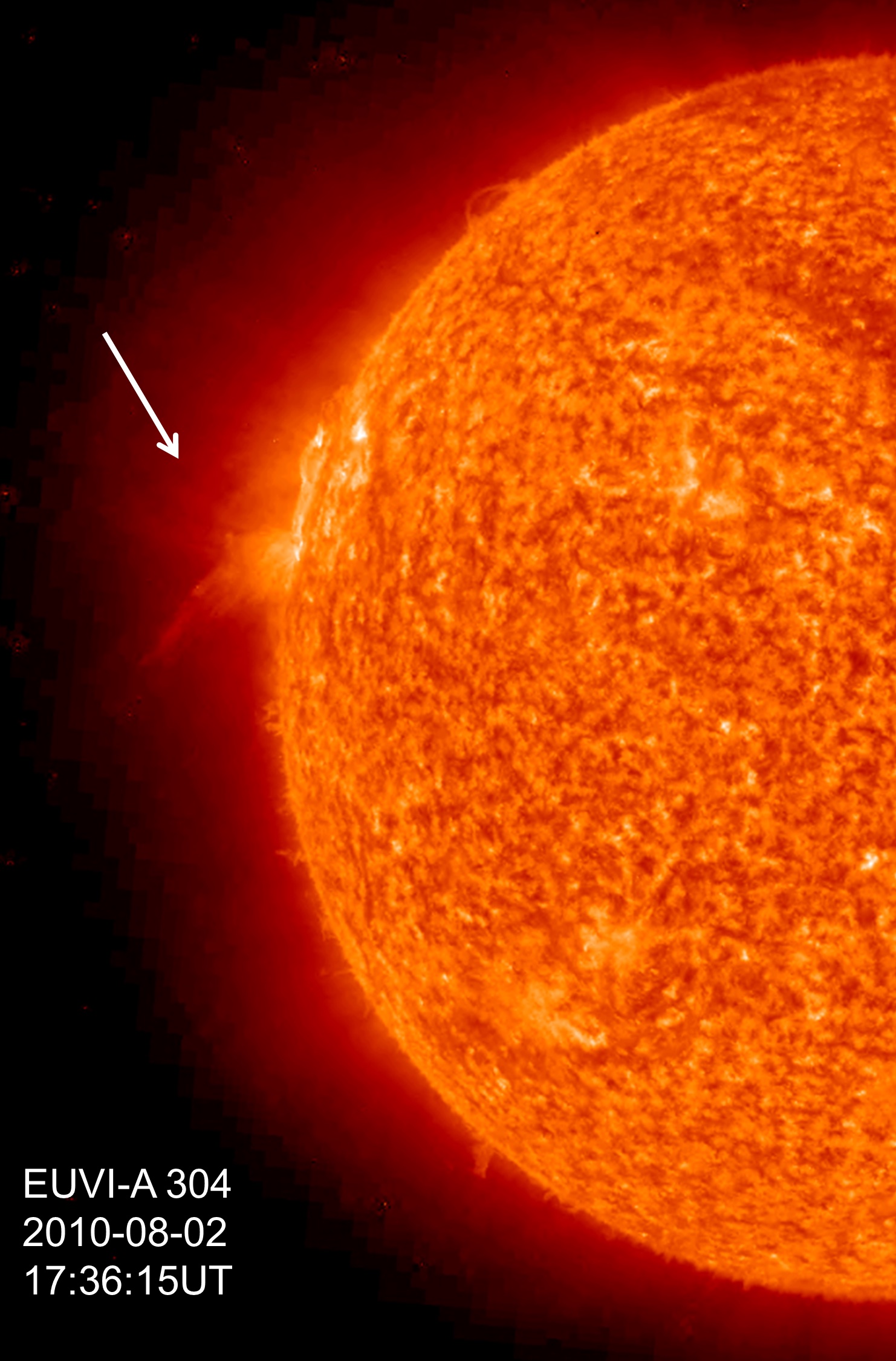}
\includegraphics[width=5.4cm]{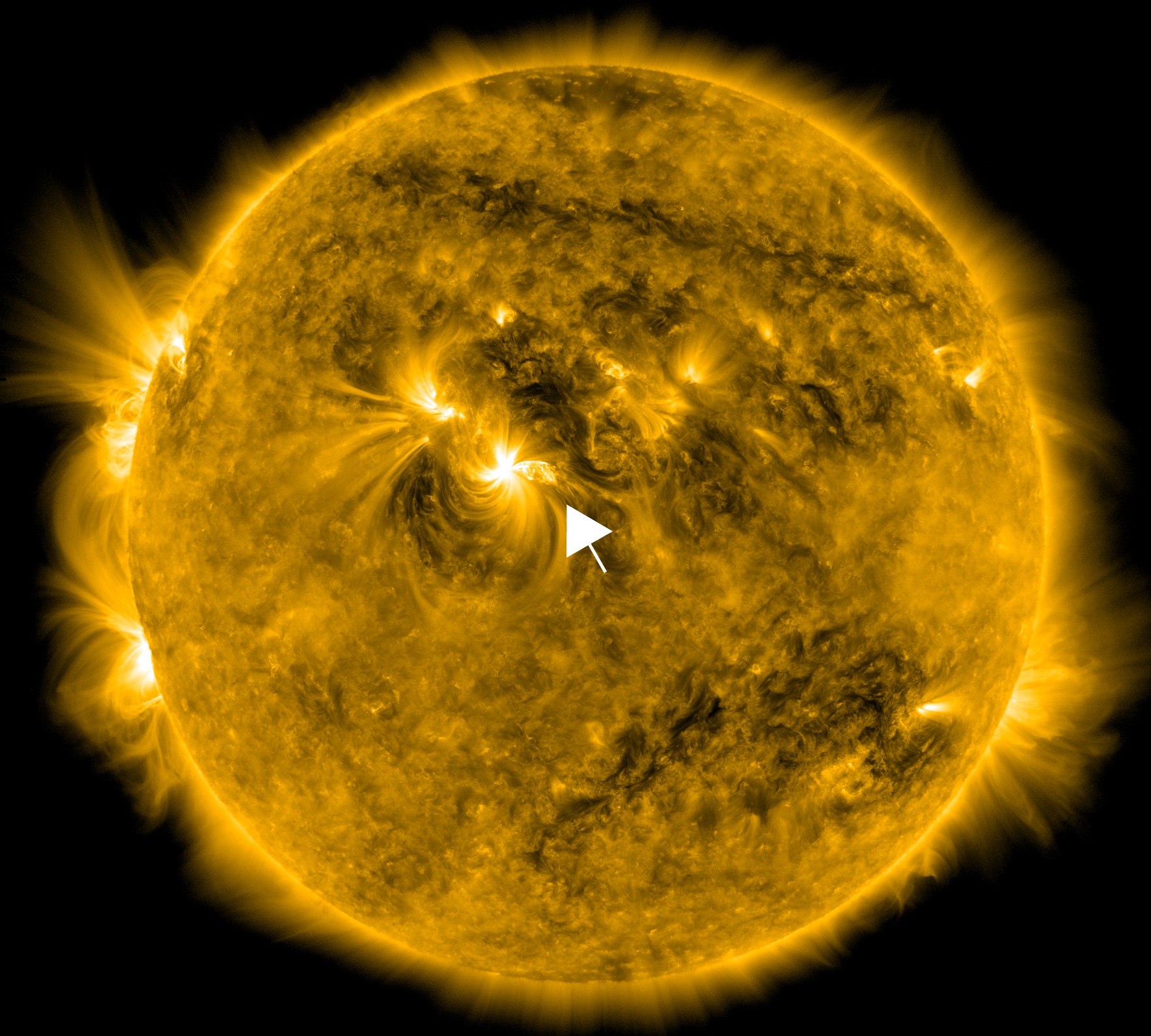}
\includegraphics[width=3.25cm]{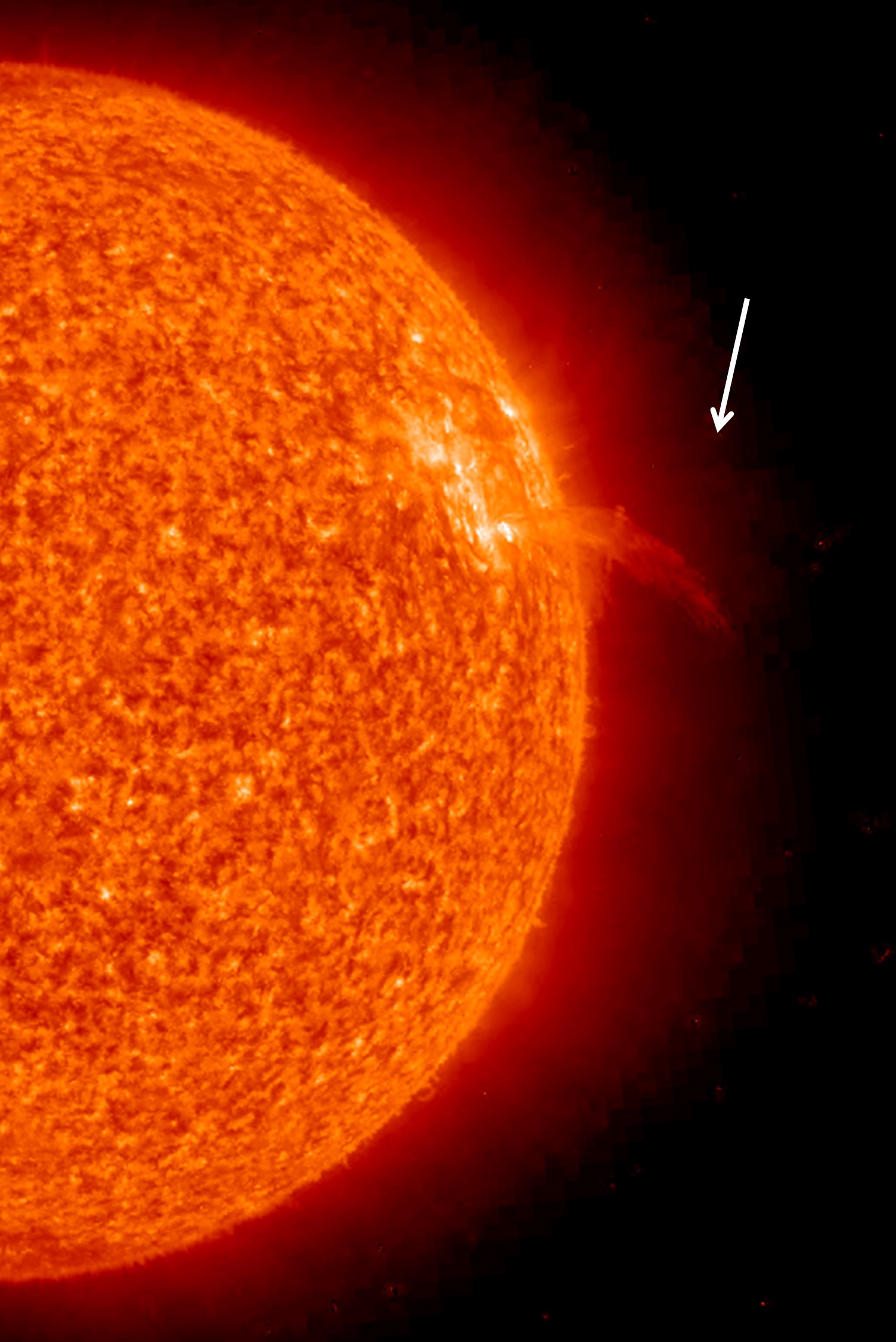}
\caption{Images taken by STEREO/EUVI A and B in the 304 channel (left, right) capturing the jet on August 2 at 17:35 UT as seen at the solar limb (white arrow). Middle: the Sun on the same day at 17:27 UT as seen by SDO/AIA 171 channel (the jet is indicated by the white arrow).}
\label{fig:sdo_context}
\end{figure*}

\begin{figure*}[!htb]
\includegraphics[width=0.335\textwidth ]{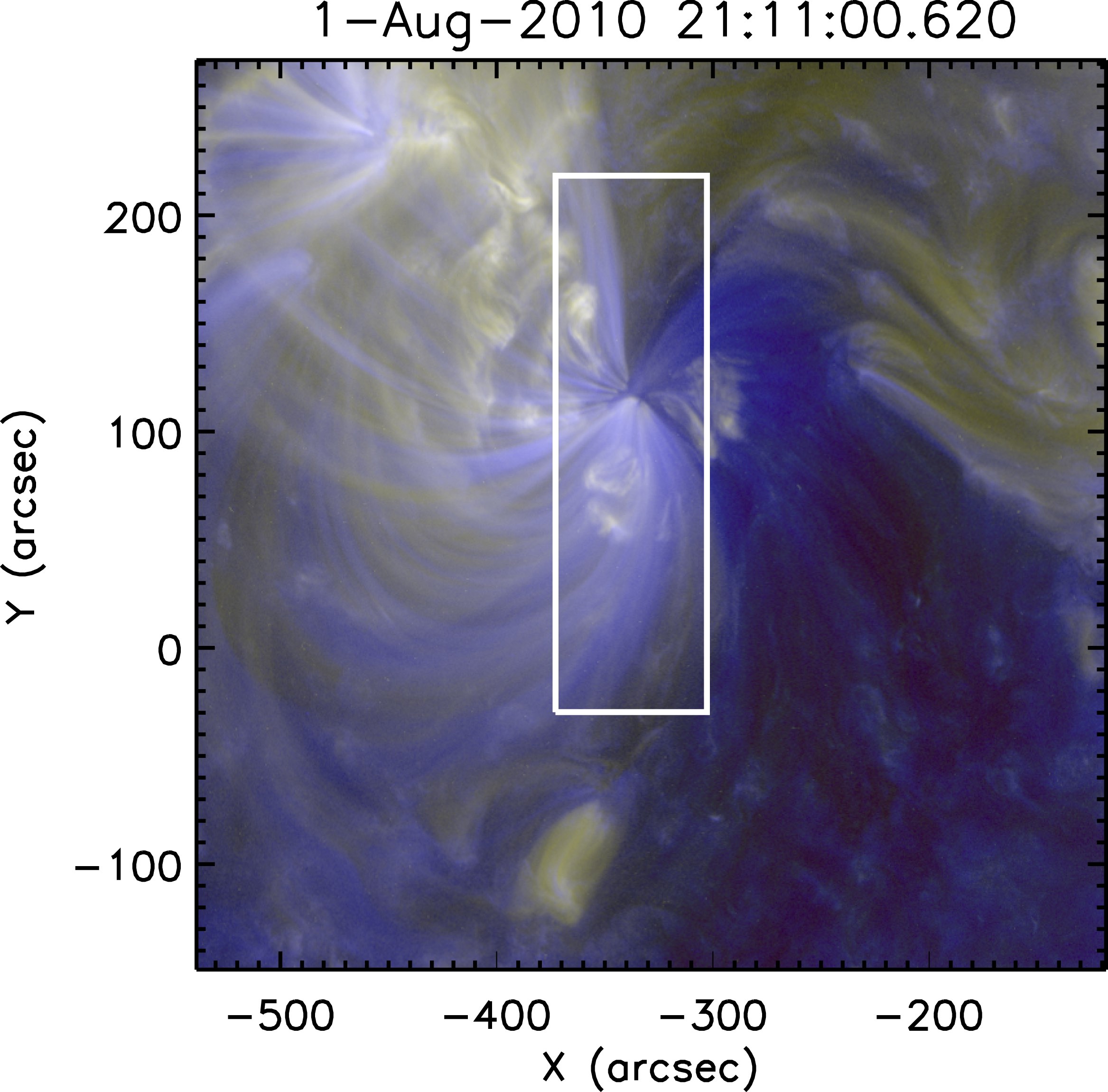}
\includegraphics[width=0.314\textwidth, trim = 1.0in 0 0 0, clip]{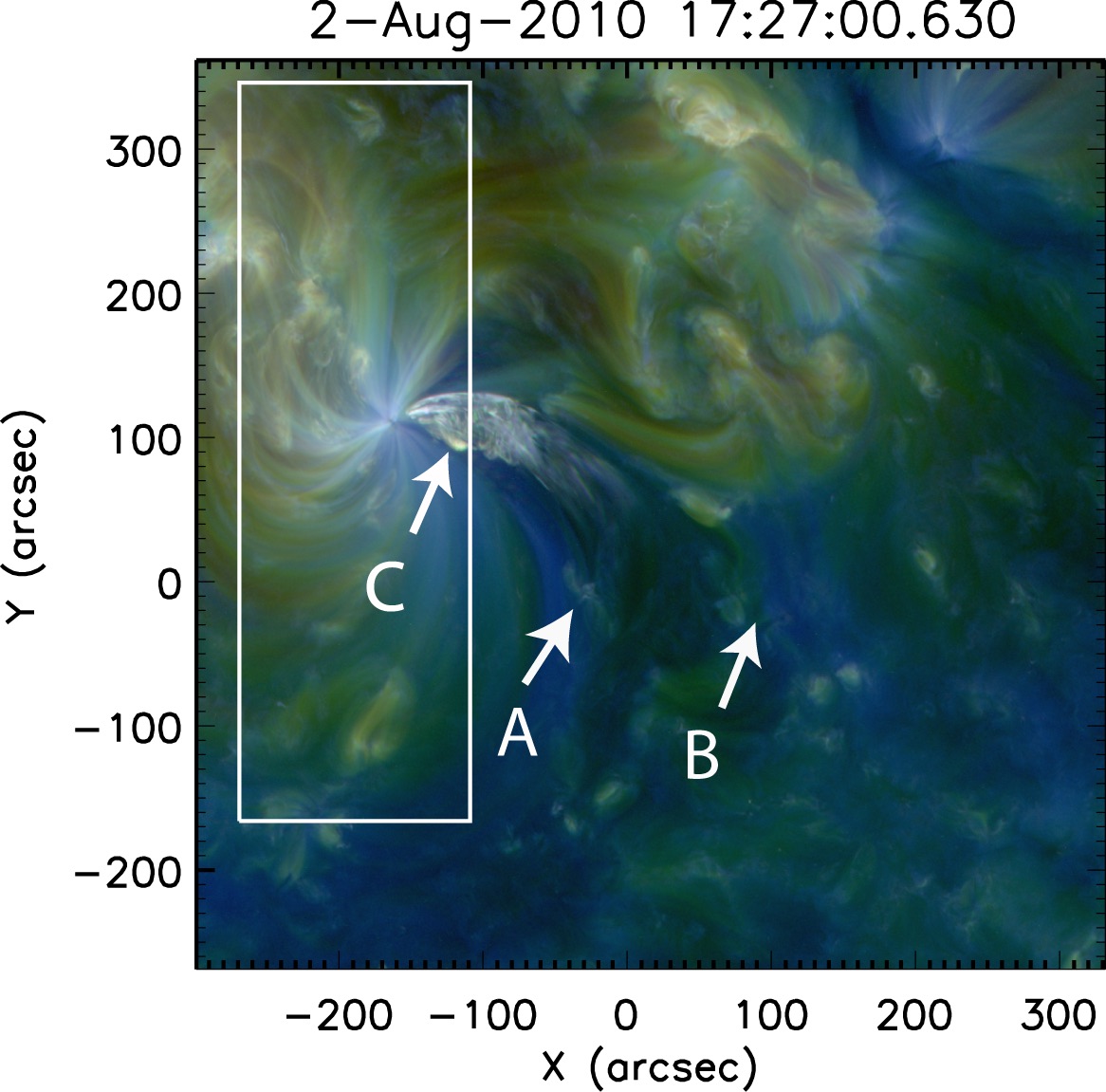}
\includegraphics[width=0.32\textwidth, trim = 1.45in 0 0 0, clip]{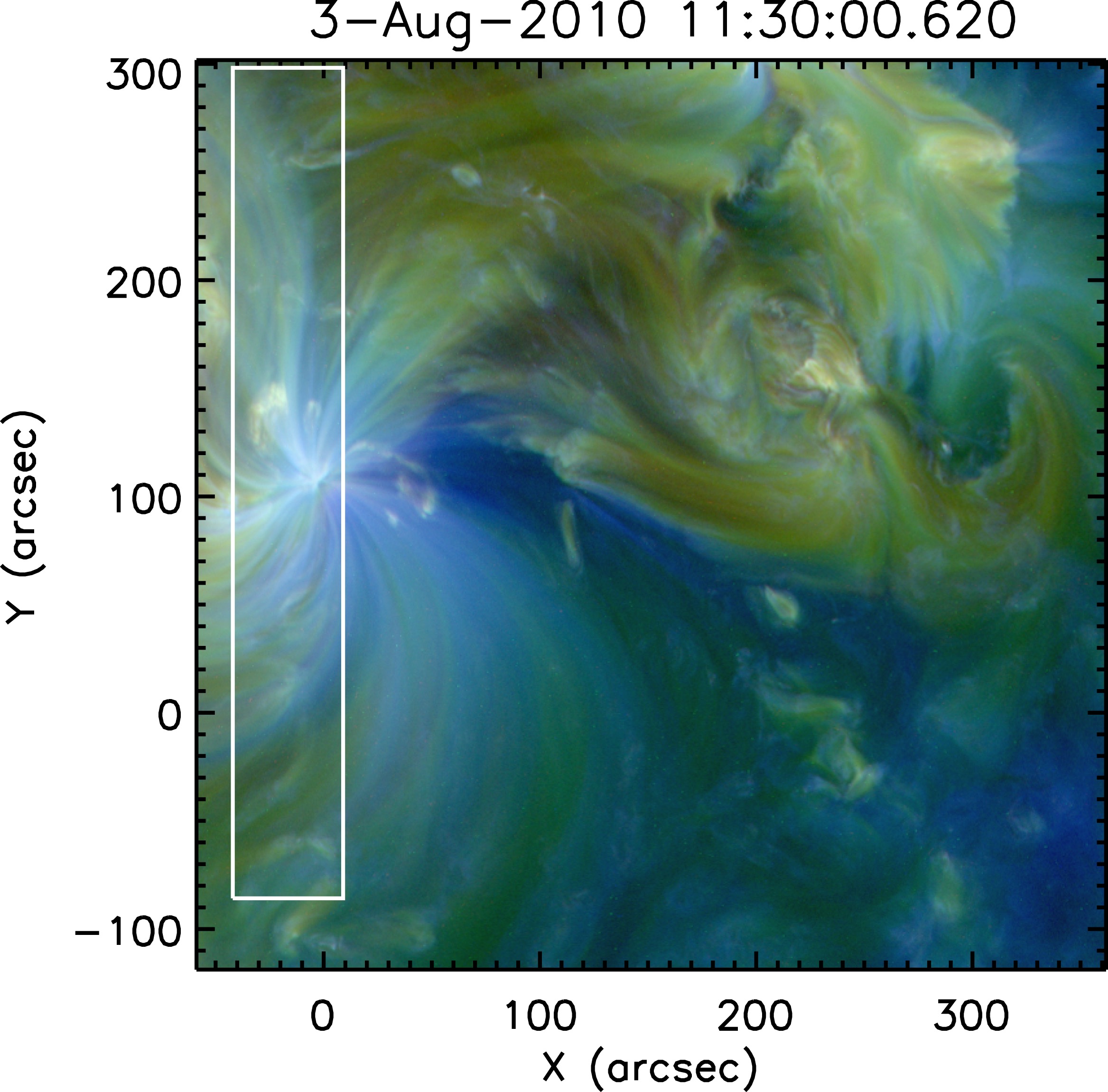}
\caption{A zoom in of the AR as seen in the superposition of the SDO/AIA 211 (red), 193 (blue) and 171 (green) channels. 
From left to right: 1st, 2nd and 3rd of August. The white boxes represent the EIS FOV of the different datasets.
Middle panel: on the right side of the jet (C arrow) is the separatrix dome surface (arrow B, see text). Labeled with A is one of the footpoints of the loops at the origin of the jet. The other footpoint is at the spot.}
\label{fig:aia_context}
\end{figure*}

{To better understand the evolution of the AR environment before, during and after the eruption, we show some context images in Figures \ref{fig:sdo_context} and \ref{fig:aia_context}. 
The jet occurred on August 2, 2010  in the AR 11092 between 17:10 and 17:35 UT, as seen in the AIA 171 channel and STEREO (Solar TErrestrial RElations Observatory) EUVI (Extreme UltraViolet Imager) (Figure \ref{fig:sdo_context}). 

The pre--erupting corona is instead shown in the left panel of Figure \ref{fig:aia_context}. Here we superposed three images from AIA, representing roughly the corona at 0.9 MK (green, 171 channel), 1.5 MK (blue, 193 channel) and 2 MK (red, 211 channel).
The image shows the AR the day before the eruption at close time when two of the EIS rasters used for our analysis were taken ('A' and 'B', see Table \ref{tab:uv_data}).
The multi color image reflects the multi-temperature property of the AR. The half left side of the AR is made of  low lying 'hot' ($\mathrm{T = 3 ~MK}$) loops and 'warm' ($\mathrm{T = 1 ~MK}$) loops arcade above them. The north west side is made of mostly hot loops while the mid-south part contains low, large scale warm loops. This latter is probably a low density and low temperature area with a magnetic structure which opened up during the jet. This is the escaping channel of the jet.

This jet was analysed in detail by \cite{mulay16} who studied its possible relation with the observed interplanetary radio type-III burst using WIND data. }
The authors identified the footpoints of the jet in the spot and in one of the small loops just at the west side of it (labeled 'C' in Figure \ref{fig:aia_context}). Among the observed signatures of the dynamics in this area, they noticed flux emergence and cancellation {prior to the eruption}, cool H$\alpha$ material ejected at the same time of the coronal material, and hard HXR emission from RHESSI (Reuven Ramaty High Energy Solar Spectroscopic Imager, see their Figure 5). These are clear signatures of ongoing magnetic reconnection and plasma ejection.

Figure \ref{fig:aia_context} right shows the AR the day after the jet, at the time when the EIS raster named 'D' was taken. We notice on the west side that the area which had a void and small scale loops has changed and filled by warm loops. We will show in Section \ref{sec:loc1-3} that some plasma is escaping from this open magnetic field area. The image also shows that the dome structure on the right side of the active region (structure 'B' in the Figure) is still there after the jet. 

On 2 August, some activity in the AR started few hours before the jet.  
This was in the form of plasma flows and brightening along the loops having the footpoints at the spot in the area labeled 'A' in Figure \ref{fig:aia_context}. During the eruption, the {southern} part
of the jet moved and expanded  toward north-west, interacting with the dome-shaped loops system and its spine (arrow 'B' in the figure). This can also be observed from the vantage point of STEREO EUVI-A and B images, as shown in Figure \ref{fig:sdo_context}, where the material is released near the limb. In this configuration, STEREO A and B were at separation angles of 71 and 78 degrees, respectively, from the Earth. Due to the location of the jet, the observed opening of field lines, and subsequent release of material, this jet was selected as a good candidate to possibly reach into the heliosphere and be detected by the {\it in--situ} instruments.

In the following sections we will describe our attempt to fully understand if this jet material reached the interplanetary medium. We discuss the plasma diagnostics of the AR performed during the period 1-3 August 2010 and the analysis of the magnetic field configuration before and after the jet, at small and large scales. As part of our comprehensive effort to connect the jet, we also modeled the element fractionation within the open and closed regions in order to connect remote sensing and {\it in--situ} measurements. We will discuss the search for its signatures within the {\it in--situ} measurements in Section \ref{res:insitu}.

\section{Methods: Building Blocks for Connectivity}
\label{sec:methods:gen}

In this section we present the modeling and observational methods used in our attempt to establish the connection between the Sun and the heliosphere and track distinct coronal plasma from its solar origin into interplanetary space. 
In  Sect.~\ref{methods:field}, we review different approaches to modeling the magnetic field in the extended solar atmosphere, including the potential field source surface (PFSS) model, non-linear force free field (NLFFF) modeling, and steady-state magnetohydrodynamic (MHD) modeling. 
In  Sect.~\ref{methods:solarwind:fip} we present the model for elemental abundance variations.
In Sect.~\ref{sec:UVgen}, we review UV spectroscopic techniques and the various plasma diagnostics to determine densities, temperatures, and elemental abundances in the solar atmosphere. 
In Sect.~\ref{methods:insitu}, we discuss the {\it in--situ} observational techniques for measuring ionic and elemental abundances in the solar wind and the heliospheric back-mapping procedure to estimate their solar origin.

\subsection{Modeling the Coronal Magnetic Field Structure}
\label{methods:field}


\subsubsection{Global PFSS Modeling from Synoptic Line-of-sight GONG and SDO/HMI Magnetograms}
\label{methods:field:pfss}

The potential field source surface (PFSS) model has been one of the most reliable and robust tools for estimating the large-scale magnetic structure of the solar corona \citep{Altschuler1969, Schatten1969, Hoeksema1991, WangYM1992}. In the potential field approximation the magnetic field is derived from
\begin{eqnarray}
\nabla \times {\bf B} & = & {\bf 0}, \\
\nabla \cdot{\bf B}    & = &         0,
\end{eqnarray}
which yields a solution of the form ${\bf B} = -\nabla {\rm \Psi}$ where $\nabla^2 {\rm \Psi} = 0$. The solution is obtained for $R_{\odot} \le r \le R_{\rm ss}$ where the radial field at $r=R_\odot$ is taken from photospheric magnetogram observations and the source-surface, $R_{\rm ss}$, is the distance where the magnetic field is assumed to have become purely radial, typically $R_{\rm ss}=2.5R_\odot$ {\citep{Hoeksema1983}}. Line-of-sight photospheric observations are used to approximate the radial field at the central meridian with the $B_r = B_{\rm los}/\sin{\theta}$ relation. The scalar potential function ${\rm \Psi}$ can then be solved via a number of techniques and is often represented with a spherical harmonic expansion with the coefficients calculated in the usual fashion from the $B_r(R_\odot,\theta,\phi)$ boundary condition.

While the real solar corona is realistically not current-free, the PFSS model continues to be extremely useful and widely used for estimating the large-scale magnetic field structure of the corona \citep[e.g.][]{Hill2018, DeRosa2018}, detailed topological analyses of magnetic flux systems over a wide range of spatial scales \citep{Antiochos2011, Titov2011,  Scott2018}, for analyzing active region energization \citep{Welsch2016}, as the backbone of semi-empirical solar wind descriptions \citep{Arge2004}, and as inputs for solar and heliospheric MHD modeling \citep{Odstrcil2004, Toth2011, Pomoell2018} as well as to serve as initial condition for global NLFFF modeling (see section \ref{tw_nlfff}).

{ 
In this work, each of the PFSS extrapolations are calculated from their respective magnetogram synoptic maps using the formulas given in \citet{ZhaoXP1993} for the current-free case. We use a spherical harmonic expansion through degree $\ell = 25$ on the MHD grid (described below) and interpolated to a uniform $256 \times 512 \times 1024$ grid over $r/R_\odot \in [ 1, 3 ]$, $\theta \in [ 11.25^\circ, 168.75^\circ ]$, $\phi \in [0^\circ, 360^\circ ]$. We have used the standard source surface height of $R_{ss}=2.5R_\odot$ herein, but we note that changing the $R_{ss}$ value can modify the PFSS open flux regions. Several recent studies have investigated this issue when comparing PFSS results with the observed locations and/or sizes of coronal holes \citep[e.g.][]{Linker2017,Riley2019,Badman2020}.}
{The results of our PFSS extrapolations are presented in Sect. \ref{sec:GLB_res}.}

\subsubsection{Global NLFFF Modeling from SDO/HMI Synoptic Vector Magnetograms}
\label{tw_nlfff}

The solar corona is considered to be force-free, see
\cite{2012LRSP....9....5W, wiegelmann2021} for a review. This means that,
due to the low plasma $\beta$ in the solar corona,
plasma forces can be neglected in lowest order and the Lorentz force vanishes.
In this approximation one has to solve
the nonlinear force-free field (NLFFF) equations
\begin{eqnarray}
(\nabla \times {\bf B }) \times{\bf B} & = & {\bf 0}, \\
\nabla \cdot{\bf B}    & = &         0,
\end{eqnarray}
which we solve in spherical geometry by minimizing the functional:
\begin{eqnarray}
L({\bf B}) & = & \int_{V}   \left[B^{-2} \, |(\nabla \times {\bf B}) \times {\bf
B}|^2 +|\nabla \cdot {\bf B}|^2\right] \, r^2 \, \sin \theta \, dr \, d \theta \, d \phi \nonumber \\
& & + \int_{S}\, ({\bf B} - {\bf B_{\rm obs}}) W(\theta, \phi) ({\bf B} - {\bf B_{\rm obs}})
\, r^2 \, \sin \theta d \theta \, d \phi
\label{def_L},
\end{eqnarray}
where the surface integrals contains the observed vector magnetogram
${\bf B_{\rm obs}}$, here synoptic vector maps from SDO/HMI (Solar Dynamics Observatory, Helioseismic and Magnetic Imager).
 We developed an optimization code to minimize the functional (\ref{def_L}).
As initial condition for the iterative minimization of the functional  a
PFSS model (see section \ref{methods:field:pfss}) is used.
The method has been originally proposed in Cartesian geometry by
\cite{2000ApJ...540.1150W}. Here we use the spherical optimization code
as originally developed in \cite{2007SoPh..240..227W} and adjusted
for the use of synoptic vector maps in \cite{2014A&A...562A.105T}.
As a further refinement we use a 3-level multi-scale approach here
to minimize (\ref{def_L}). More details on the application of the code
to synoptic vector magnetograms and a comparison with other global coronal
magnetic field models can be found in \cite{2018SSRv..214...99Y}.
{ 
For our simulations the computational grid is $180 \times 280 \times 720$ in $(r,\theta,\phi)$ over the domain $r/R_\odot \in [1, 2.5]$, $\theta \in [ 20^\circ, 160^\circ ]$, $\phi \in [ 0^\circ, 360^\circ ]$. The results are presented in Sect. \ref{sec:GLB_res}.
}


\subsubsection{Global Steady-state MHD Modeling}
\label{methods:field:mhd}

For the global steady-state MHD modeling, we use the Adaptively Refined MHD Solver (ARMS), developed by \citet{DeVore2008} and collaborators. ARMS calculates solutions to the 3D nonlinear, time-dependent MHD equations that describe the evolution and transport of density, momentum, and energy throughout the system, and the evolution of the magnetic field and electric current. The numerical scheme used is a finite-volume, multidimensional flux-corrected transport algorithm \citep{DeVore1991}. The ARMS code is fully integrated with the adaptive-mesh toolkit PARAMESH \citep{MacNeice2000}, to handle dynamic, solution-adaptive grid refinement and enable efficient multiprocessor parallelization.
ARMS has been used to perform a wide variety of
numerical simulations of dynamic phenomena in the solar atmosphere,
including CME initiation \citep{Karpen2012, Dahlin2019, Masson2019}, coronal jets \citep{Pariat2015, Wyper2017, Karpen2017} 
and interchange reconnection dynamics \citep{Edmondson2010a,Lynch2014,Edmondson2017,Higginson2017a,Higginson2017b,Higginson2018}.

ARMS solves the ideal MHD equations in spherical coordinates,

\begin{eqnarray}
    \frac{\partial \rho}{\partial t} + \nabla \cdot \left( \rho
    \bf{V} \right) & = & 0 \; , 
    \label{eq1} \\  
    \frac{\partial}{\partial t} \left( \rho \bf{V} \right)
    + \nabla \cdot \left( \rho \bf{V} \bf{V} \right)
     & = & \frac{1}{4\pi}\left( \nabla \times \bf{B}
    \right) \times \bf{B} - \nabla P - \rho \bf{g} \; , 
    \label{eq2} \\
    \frac{\partial T}{\partial t} + \nabla \cdot \left( T
    \bf{V} \right) & = & (2 - \gamma) T \nabla \cdot {\bf V} \; , 
    \label{eq3} \\
    \frac{\partial \bf{B}}{\partial t} & = & \nabla \times
    \left( \bf{V} \times \bf{B} \right) \;, 
    \label{eq4}
\end{eqnarray}
where all the variables retain their usual meaning, solar
gravity is ${\bf g} = g_\odot (r/R_\odot)^{-2} {\bf \hat{r}}$
with $g_\odot = 2.75\times10^4$~cm~s$^{-2}$, and we use the ideal
gas law $P=2(\rho/m_p)k_BT$.  
To obtain a basic isothermal solar wind outflow, we set $\gamma=1$ and do not solve the temperature equation. The plasma
temperature remains uniform throughout the domain for the duration
of the simulation ($\partial T/\partial t =0$).
The solar wind is initialized in ARMS by first solving the
one-dimensional \citet{Parker1958} equation for an isothermal coronal
atmosphere,
\begin{equation}
    \frac{V_{\rm sw}^2}{c_0^2} - \ln\left( \frac{V_{\rm sw}^2}{c_0^2}
    \right) = -3 + 4 \ln\left( \frac{r}{r_c} \right) + 4 \frac{r_c}{r},
\end{equation}
which is characterized by the base number density $n_0$, pressure $P_0$, and temperature $T_0$.
Thus, 
$c_0 = (2 k_B T_0 / m_p)^{1/2}$ 
is the thermal velocity at $T_0$ and the location of the critical point is
$r_c = G M_\odot / 2 c_0^2$.
The MHD simulation starts with the PFSS initial state for magnetic field everywhere in domain, an initial spherically-symmetric density structure of the form $\rho(r) = \rho_0 (r/R_\odot)^{-\mu}$ where $\mu = m_p g_\odot R_\odot/2 k_B T_0$. We then apply the Parker solution outflow and let MHD system relax toward a quasi steady-state  \citep[see description in][]{Lynch2016b}. 
This yields an open--closed magnetic field structure that is more physical than the PFSS extrapolations because of the force balance between the solar wind plasma outflow and the magnetic field tension in the closed flux systems. However, we note that {the isothermal $\gamma=1$ condition generates} the simplest possible MHD wind scenario, {which is done to try to minimize the required computational resources. There are other,} more sophisticated {thermodynamic MHD treatments that include} sources and sinks in the internal energy equation such as thermal conduction, radiative losses, and various coronal heating parameterization \citep[e.g.][]{Lionello2014,vanderHolst2014,Oran2017}.

{ 
For our simulation 
the isothermal solar atmosphere was initialized with a spherically symmetric density profile with a base density of 
$n_0 = \rho_0/m_p = 1.0 \times 10^9$~cm$^{-3}$, 
$P_0 =0.276$~dyn~cm$^{-2}$, and 
$T_0 = 1.0\times10^6$~K.
This yields a sound speed of
$c_0 = 128.5$~km~s$^{-1}$, the critical point of the initial Parker solar wind at
$r_c = 5.74R_\odot$, and a solar wind speed at the outer boundary of $V_{\rm sw}(30R_\odot) \sim 350$~km~s$^{-1}$. The MHD computational grid utilizes block decomposition in logarithmic spherical coordinates with three levels of refinement above the level~1 grid of $8 \times 8 \times 16$ blocks over $r/R_\odot \in [1, 30]$, $\theta \in [ 11.25^\circ, 168.75^\circ ]$, and $\phi \in [ 0^\circ, 360^\circ ]$. Each computational block contains $8^3$ cells. The level~3 grid, therefore, represents a $256 \times 256 \times 512$ resolution in $(r,\theta,\phi)$ over $r/R_\odot \in [1, 5.48]$, $\theta \in [ 11.25^\circ, 168.75^\circ ]$, and $\phi \in [ 0^\circ, 360^\circ ]$. The level~4 grid covers the location of the August~02 jet, $r/R_\odot \in [1, 2.34]$, $\theta \in [ 50^\circ, 114.75^\circ ]$, and $\phi \in [ 0^\circ, 90^\circ ]$ with an effective resolution of $512 \times 512 \times 1024$ over that region.  
Results from {our MHD modeling} are used in Sec. \ref{sec:GLB_res}.
}


\subsection{Modeling the FIP Fractionation and Elemental Composition} 
\label{methods:solarwind:fip}

A difference in the elemental composition of the solar corona compared to that of the underlying photosphere was first observed by \citet{pottasch63}. 
As presented in Section \ref{intro}, since this first
observation, this so-called ``FIP Effect'' has also been observed in various regimes of the
solar wind \citep[e.g.][]{bochsler07}, in solar energetic particles (SEPs) \citep[e.g.][]{reames18}, and in the coronae of other late-type stars \citep[e.g.][]{drake97}.
The elements that are enhanced are those that are predominantly ionized in the solar
chromosphere, which indicates the likely location of the ion neutral separation.

Since the original observation of FIP fractionation, a number of key features of its variation 
with solar coronal region or solar wind regime have become established, allowing the abundance
pattern a certain amount of diagnostic potential. Coronal holes, and the fast solar wind
emanating from them, have a relatively small FIP fractionation, by a factor typically of 1.5.
Slow solar wind and the closed loop solar corona {usually} have larger fractionations, as mentioned above by a factor $\sim 3-4$ {\citep[e.g.][]{bochsler07}}. However subtle differences remain. The closed loop solar corona
does not show an appreciable fractionation of S, whereas frequently the slow speed solar wind does. The element \ion{S}{} (FIP 10.36 eV) is formally a high FIP element, but among high FIP elements has the lowest FIP. This difference also shows up in abundances measured in gradual SEP events compared to those measured in accelerated particles in co-rotating interaction regions
\citep{reames18}, with the implication that seed particles for shock acceleration in gradual SEP events originate in the solar corona, whereas those for particle acceleration in 
co-rotating interaction regions are swept directly out of the ambient solar wind.

Further variations in fractionation are also seen in the coronae of active stars, and more recently in some solar flares, where the FIP effect is inverted. This depletion of the low-FIP element abundances has become known as the Inverse FIP or I-FIP Effect, and posed a particular
challenge to understanding the phenomenon. While various mechanisms might be able, qualitatively at least, to accelerate chromospheric ions into the corona in preference to
chromospheric neutrals, the reverse is more difficult to envisage. One mechanism of fractionation
that does offer the possibility to act in either direction, and which also captures the variations
in fractionation outlined above is that due to the ponderomotive force \citep[e.g.][]{laming04,laming09,laming12,laming15}. This arises due to the effects of wave reflection and refraction as Alfv\'en waves propagate through the chromosphere. The chromospheric wave field depends on the 
coronal magnetic field structure above it, hence the difference between open and closed fields, but the fractionation also depends on local chromospheric physics.

The instantaneous ponderomotive acceleration, $a$,
acting on an ion is evaluated from the general form \citep[see e.g. the appendix
of][]{laming17}
\begin{equation}
a={c^2\over 2}{\partial\over\partial z}\left(\delta E^2\over B^2\right)
\end{equation}
where $\delta E$ is the wave (transverse) electric field, $B$ the ambient (longitudinal) magnetic field,
$c$ the speed of light, and $z$ is a coordinate along the magnetic field.

Given the ponderomotive acceleration, element fractionation is calculated
using input from the chromospheric model \citep{laming17}
\begin{eqnarray}
\nonumber f_k&=&{\rho _k\left(z_u\right)\over\rho _k\left(z_l\right)}\\ &=&\exp\left\{
\int _{z_l}^{z_u}{2\xi _ka\nu _{kn}/\left[\xi _k\nu
_{kn} +\left(1-\xi _k\right)\nu _{ki}\right]\over c_0^2/A_k+v_{||,osc}^2+2v_{sw,k}^2}dz\right\}.
\end{eqnarray}

This equation is derived from the momentum equations for ions and neutrals in
a background of protons and neutral hydrogen. Here $\xi _k$ is the element
ionization fraction, $A_k$ the element atomic mass number, $\nu _{ki}$ and $\nu _{kn}$ are collision frequencies of
ions and neutrals with the background gas \citep[mainly hydrogen and protons,
given by formulae in][]{laming04}, $c_0^2/A_k=2k_{\rm B}T/m_k \left(
=2v_z^2\right)$ represents twice the square of the element thermal velocity along
the $z$-direction, $v_{sw,k}$ is the upward flow speed and $v_{||,osc}$ a
longitudinal oscillatory speed, corresponding to upward and downward
propagating sound waves. At the top of the
chromosphere where background \ion{H}{} is becoming ionized $\nu _{ki}>>\nu _{kn}$,
and small departures of $\xi _k$ from unity can result in significant
decreases in the fractionation. This feature is important in inhibiting the
abundance enhancements of \ion{S}{} and \ion{P}{} at the top of the chromosphere
\citep{laming19}. These are the high-FIP elements with the lowest FIP (10.36 and 10.49 eV respectively) and
reach high levels of ionization ($\sim 0.8$) in this region. Lower down where
the H is neutral this inequality does not hold, and these elements can become fractionated.
Results from this model are presented in Section \ref{sec:res_fip_charge}. 

%
\subsection{Coronal Observations and EUV Plasma Diagnostics}
\label{sec:UVgen}

The solar conditions for the three days period of interest were investigated using Hinode/EIS (EUV imaging spectrometer) UV data \citep{culhane07}. {Several  observations are summarized in Table \ref{tab:uv_data}}. The EIS studies had different properties but they were mostly centered in the spot, as shown in Figures \ref{fig:aia_context}, \ref{fig:fe12_i}, \ref{fig:aia_eis_full} and \ref{fig:eis_full}.

For 1 August, the analysis of two different datasets has provided a quite complete picture of the pre-eruption AR conditions. In particular, a full-spectrum (dataset B) has been used for the FIP bias diagnostics. The dataset taken on 2 August (dataset C), contains a very limited number of lines which could only be used for Doppler maps in Fe XII. These, however, provide us with temporal information. Density and Doppler maps could be obtained also for 3 August (dataset D). 

The EIS data have been processed using the standard software available on the $Solarsoft$ EIS database, unless differently stated. The spectral line profiles and total intensities were measured assuming a Gaussian profile using the \cite{delzanna13} radiometric calibration.

The plasma diagnostics were done using the CHIANTI v.8 atomic and software database \citep{dere97, dere19}, unless  differently mentioned. Synthetic line intensities
were calculated assuming ionization equilibrium and photospheric composition \citep{asplund09}.
Density maps of the observed regions were obtained applying the line ratio technique to \ion{Fe}{xiii}  \citep{parenti15} diagnostics (see Section \ref{sec:uv_res}). The Differential Emission Measure distribution (DEM) was derived using the (Del Zanna, 1999) method which provides the thermal structure distribution of the plasma integrated along the line of sight. The FIP bias analysis was done using two different methods: the DEM and LCR (see Sections \ref{sec:eis_full}, \ref{se:LCR}).

\begin{table*}[th]
    \centering
    \begin{tabular}{|c|c|c|c|c|c|c|}
    \hline
     Inst.& Date & Start  & End & Set & Center & FOV\\ 
          &      &  [UT]  & [UT] &    &  [x$^{\prime\prime}$, $y^{\prime\prime}$] &[$x^{\prime\prime}$, $Y^{\prime\prime}$]\\
    \hline    
      EIS & 1/8/2010 & 21:05:24 & 21:15:05 &A &   -337.8,  94.3 & 69.9, 248.0\\
      EIS & 1/8/2010 & 23:39:35 & 00:41:28 &  B& -315, 93.3 & 60.0, 512.0\\
      EIS & 2/8/2010 & 10:51:02 & 11:36:08&  C & -219.3, 90.0  & 159.7,512.0 \\
       EIS & 2/8/2010 & 14:13:01&14:58:06 & C & -189.8, 89  &  159.7, 512.0\\
      EIS & 3/8/2010 & 11:28:21 & 11:46:44 &  D & -16.5,  106.2 & 50.9, 384.0\\
     \hline 
    \end{tabular}
    \caption{EUV HINODE/EIS data used in the paper.}
    \label{tab:uv_data}
\end{table*}

%
%

%
\subsection{Connecting Solar to the {\it in--situ} Observations}
\label{methods:insitu}

In this section we describe the technique we used for another important block for our Heliophere-Sun connectivity, that is the  procedure which backs-map the {\it in--situ} measures to a location on the solar surface. 

The heliospheric structure of the solar wind's Parker spiral \citep{Parker1958} can be estimated through integrating the velocity streamlines associated with the observed radial speed measurements at 1~AU. Assuming the observed radial speed is constant, this yields
\begin{eqnarray}
v_r(r) & = & V_r(1~AU) \\
v_\phi(r) & = & -\Omega_\odot \left( r - R_\odot \right)
\end{eqnarray}
where the solar rotation rate $\Omega_\odot = 2.87 \times 10^{-6}$ rad~s$^{-1}$. This ballistic back-mapping projects the time series of plasma observed at 1~AU (which has an equivalent longitude in each Carrington Rotation) backwards in time to an earlier radial location, taken here to the be potential field source surface at $r=2.5R_\odot$. The resulting Carrington longitude of the velocity streamline origin has an equivalent date and time corresponding to the time the plasma parcel was released from the Sun. 
From the inner boundary of the ballistic back-mapping, a coronal magnetic field model or extrapolation (e.g. PFSS, NLFFF, MHD) can then be used to continue the mapping from the corona to the solar surface.
This ballistic mapping method has been applied to Ulysses data and discussed by \citet{Neugebauer2002, Neugebauer2004} and variations of this approach continue to be used to analyze solar wind source regions \citep[e.g.][]{Gibson2011, ZhaoL2013a, ZhaoL2013b, ZhaoL2017a, Badman2020}. Note that the overall uncertainty in the position of the footpoint of the magnetic field lines as mapped by these techniques are within approximately 10$^\circ$ \citep{Neugebauer2002, Leamon2009}. Results for this analysis are given in Section \ref{res:insitu}.


\section{Results}
\label{sec:results}

\subsection{Global Magnetic Field Configuration for 2010~August~02}
\label{sec:GLB_res}

\begin{figure*}[!htb]
    \includegraphics[width=0.99\textwidth]{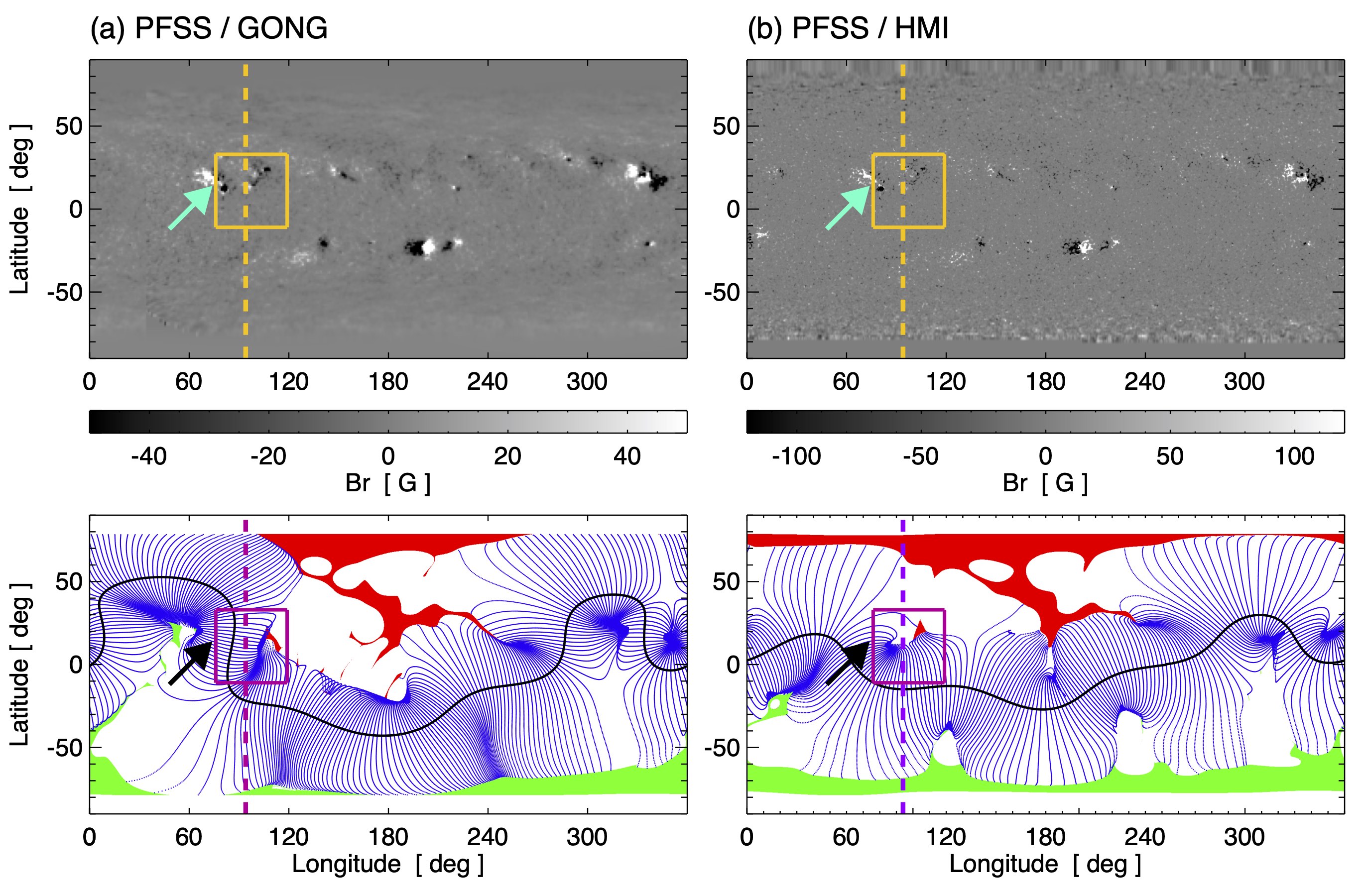}
    \includegraphics[width=0.99\textwidth]{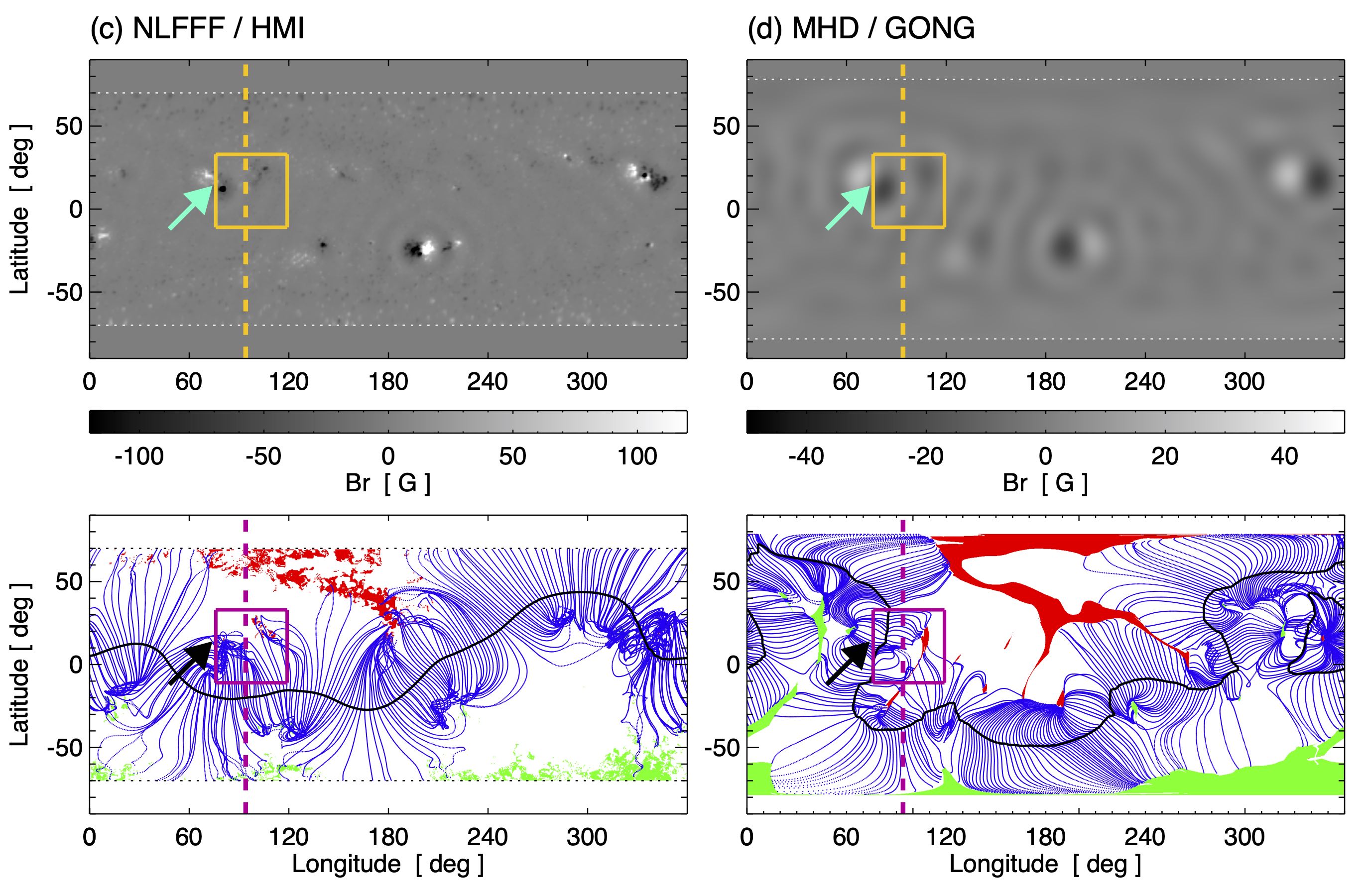}
    \caption{Comparison of the global magnetic structure during CR2099. Panel (a) PFSS extrapolation based on the GONG synoptic map. Panel (b) PFSS extrapolation based on the HMI radial field synoptic map. Panel (c) NLFFF extrapolation from HMI vector data. Panel (d) steady-state MHD solution initialized from the GONG PFSS extrapolation. The dashed line indicates the central meridian on 2~Aug 2010 {while the rectangle shows the field of view in Figure~\ref{fig:local_mag}. The arrow indicates AR 11092}. The upper plot in each panel shows either the observed or modeled radial field distribution at $r=R_\odot$ and the lower plot shows the map of open field regions (red negative polarity, green positive polarity) and structure of the helmet streamer belt (blue field lines). The black line is the $B_r=0$ contour at $r=2.5R_\odot$.}
    \label{fig:global_mag}       
\end{figure*}

We apply each of the magnetic field modeling techniques described in Section~\ref{methods:field} to various synoptic magnetogram data during Carrington Rotation 2099 (observed from 13~July through 08~August) which encompasses the 2010~August~02 coronal jet.
Figure~\ref{fig:global_mag} shows the results of the global PFSS, NLFFF magnetic extrapolations and steady-state MHD modeling, co-aligned such that August~02 (yellow dashed line) corresponds to a Carrington Longitude of $\phi=94^\circ$ in each panel.

The top panel of Figure~\ref{fig:global_mag}(a) shows the GONG synoptic map\footnote{ftp://gong2.nso.edu/oQR/zqs/201008/mrzqs100802/mrzqs100802t1154c2099\_034.fits.gz} interpolated to a uniform $180 \times 360$ grid in $(\theta, \phi)$.
The bottom panel of Figure~\ref{fig:global_mag}(a) plots the open field footpoints in green (positive polarity) and red (negative polarity) along with the global helmet streamer structure (blue field lines) and the $B_r=0$ neutral line at $r=2.5R_\odot$ representing the cusp of the global helmet streamer belt.
Figure~\ref{fig:global_mag}(b) shows the HMI radial field synoptic map\footnote{http://jsoc.stanford.edu/data/hmi/synoptic/hmi.Synoptic\_Mr\_small.2099.fits} on a $360 \times 720$ $(\theta, \phi)$ grid and the corresponding open field regions and streamer belt configuration.
Figure~\ref{fig:global_mag}(c) shows the results for the global, spherical NLFFF modeling ($\S$\ref{tw_nlfff}) based on the {HMI synoptic vector magnetograms\footnote{for an overview http://hmi.stanford.edu/hminuggets/?p=1689} for CR~2099.} 
Here, the upper panel shows the NLFFF $B_r(\theta,\phi)$ solution at $r=R_\odot$ (rather than the observed synoptic map), and the lower panel plots the open field foot points colored by their polarity, representative streamer belt field lines, and the $B_r=0$ contour at $2.5R_\odot$ as the black line to indicate the streamer cusp and heliospheric current sheet location. 
Figure~\ref{fig:global_mag}(d) shows the steady state MHD solar wind outflow ($\S$\ref{methods:field:mhd}) based on the GONG PFSS magnetic field extrapolation for the initial magnetic state using $\ell = 16$ degrees of the spherical harmonic expansion. As in panel (c), the top panel shows the PFSS solution at $r=R_\odot$, and the bottom panel shows the open field regions and streamer structure at $t=100$~hr into the solar wind relaxation.

Figure~\ref{fig:global_mag} illustrates both similarities and differences in each of the global magnetic field models. In a rough, qualitative sense, the open field regions of the large-scale coronal holes are essentially in agreement. 
The northern negative polarity coronal hole has an extended ``elephant trunk'' structure extending to low-latitudes and the global helmet streamer belt shape has significant excursions above and below the ecliptic plane, as characterized the $B_r=0$ contour at 2.5$\, R_\odot$ in black and the spatial coverage of the blue streamer belt field lines.
While the GONG-based modelling shows the helmet streamer belt extending to greater latitudes than the HMI-based modelling, in general, both the NLFFF results (based on HMI synoptic map show) and the steady-state MHD results (based on the GONG synoptic map) maintain their qualitative agreement to the global configurations of the PFSS extrapolations based on their respective source synoptic maps.
We note that the PFSS/GONG extrapolation does not capture the open field region adjacent to AR 11092 that is present in both of the HMI-based models at a Carrington longitude of $\sim$84$^\circ$ {just west of the arrow (see also Figure~\ref{fig:local_mag})}. However, the MHD configuration does recover this AR-adjacent open field due to the restructuring of the open-closed flux boundary and establishment of the force-balance of the steady-state solar wind outflow.

A quantitative characterization of the differences in the modelling results is presented in Table~\ref{tab:cr2099} where we compare properties of the global magnetic field distribution in each of the model cases. 
The four columns represent each of the models shown in Figure~\ref{fig:global_mag}. The first three rows are the minimum and maximum ranges of $(B_r, B_\theta, B_\phi)$ on the $r=R_\odot$ lower boundary, followed by the total unsigned radial magnetic flux, $\Phi(r) = \int |B_r| \; dA$, at radial heights of $r/R_\odot = \{1.0,1.1,1.5,2.5\}$, and the total magnetic energy $E_M = \int B^2/(8\pi) \; dV$. For the MHD simulation, the radial fluxes and total magnetic energy values are at $t=100$~hr, i.e. at the end of the steady-state solar wind relaxation.
{The differences in $\Phi(r)$ between $2.5R_\odot$ and $1.5R_\odot$ show the same level of variation seen between the different models. The fact that three of our four cases have regions of open flux adjacent to our jet event location ensures this is a robust feature of the large-scale topology. We also refer the reader to the study by \citet{Riley2019} that examines the impact of varying the PFSS source surface height on the area and shape of open flux regions.}

\begin{table*}[!t]
\centering
\begin{tabular}{|c|c|c|c|c|}
  \hline
  &  PFSS (GONG)  &  PFSS (HMI)   &   NLFFF (HMI) & MHD (GONG) \\
  \hline
  $\min, \max \; B_r$ [G] 	& 	$-65.2, \; +57.3$ 		
                            & 	$-72.7, \; +70.7$  	
                            & 	$-820.5, \; +381.4$
                            &   $-26.0, \; +24.8$ \\
  $\min, \max \; B_\theta$ [G] 	& 	$-25.7, \; +24.3$ 		
                            & 	$-31.6, \; +33.0$ 		
                            &   $-173.2, \; +213.0$
                            &   $-10.8, \; +10.7$\\
  $\min, \max \; B_\phi$ [G] 	& 	$-55.5, \; +38.7$ 		
                            & 	$-64.9, \; +41.3$ 		
                            & 	$-421.3, \; +291.9$
                            &   $-17.4, \; +21.7$ \\
  \hline 
  $\Phi (r=1.0)$ [Mx]   &  	$1.360 \times 10^{23}$ 	
                        &	$1.669 \times 10^{23}$ 	
                        &	$1.893 \times 10^{23}$
                        &   $1.209 \times 10^{23}$ \\
  $\Phi (r=1.1)$ [Mx]   & 	$6.094 \times 10^{22}$ 	
                        & 	$7.282 \times 10^{22}$ 	
                        &	$7.676 \times 10^{22}$
                        &   $5.887 \times 10^{22}$ \\
  $\Phi (r=1.5)$ [Mx]   & 	$2.242 \times 10^{22}$ 	
                        &	$3.124 \times 10^{22}$ 	
                        &	$2.485 \times 10^{22}$
                        &   $2.151 \times 10^{22}$ \\
  $\Phi (r=2.5)$ [Mx]   & 	$1.479 \times 10^{22}$ 	
                        & 	$2.381 \times 10^{22}$     
                        &	$1.737 \times 10^{22}$
                        &   $1.494 \times 10^{22}$ \\	
  \hline
  $E_M$ [erg]	    & 	$4.197 \times 10^{32}$ 	
                    &	$6.048 \times 10^{32}$   
                    &	$1.177 \times 10^{33}$
                    &   $2.674 \times 10^{32}$ \\
  \hline
\end{tabular}
\caption{{Comparisons of the magnetic field model extrema at the lower $r=1R_\odot$ boundary and total unsigned magnetic flux at increasing heights.}}
\label{tab:cr2099}
\end{table*}

\begin{figure*}[!thb]
    \includegraphics[width=1.0\textwidth]{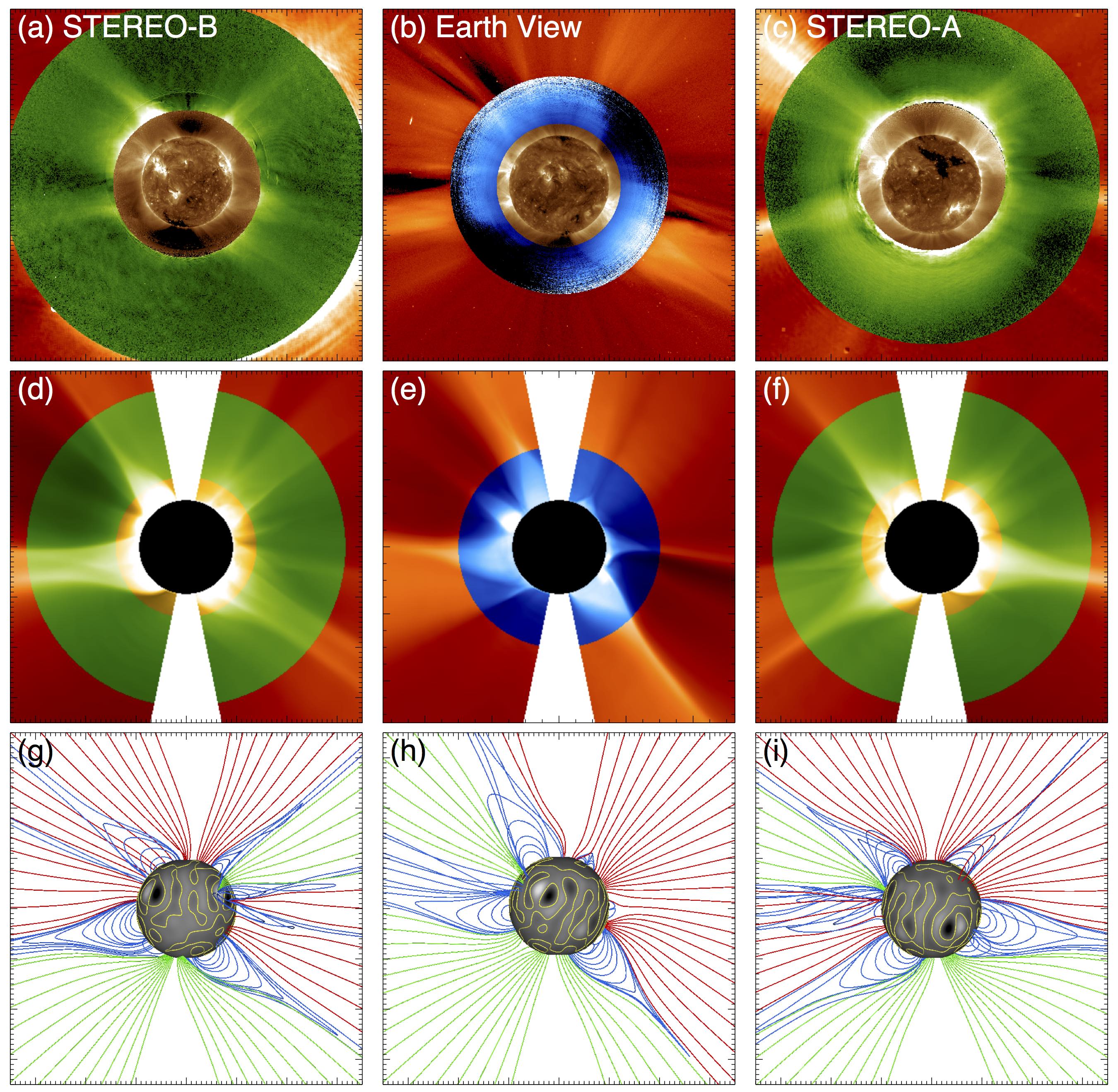}
    \caption{Comparison of the steady-state MHD coronal structure to white light coronagraph data. Panels (a)--(c): STEREO/SECCHI and SOHO/LASCO with SDO/AIA data from 2010 Aug 02. Panels (d)--(f): synthetic white light images from MHD density structure from each viewpoint. Panels (g)--(i) Representative magnetic field lines in the plane of the sky showing extended streamer structures.}
    \label{fig:global_mhd}      
\end{figure*}

Figure~\ref{fig:global_mhd} presents the comparison between the multi-viewpoint white-light coronagraph observations with the steady-state MHD wind outflow and its resulting streamer structure shown in Figure~\ref{fig:global_mag}(d). The top row of Figure~\ref{fig:global_mhd} contains the (a) STEREO-B, (b) Earth view, and (c) STEREO-A perspectives as composite plots of EUVI, COR1, and COR2 data in panels (a) and (c), and a composite plot of AIA, the HAO MLSO MK4, and LASCO C2 data in panel (b). The middle row of Figure~\ref{fig:global_mhd}, panels (d)--(f), show the synthetic white light intensity derived from the MHD 3D density data cube, normalized by the $t=0$~hr spherically-symmetric density profile \citep[as in][]{Lynch2016b}. The bottom row of Figure~\ref{fig:global_mhd}, panels (g)--(i), show the representative open field lines colored with the same positive, negative (green, red) polarity and the closed field streamer structures (blue) in the plane of the sky for each respective viewpoint.

The qualitative agreement between the observed and modelled helmet streamer structure is quite reasonable. The MHD simulation's synthetic brightness structures (corresponding to the line-of-sight integrated white light emission from the density enhancements of the global helmet streamer belt and the largest pseudostreamers) occur at approximately the same locations with approximate the same widths as the observed coronagraph features. The agreement for the STEREO spacecraft viewpoints is a bit better than the LASCO/Earth-view. In general, the disagreement is most apparent in the simulation streamers being slightly more radial than the observed streamers. This is influenced by the nearby open field strengths and their resulting open fluxes. The representative open field lines of Panels (g)--(i) were traced from a series of points in the selected planes-of-the-sky so some overlap between these open field lines and some of the representative closed streamer field lines (blue) show that there are significant line-of-sight contributions in both the observed and synthetic white-light intensities.

\subsection{Local Magnetic Field Configuration for 2010~August~02}
\label{sec:loc02}

\begin{figure*}[!htb]
    \includegraphics[width=1.0\textwidth]{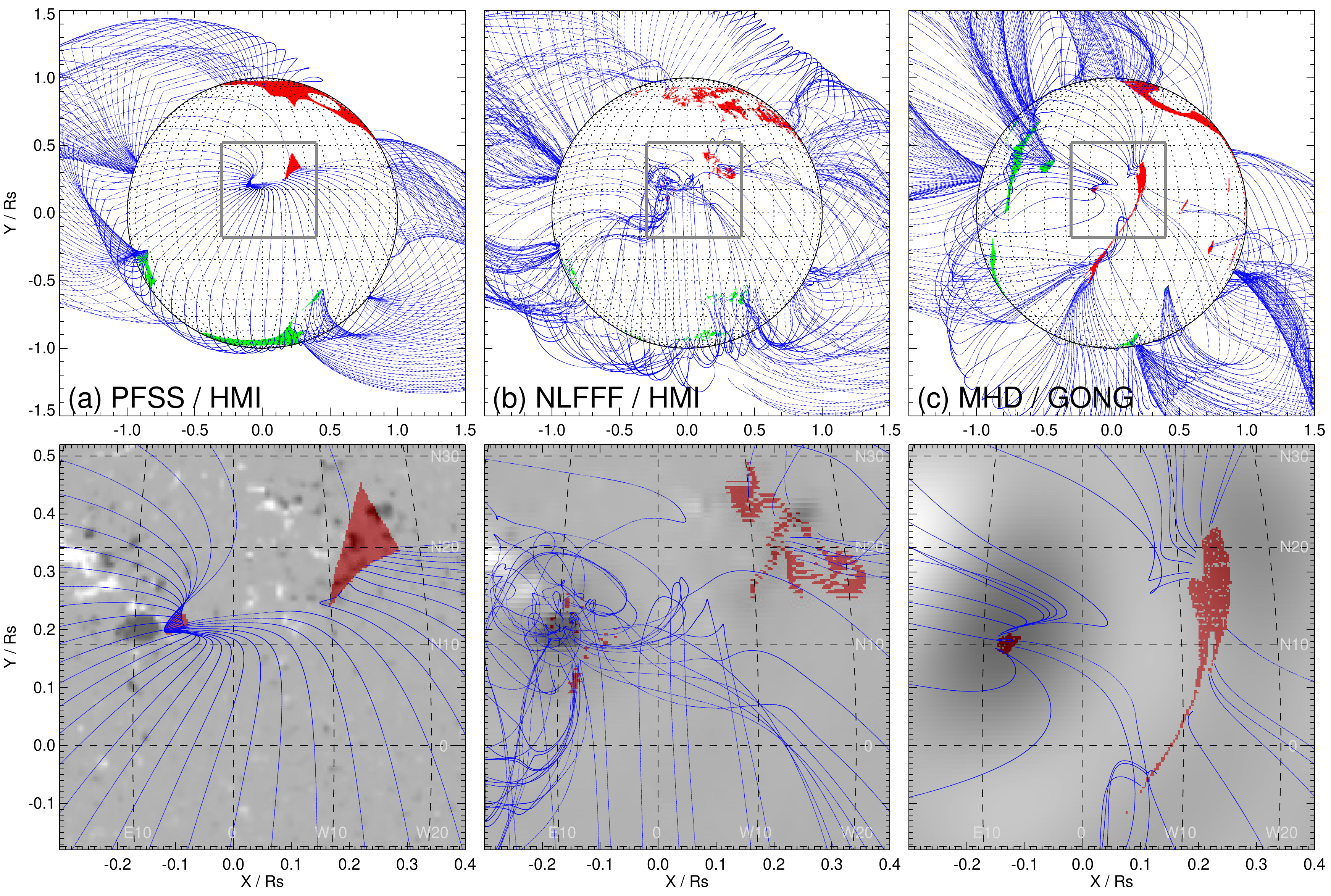}
    \caption{Top row: Representative streamer belt magnetic field lines over the open-field maps for the (a) PFSS/HMI, (b) NLFFF/HMI, and (c) MHD/GONG magnetic configurations viewed on the solar disc for 2010 August~02. Bottom row: A close up of the AR 11092 jet source region and the open field foot points and streamer boundaries in the vicinity of the AR jet plotted over the observed or modelled $B_r(R_\odot, \theta,\phi)$ distribution.}
    \label{fig:local_mag}      
\end{figure*}

In this section, we further examine the coronal magnetic field structure in the vicinity of AR 11092 and the August~02 jet source region in order to investigate the relationship between these structures and the EUV remote sensing observations of the coronal plasma.
The top row of Figure~\ref{fig:local_mag} plots the Earth-view of the large-scale coronal topology associated with the helmet streamer belt (blue field lines) and the (positive, negative) open field regions (green, red) predicted from the (a) PFSS/HMI, (b) NLFFF/HMI, and (c) steady-state MHD/GONG modeling results presented in Figure~\ref{fig:global_mag}. In each panel, we have drawn a $0.7\,R_\odot\times0.7\,R_\odot$ box on the disk face corresponding to the approximate field-of-view of the region encompassing the various EUV spectroscopy measurements.
The bottom row of Figure~~\ref{fig:local_mag} plots each of the regions indicated above ($\sim$ 670''$\times$670'') including the strong-field negative polarity spot in AR 11092 and the AR-adjacent open field regions (red shading).

There are both similarities and differences in each of the magnetic field modelling results in and around the coronal jet source region. Each model produces some small concentration of open flux associated with the western edge of AR 11092 and larger region of open flux further west, even in the PFSS and MHD cases where the magnetic field on the 1$\, R_\odot$ boundary is severely under-resolved. Based on the visual inspection, the shapes of these open flux concentrations are quite different, but their overall locations with respect to AR 11092 are in fairly good agreement. The helmet streamer field lines are representative of the open-closed field topological boundary. Additionally, in each case, the two concentrations of negative polarity open fields signify a pseudostreamer-like closed-flux system to the west of AR 11092.  This recalls the dome structure visible in Figure \ref{fig:aia_context} label as 'B'.  For the purposes of our investigation herein, the fact that each model reproduces open flux in the immediate vicinity of the observed AR jet means that there is a strong possibility for at least some AR material to end up on open or newly-opened field lines during the eruption process.

\subsection{Local Magnetic Field Configuration for the Period August 1--3}
\label{sec:loc1-3}

We inferred the temporal evolution of the magnetic field around the spot during the period August 1--3 applying the NLFFF method described in Sec. \ref{tw_nlfff}. Differently from what made for the global extrapolation {in the previous} section, here we used the daily vector magnetic field data. This would be a more representative for the temporal evolution of the smaller scale features.

We used full disk vector field for the 1st, 2nd and 3rd of August 2010 obtained in Heliographic--{Helioprojective} Cartesian (HGLN/HGLT-HPC). All the vector magnetic field data are provided by SDO/HMI. The computational grid is 256x372x512 over the domain r/R$_\odot$ $\in$ [1, 2.5], $\theta \in$ [20$^\circ$, 160$^\circ$], $\phi \in$ [90$^\circ$, 270$^\circ$].  

\begin{figure*}[!thb]
\centering
\includegraphics[width=0.25\textheight, trim = 110 10 0 0, clip]{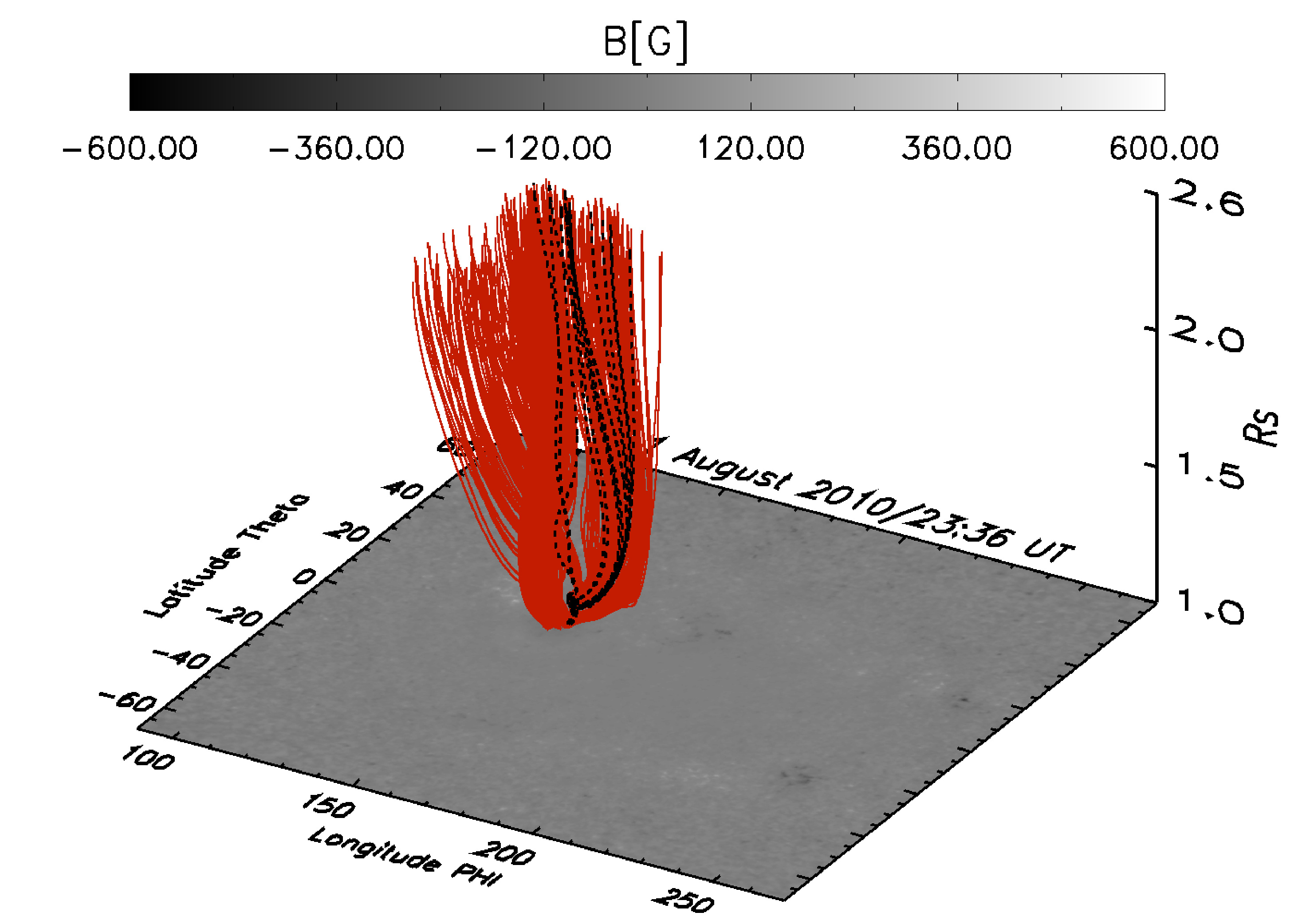}
\includegraphics[width=0.15\textheight, trim = 0 40 0 0, clip]{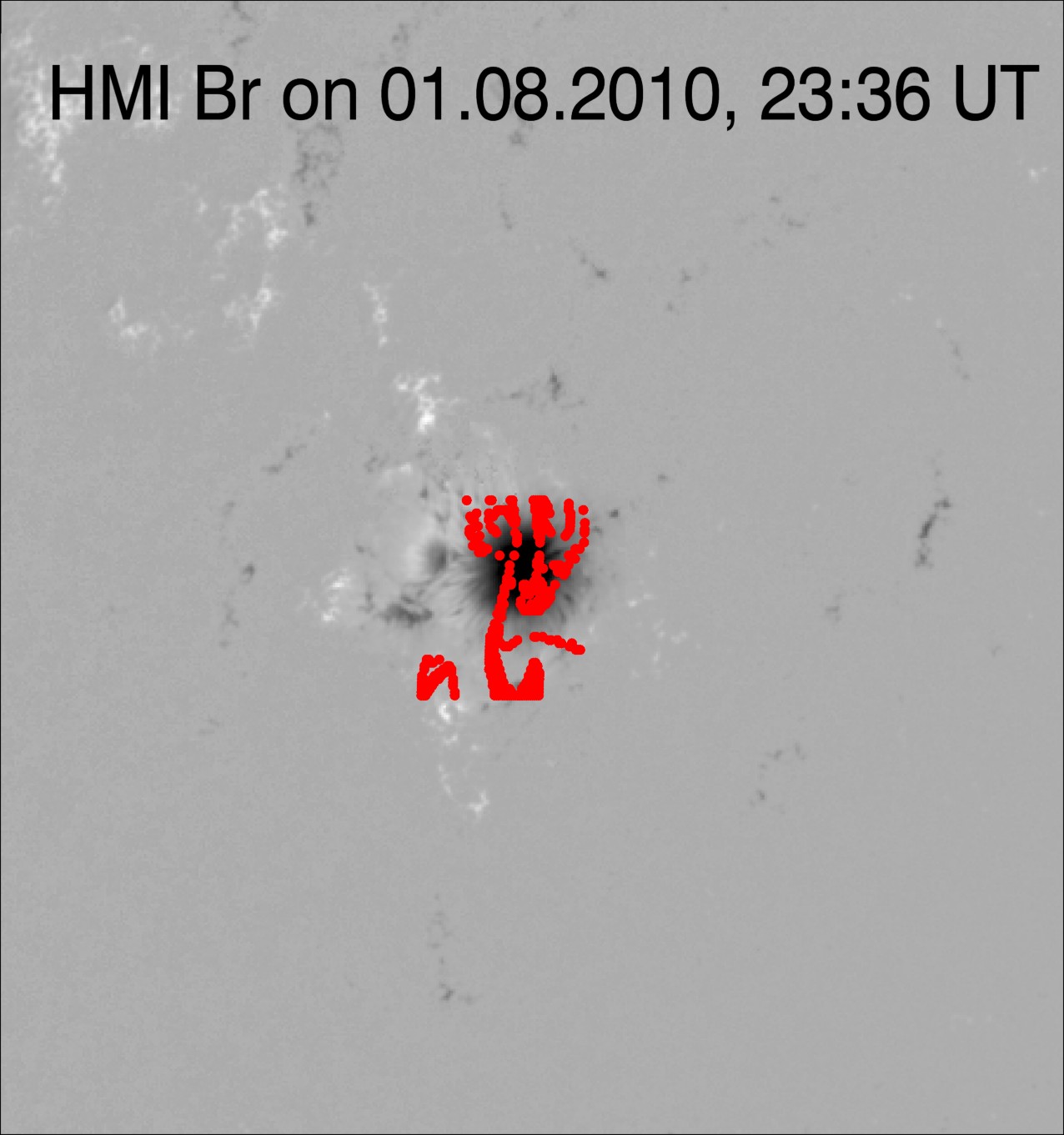}
\includegraphics[width=0.25\textheight, trim = 110 10 0 190, clip]{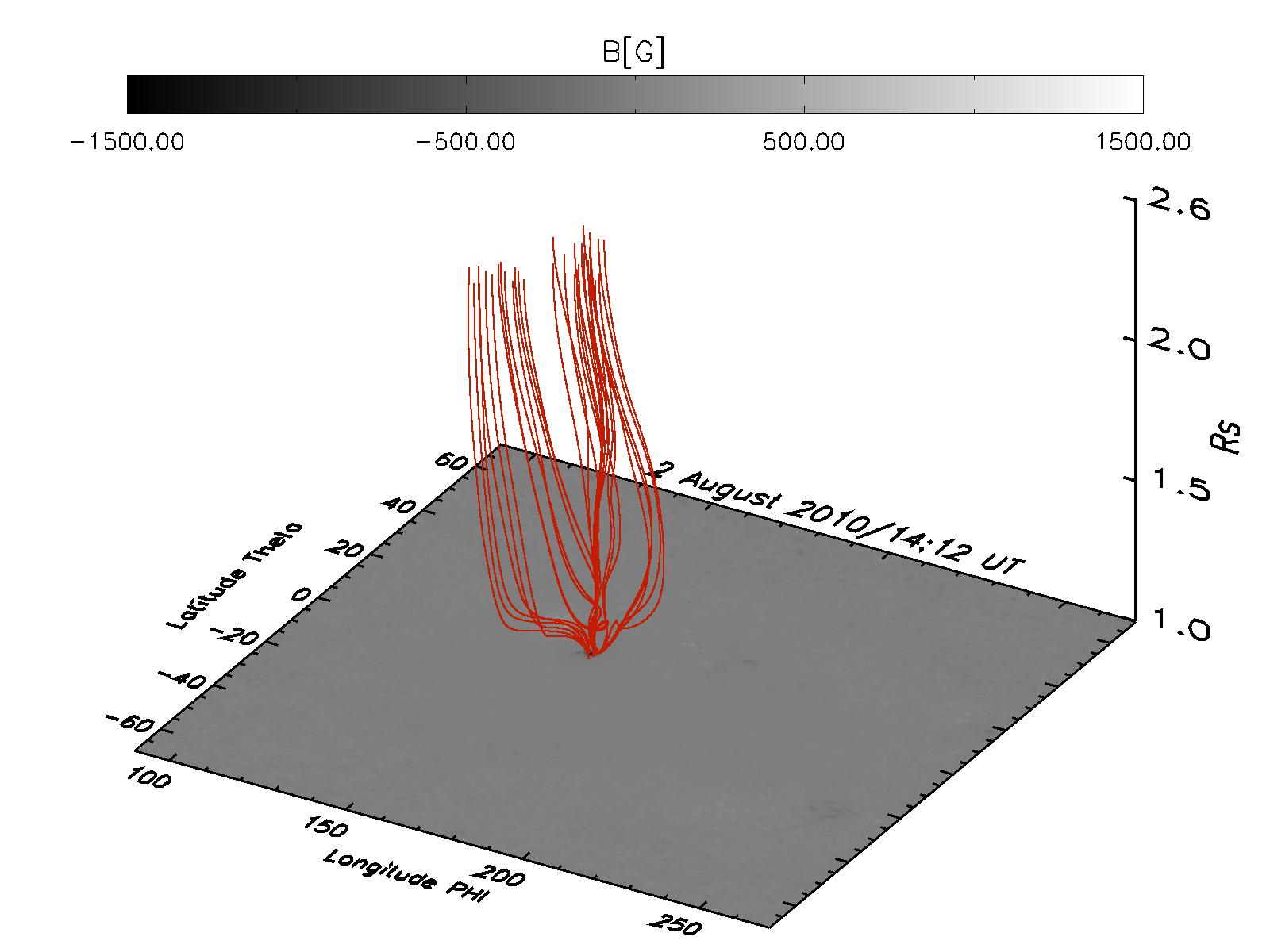}
\includegraphics[width=0.15\textheight, trim = 0 40 0 0, clip]{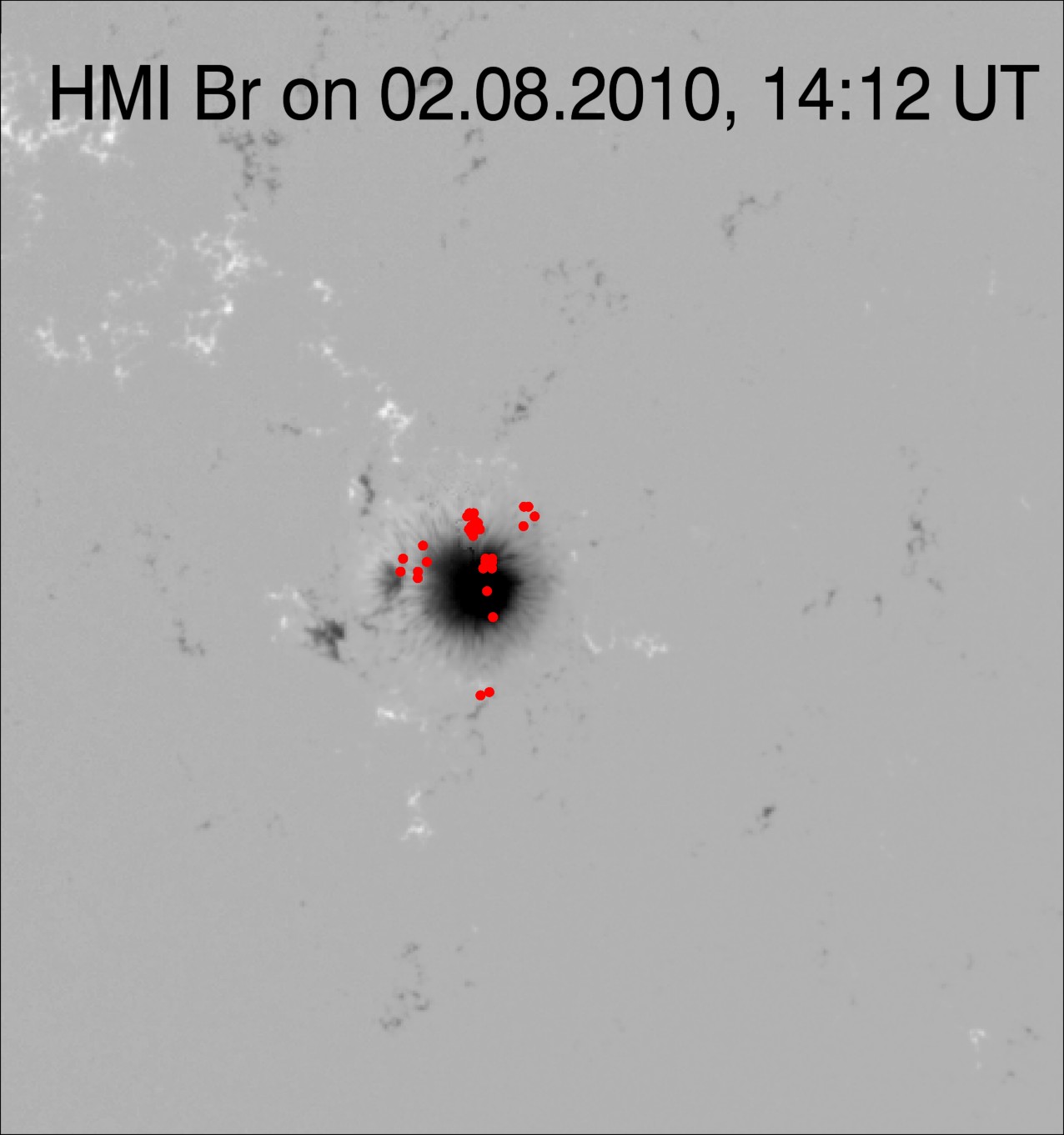}
\includegraphics[width=0.25\textheight, trim = 110 10 0 190, clip]{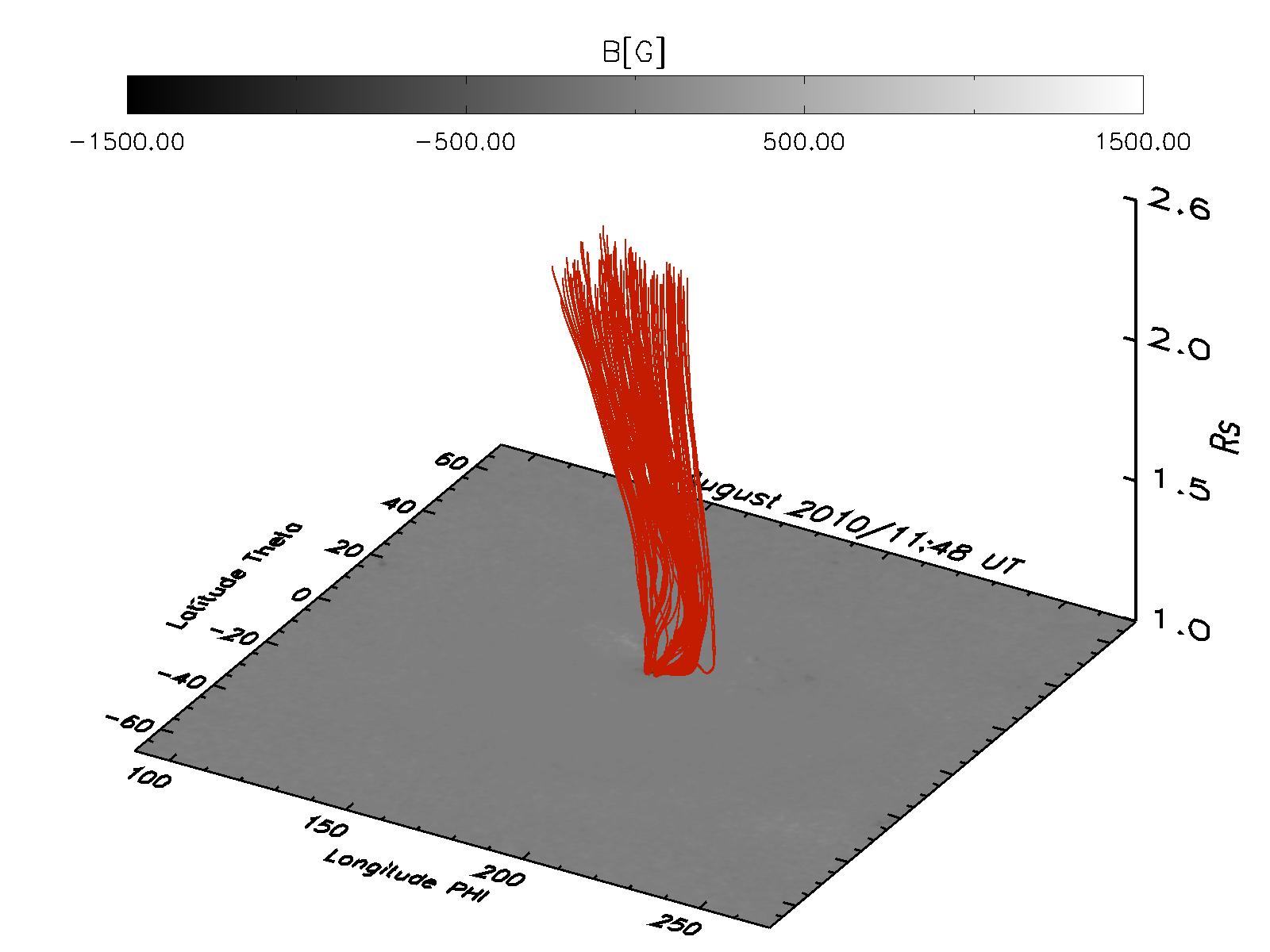}
\includegraphics[width=0.15\textheight, trim = 0 40 0 0, clip]{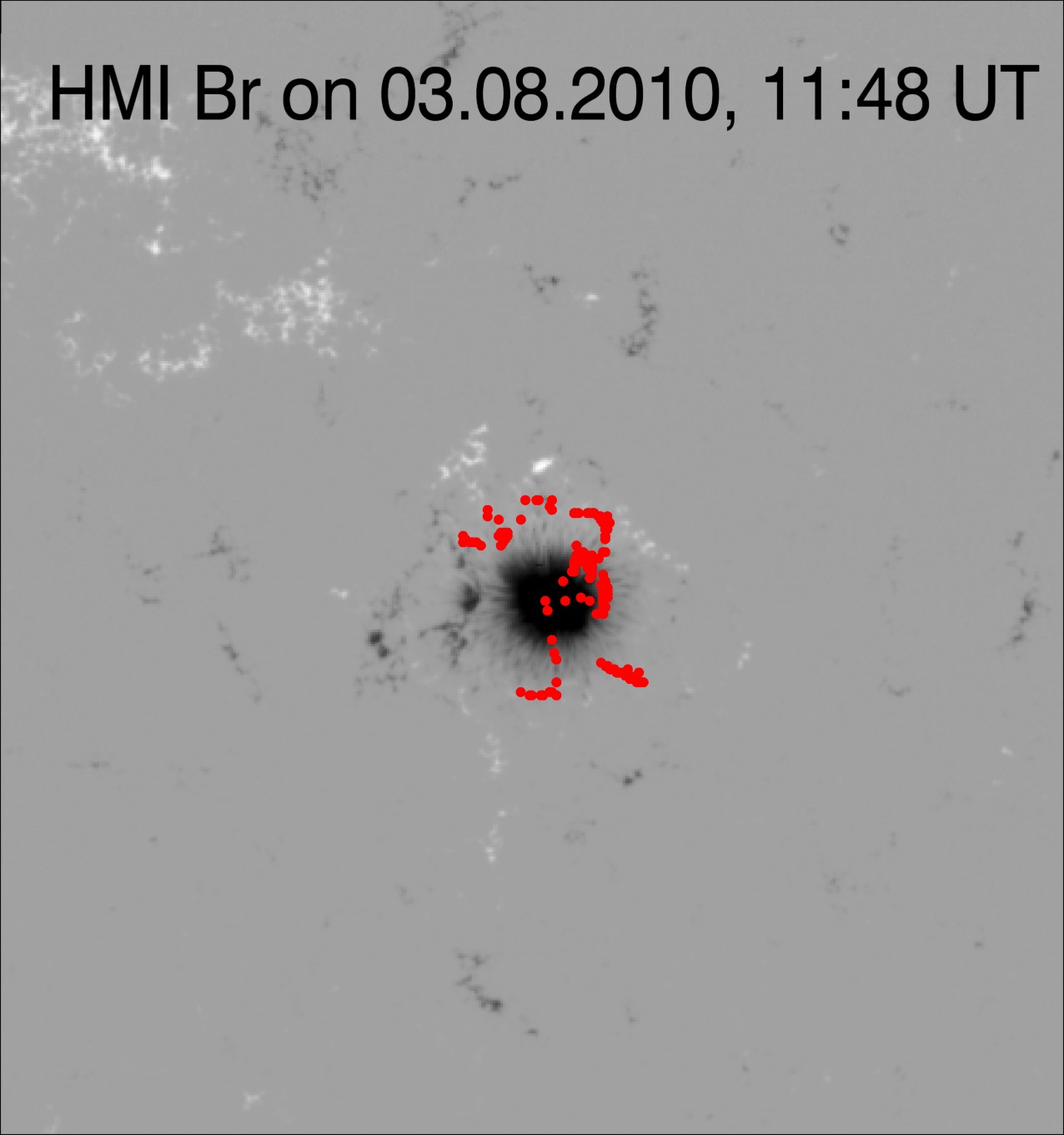}
\caption{Left column: open field lines for the 1st of August (top panel), 2nd of August (middle panel), 3rd of August (bottom panel). For the top-left plot, the black field lines are those  rooted in the CO1 region.
The bottom boundary is the solution of the extrapolation at the photospheric level. The x axis represent the longitude and the y axis shows the latitude. Right column: photospheric foot points (red crosses) of the open magnetic field lines on top of the radial component of the HMI magnetic field for the 1st of August (top panel), 2nd of August (middle panel) and 3rd of August (bottom panel).}
\label{fig:OpenField}
\end{figure*}

We applied the NLFFF model on HMI data before the jet on 1st of August 2010 at 23:36:00 UT, 2nd of August at 14:12:00 UT and after the jet on 3rd of August at 11:48:00 UT. From the solution of the extrapolation we trace the open field lines from the photosphere into the corona at 2.5 $R_\odot$ which we display in Fig. \ref{fig:OpenField}, left column. We performed the tracing of the field lines for window of 7$^\circ$ latitude and 10$^\circ$ longitude, around the sunspot. {We also traced in back the field lines rooted in the CO1 region.}
While the position of the latitude window remains the same in all of the images, the position of the longitude window changes with the solar rotation.

In Fig. \ref{fig:OpenField} right column we show the photospheric footprints of the open magnetic field lines overplotted over the HMI radial magnetic field. 
The variations in the local topology of the magnetic field are strongly visible before and after the jet. We remark a decrease of open field before the jet, probably connected with the activity visible in the SDO/AIA data in the region before the jet and indicating that some reconnection have already taken place. Our results also show a variation of the volume occupied by this open field with different expansion with height during the three days. 
Some small differences in the position of the open field footpoints for the 2nd of August is visible between this result and that shown on the middle panel of Figure \ref{fig:local_mag}. But the general agreement that the area around the spot contains open flux tubes is respected.


\subsection{Corona plasma parameters from EUV observations.}
\label{sec:uv_res}

Using the datasets listed in Table \ref{tab:uv_data} we performed the diagnostics analysis for density, temperature, Doppler motion and FIP bias in various area of the active region. Our purpose is to identify any difference within the active region that can uniquely be used to possibly match the interplanetary properties measured by WIND  at the estimated solar wind arrival time. 

\begin{figure*}[!htb]
\begin{tabular}{lll}
 \includegraphics[scale=.3, trim=-50 0 0 0, clip]{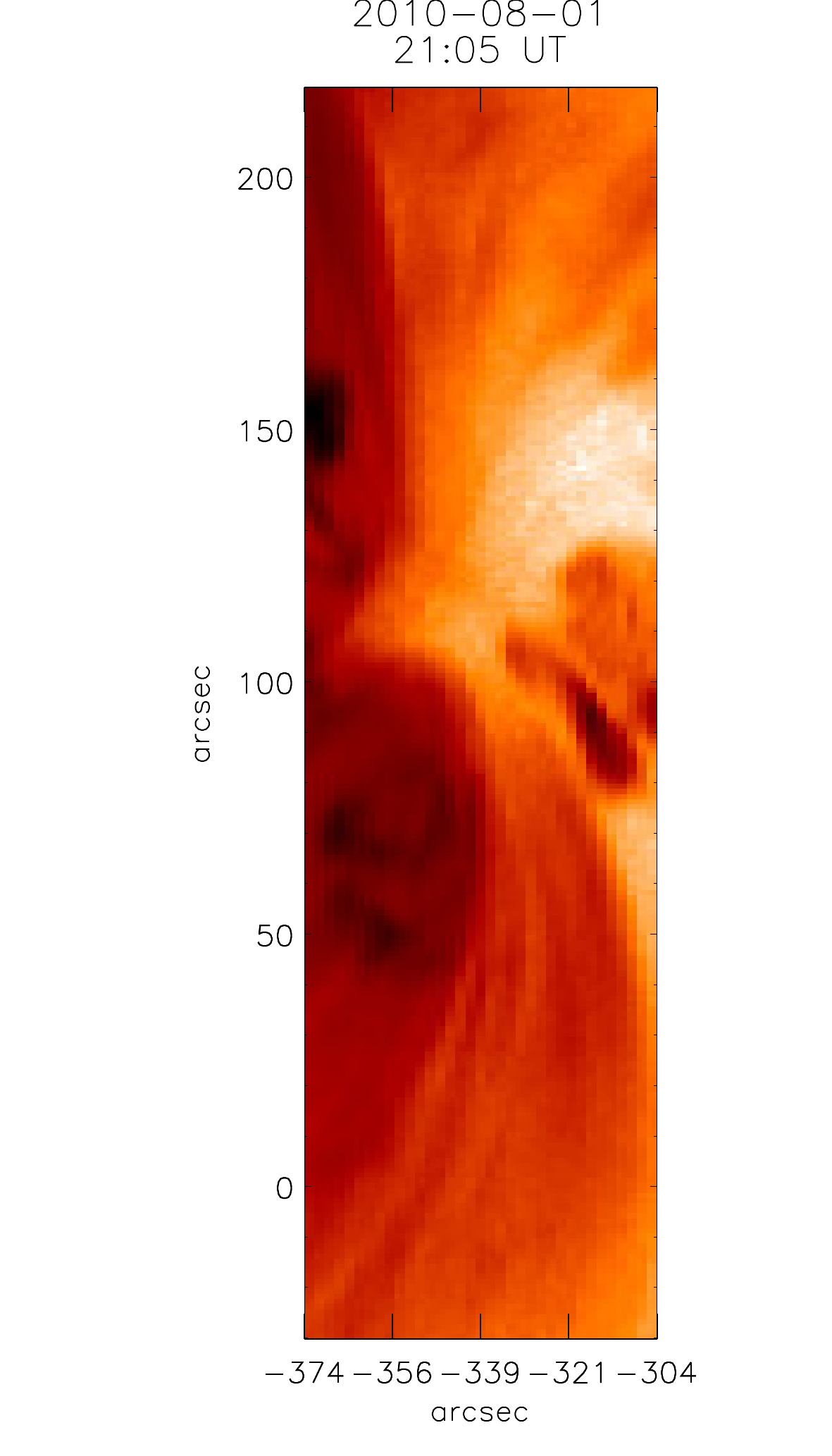}
\includegraphics[scale=.3, trim=-80 0 0 0, clip]{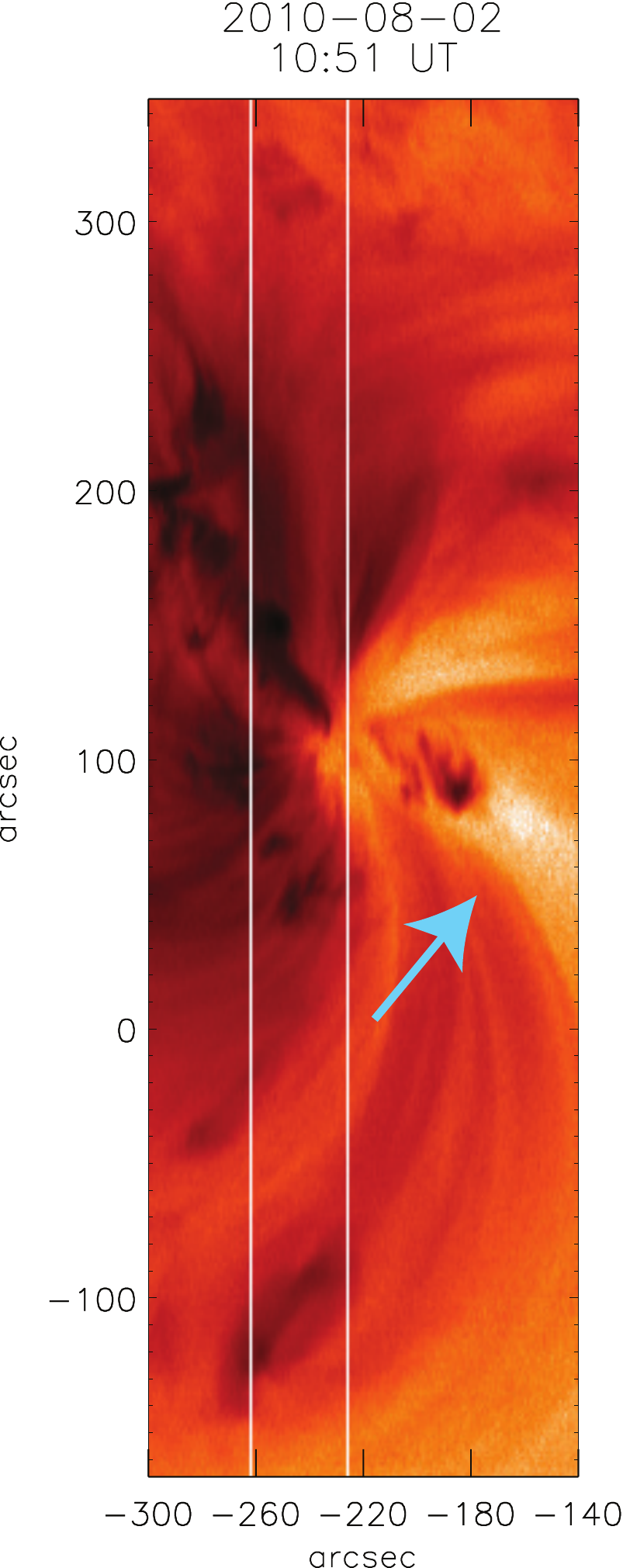}
\includegraphics[scale=.3, trim=-60 0 0 0, clip]{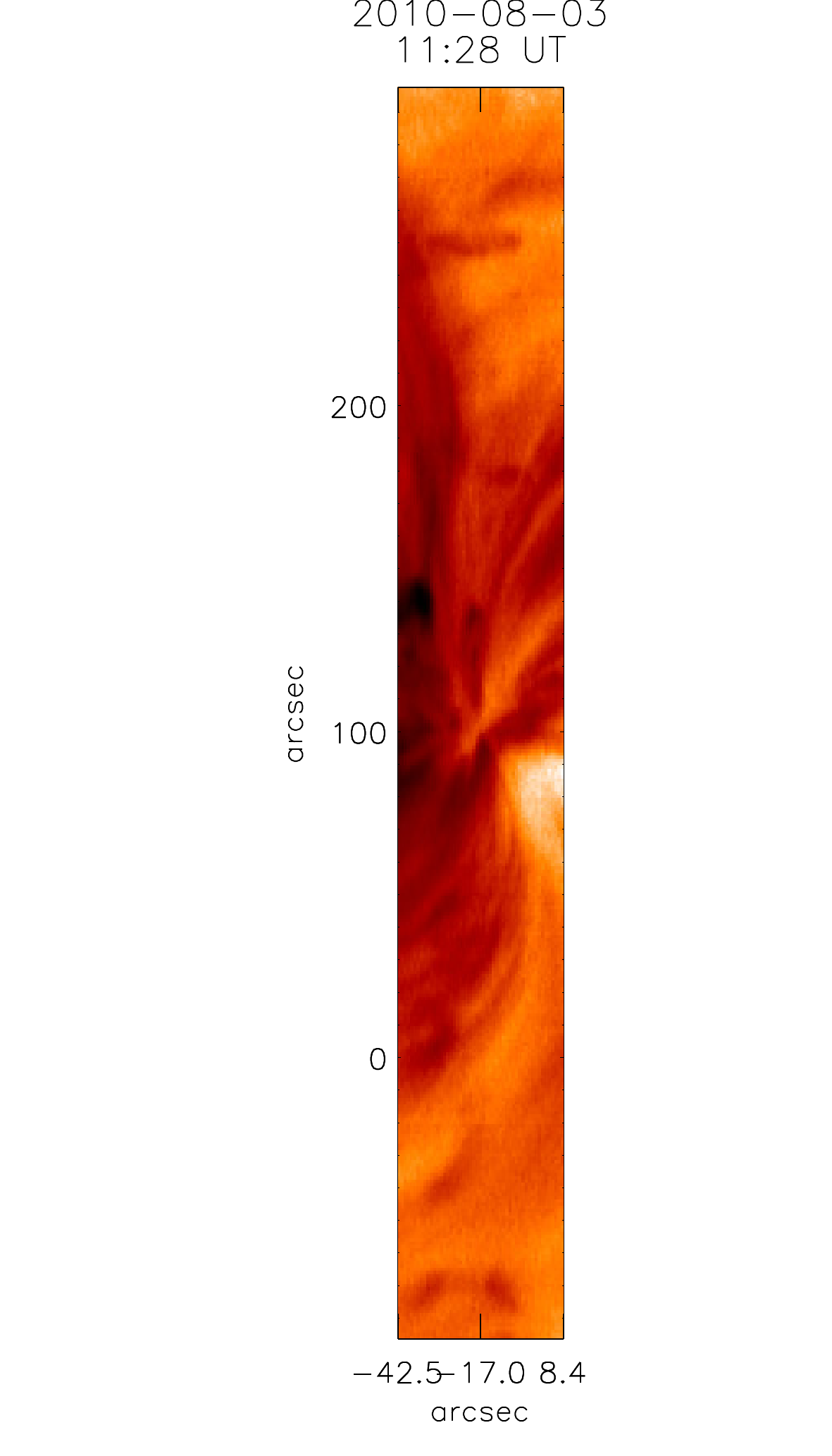} \\
 \includegraphics[scale=.37, trim=0 50 60 20, clip]{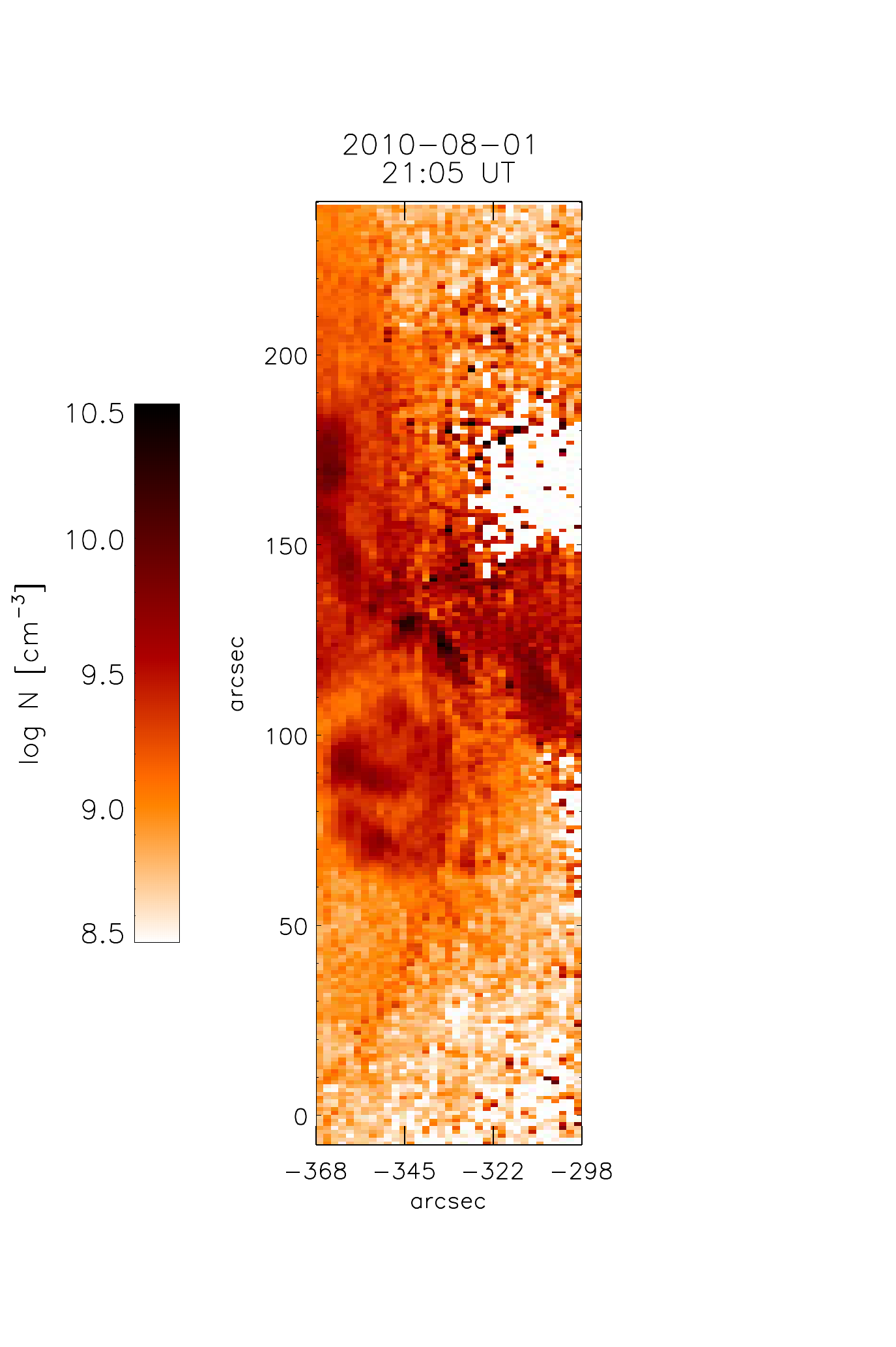}
\includegraphics[scale=.3]{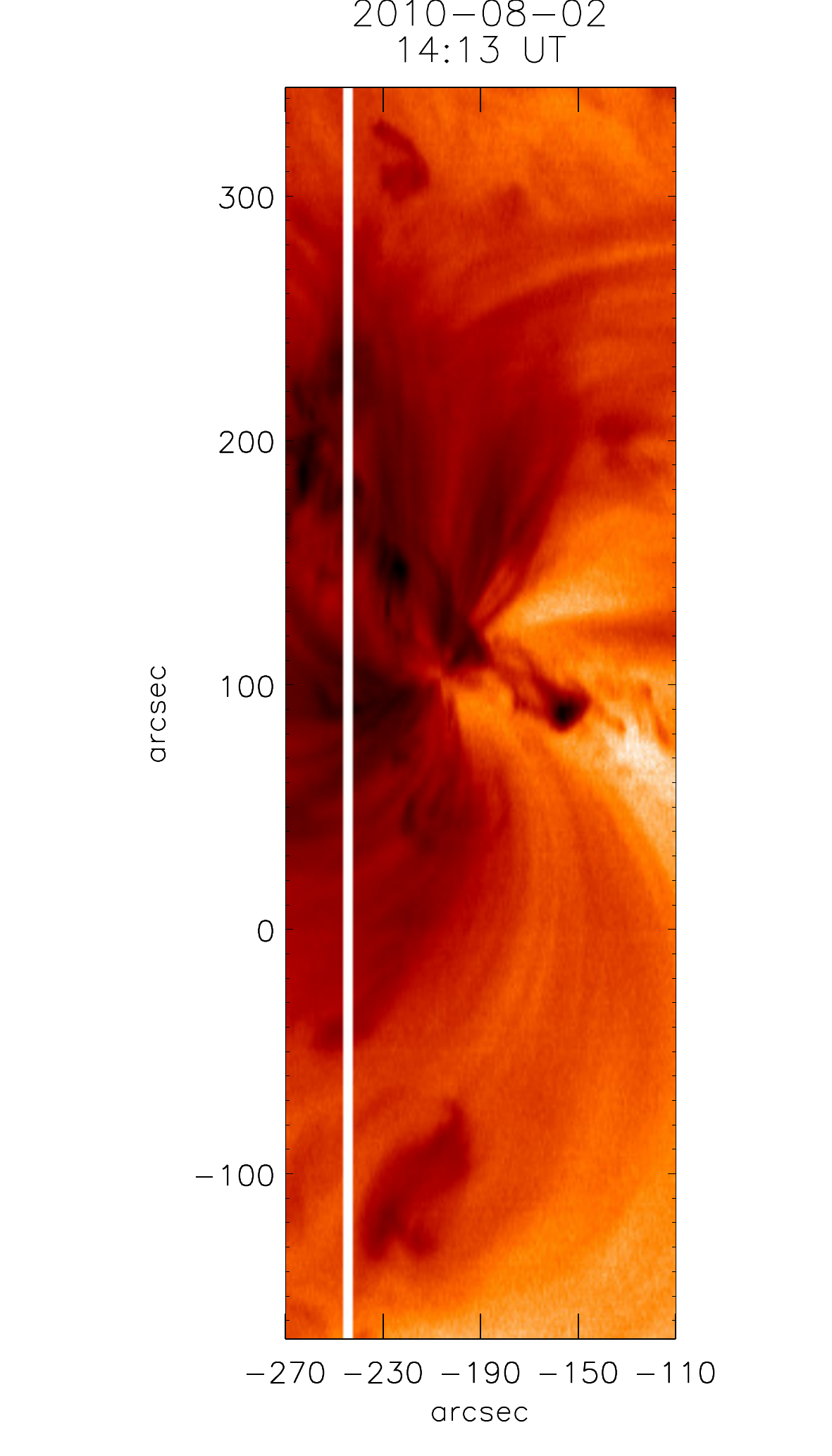}
\includegraphics[scale=.37, trim=30 50 100 0, clip]{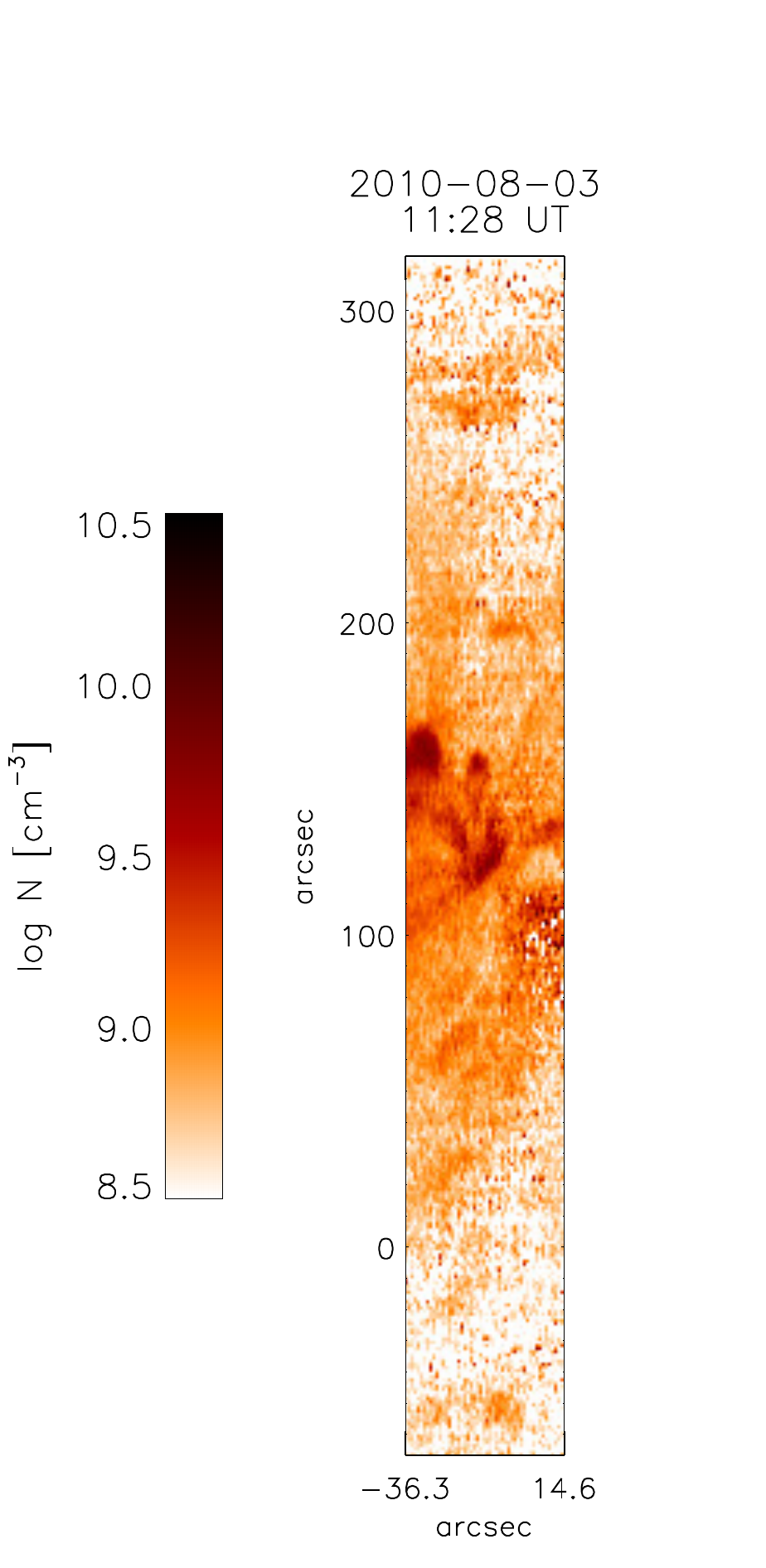}
 \end{tabular}  
\caption{Top: negative intensity maps from \ion{Fe}{xii}  195.2 \AA~ for, respectively, August 1st (dataset A), 2nd (10:51 UT, dataset C) and 3rd (dataset D). Bottom line from the left to right: electron density map (in logarithmic scale) for dataset A,  intensity maps from \ion{Fe}{xii}  195.2 \AA~ for August 2nd at 14:13 UT (dataset C), and electron density map (in logarithmic scale) from dataset D. Note that these EIS rasters have a different FOV. The arrow marks the channel location of the jet.}
 \label{fig:fe12_i}
\end{figure*}

{The top line and middle panel of the bottom line of Figure \ref{fig:fe12_i} show the \ion{Fe}{xii} 195.2 \AA~ intensity (negative) maps for the rasters listed in Table \ref{tab:uv_data}, A, C and D. We can identify the spot at the center of the FOV and the long loops in the middle-left side of it.  The south-west side of the rasters is the area interested by the jet eruption. This indeed is the area where we see the highest topological change from August 1st to the 3rd. These different structures of the AR will be investigated in Section \ref{sec:eis_full}. The first and third panels of the bottom line of the figure} show the density maps for the 1st and 3rd of August respectively, and are obtained from the line ratio technique applied to the \ion{Fe}{xii} 186.8/195.12 ~\AA~~ lines. These maps are quite noisy, but we clearly see the morphological change of the denser core of the AR. When comparing the intensity map for the 2nd of August with Fig. \ref{fig:local_mag}, we see   consistency in the morphology with both {zoomed} panels (a) and (b), even though panel (b) shows the open channel and more complex field distribution in the jet location.

\subsubsection{Pre-eruption AR from the EIS full spectrum on August 1st}
\label{sec:eis_full}

\begin{figure*}[!htb]
  \includegraphics[scale=.42,trim=13 0 0 0, clip]{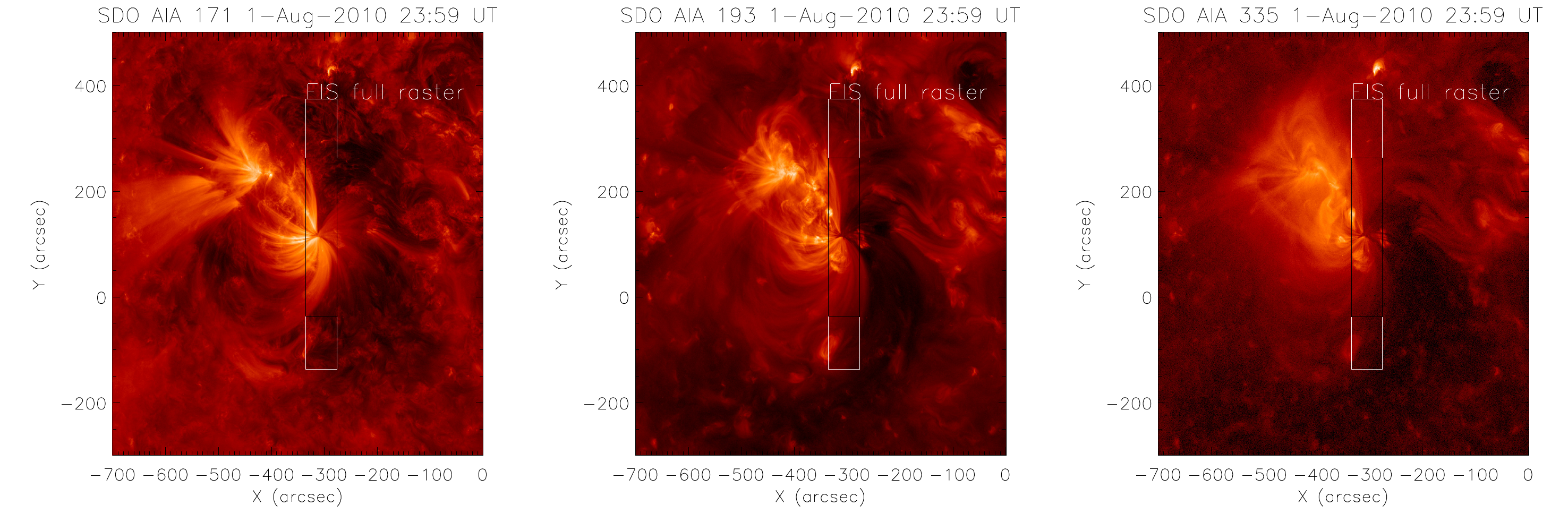}
\caption{SDO/AIA 171, 193 and 335~\AA\ images showing the AR the day before the jet. The EIS FOV  of the full spectrum raster is superimposed (dataset B, white box). The black box shows the inner region selected for further analysis and shown in Figure~\ref{fig:eis_full}.}
\label{fig:aia_eis_full}     
\end{figure*}

On August 1st EIS rastered the jet region with the 1" slit in about 1 hour. It recordered the full spectrum (dataset B in Table \ref{tab:uv_data}).   Figure~\ref{fig:aia_eis_full} shows SDO/AIA images of the AR on August 1st, with the FOV of the full EIS raster superimposed. The observation was about 17 hours before the jets observed on August 2nd. The EIS FOV was centred on the western sunspot region. The AIA 195 and 335~\AA\ images, dominated by 1.5--3 MK plasma (\ion{Fe}{xii} -- \ion{Fe}{xvi}) in the AR core, show the 'dark' channel, site of coronal outflows and the jet evolution (see also Figure \ref{fig:aia_context}, arrow 'C'). 

The AIA 171~\AA\ shows 1 MK plasma (\ion{Fe}{ix}). It is clear that a fan system of cool loops extending towards the west is superimposed on the coronal outflow channel. Various other system of loops extending towards the east are the brightest features in the image. Finally, the usual dark channel is clearly visible as surrounding the AR core.

\begin{figure*}[!htb]
\centering  \includegraphics[scale=.6]{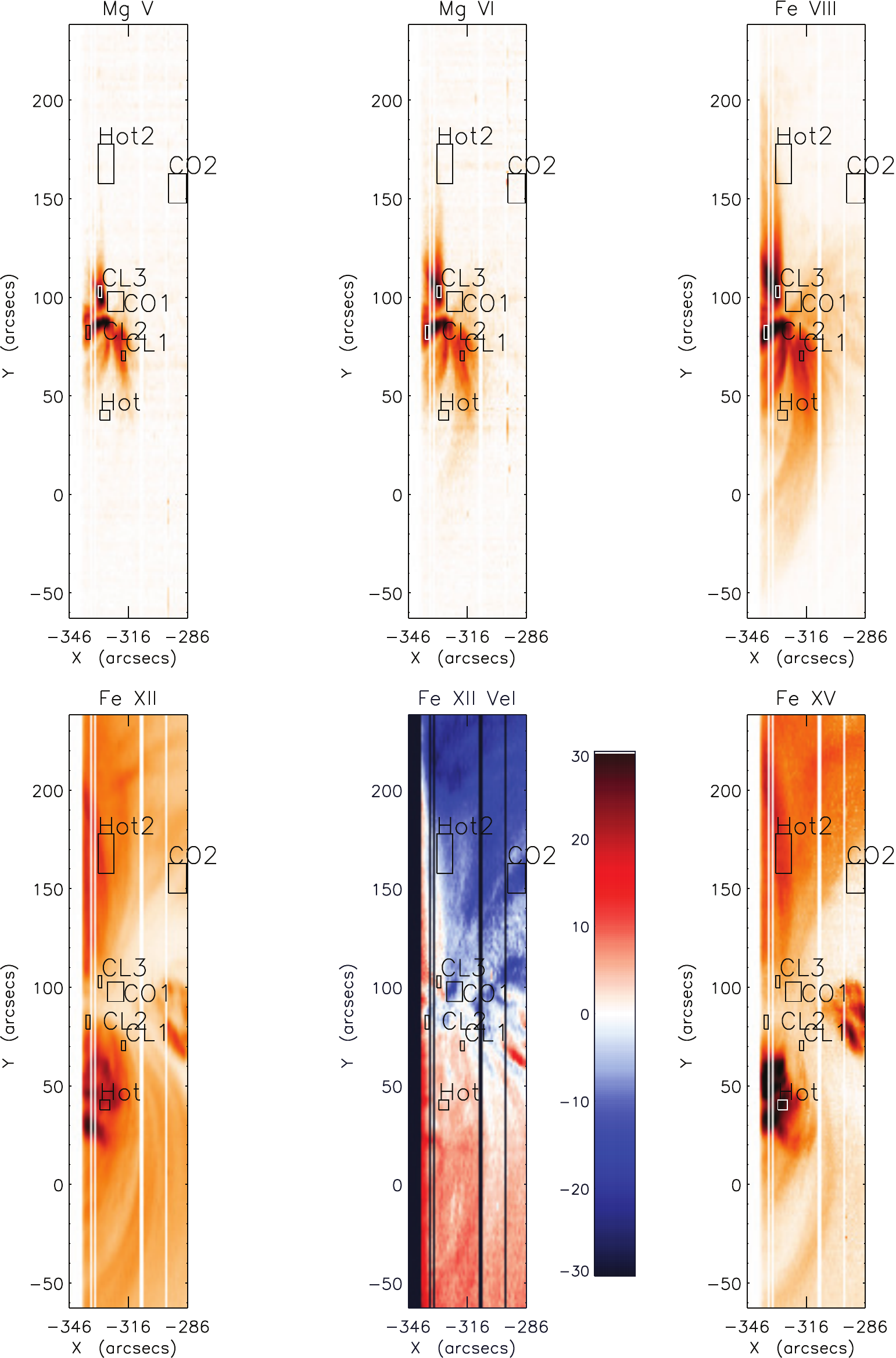}
\caption{Integrated intensities (negative images) maps at different temperatures from the EIS raster at 23:39UT August 1st (dataset B). 
 \ion{Mg}{v} 276.581 \AA~ (logT=5.5),  \ion{Mg}{vi} 268.991 \AA~  (logT=5.6),  \ion{Fe}{viii} 185.216 \AA~(logT=5.8), \ion{Fe}{xii} 195.119 \AA~ (logT=6.2), \ion{Fe}{xv} 284.163 \AA~(logT=6.3),
The velocity map from \ion{Fe}{xii} is also shown (with the limits at +/- 30 km/s). The various regions selected for further analysis are indicated.}
\label{fig:eis_full}       
\end{figure*}

The black box in Figure~\ref{fig:aia_eis_full} shows the inner region of the EIS FOV, displayed in  Figure~\ref{fig:eis_full}, which shows the intensities in a selection of spectral lines, as well as the Doppler image in \ion{Fe}{xii}. The EIS data were processed using custom-written software by one of the authors (GDZ)  which essentially reproduce the steps in the $SolarSof$t routines, with the difference that bad pixels are replaced with interpolated values and the pseudo-continuum (bias) is removed when fitting the spectral lines. The Doppler image was obtained by removing the orbital variations and the expected variations along the slit, as described in the EIS software note No. 23. For the rest wavelength, we used a region in the bottom of the slit, outside from the FOV in the image, and from the AR core. The EIS spectra are dramatically different depending on the source region. Despite the  60s exposures, spectra needed to be averaged over some regions, especially where the coronal signal is low. The regions selected for further analysis are shown in Figure~\ref{fig:eis_full}. `CO1' is the main region, close to the Sunspot and the origin of the jet, within a coronal outflow region (blueshift in the Doppler map). `CO2' is a similar region. `Hot' and `Hot2' are two regions with strong emission around 3 MK, and little emission at lower temperatures.  `Hot' is on a system of short loops, while `Hot2' is on a system of long loops. Finally, three regions on the cool loops extending towards the west were selected (CL1, CL2, CL3). Again, the loops have different lengths. Little hotter emission is present along the line of sight.

We obtained averaged spectra and carefully fitted all the lines. The most important lines for chemical abundance diagnostics are very weak. We then performed a DEM analysis \citep{delzanna_thesis:1999} to obtain the temperature distribution and the relative chemical abundances, using CHIANTI version 8 atomic data \citep{delzanna15}  although we note that changes in the latest version 10 only introduce small (10--20\%) differences in a few ions. We used a constant density of 2$\times$10$^9$ cm$^{-3}$ for all the regions (consistently with the values found by applying the line ratio technique in Section \ref{sec:uv_res}, Figure \ref{fig:fe12_i}) to calculate the contribution functions $G(T,N)$ of the lines.

\begin{figure*}[!htb]
\centering
  \includegraphics[scale=.45]{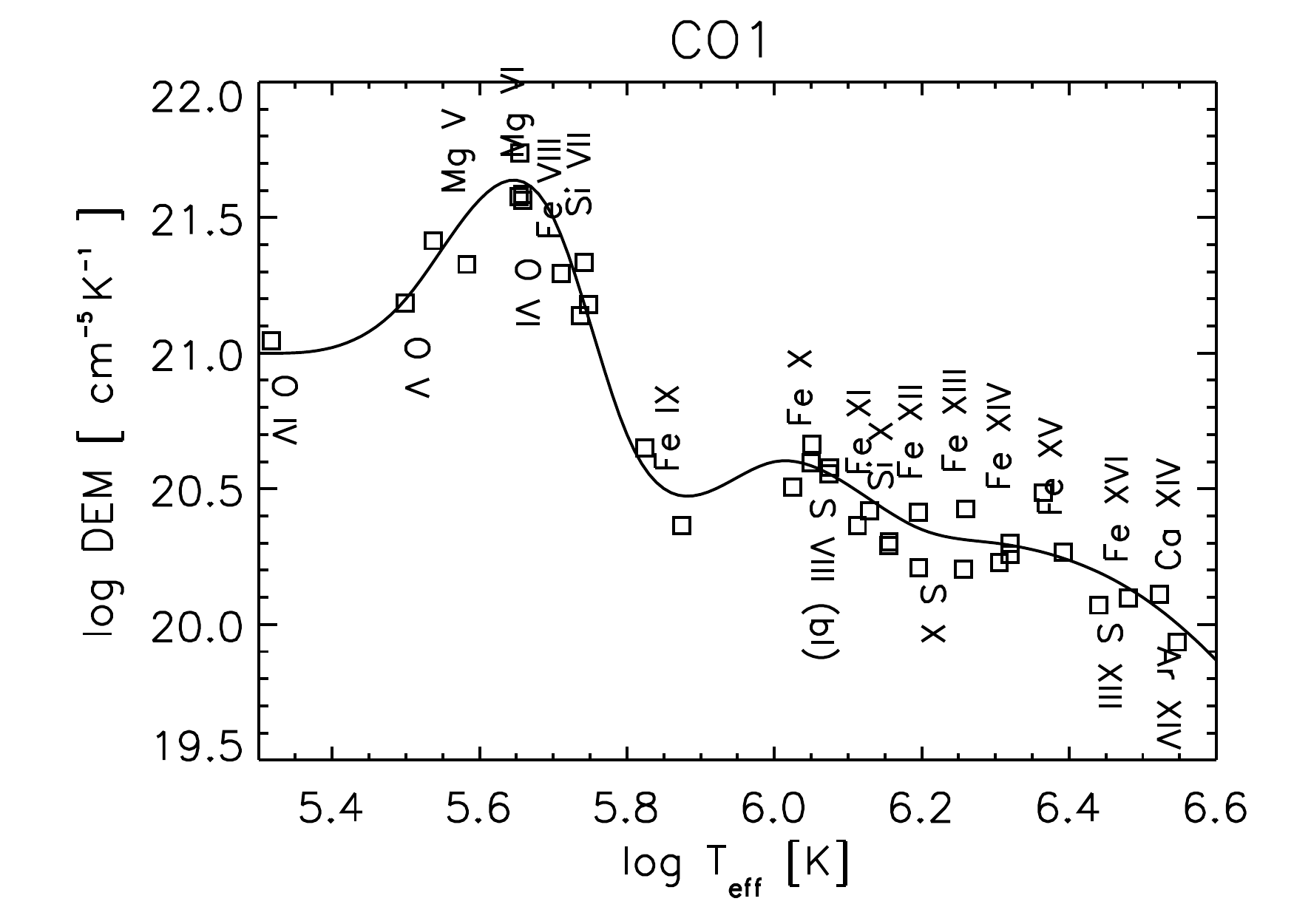}
\caption{DEM as function of temperature of the coronal outflow region CO1.
The square indicate the ratio of the predicted
vs. observed radiance, multiplied by the DEM value at
the effective temperature. }
\label{fig:dem_co1}
\end{figure*}

An example is shown in Figure~\ref{fig:dem_co1}, for the coronal outflow region CO1. The  points indicate the ratio of the predicted vs. observed radiance, multiplied by the DEM value at an effective temperature (i.e. an average temperature weighted by the DEM distribution). The emission in the 2--3 MK region is an order of magnitude lower than the typical emission in active regions. The  FIP effect can only be constrained by the \ion{Fe}/\ion{S}{}  relative abundance. Lines from \ion{S}{viii}, \ion{S}{x}, \ion{S}{xiii} were clearly observed. There are plenty of strong \ion{Fe}{} lines to constrain the DEM. The coronal abundances  discussed in \cite{delzanna18}, where \ion{Fe}{}/\ion{S}{} is enhanced by a factor of 3.2 over its photospheric value \citep{asplund09}, were used.  \ion{Ar}{xiv} is at the limit of detectability, which further confirms that the abundances were not close to photospheric, otherwise the high-FIP Ar line would have been much stronger and easily observed. The weak Ar emission is consistent with this FIP effect of 3.2. Note that the emission in the coronal outflow region is strongly multi-thermal. Also note that the peak of the DEM, at much lower temperatures, refers to the cool fan loops along the line of sight that are clearly visible in the EIS \ion{Fe}{viii} (and in the AIA 171~\AA\ image, which we have mentioned).   

Other coronal outflow regions such as CO2 showed the same coronal abundances. The hot loop regions also had the same abundances, in  agreement with the AR core results reviewed by \cite{delzanna18}. In this case,  additional lines from \ion{S}{xi} and \ion{S}{xii} could be used. 

On the other hand, all the cool loop legs (CL1, CL2, CL3) indicated  a much smaller FIP effect, with abundances closer to the photospheric values of \citep{asplund09}. The DEM is well constrained by several strong lines from low FIP ions: \ion{Mg}{v}  Mg V, \ion{Mg}{vi}, \ion{Mg}{vii}, \ion{Si}{vi}, \ion{Si}{vii}, and \ion{Fe}{viii}. The available high-FIP ions are \ion{O}{v} and \ion{O}{vi} only.  Using the CHIANTI v.8 ionization equilibrium we obtain relatively good agreement between \ion{O}{v}, \ion{O}{vi} and the low-FIP lines with an FIP bias of 1.8, for CL1 and CL3. For CL2, the intensities of the  \ion{O}{v} lines are consistent with photospheric abundances, while those of \ion{O}{vi} are consistent with an FIP bias of 1.6. As \ion{O}{vi} is one of the `anomalous' ions, where typically predicted intensities in the quiet Sun are a factor 3-5 lower than observed \citep{delzanna18}, we think that the \ion{O}{v} results are more reliable. Density-dependent effects in the ionization equilibrium are known to improve predictions. For a recent example on the quiet Sun, see \cite{parenti19}, and for  recent calculations on  \ion{C}{} and \ion{O}{} ions \citet[see][]{dufresne19}  and \cite{dufresne20}. Such effects are also taken into account within ADAS\footnote{https://www.adas.ac.uk; https://open.adas.ac.uk} (Atomic Data and Analysis Structure, see, e.g. \citet{Summers2006, Giunta2015}),  which deals with a full density dependent collisional-radiative modelling applied to astrophysical and controlled fusion plasmas. As significant uncertainties in  the ion abundances are present for some of the elements considered here, we have  not included these effects, which in general tend to shift the temperature of formation of these ions to lower temperatures, {but do not affect significantly the relative abundances. The main uncertainties are in the EIS relative calibration and in the ionization equilibrium. They are not simple to estimate, but are of the order of 30\%. The summary of our results for the FIP bias are listed in Table \ref{tab:FIP_resultsLCR} bottom line.}

\subsubsection{Estimating the FIP bias using an alternative diagnostic method}
\label{se:LCR}

In addition to the previous analysis, we determined relative FIP bias maps for these regions of interest at four different temperature ranges using the linear combination ratio (LCR) method \citep{ZambranaPrado2019}. This method allows us to determine relative FIP biases in the corona from spectroscopic observations in a way that is in practice independent from DEM inversions. We optimize linear combinations of spectral lines of low-FIP and high-FIP elements so that the ratio of the corresponding linear combinations of radiances yields the relative FIP bias with a good accuracy. We developed a Python module, which is available at \href{https://git.ias.u-psud.fr/nzambran/fiplcr}{fiplcr}, to compute the optimal linear combinations of spectral lines and to use them to compute relative FIP bias maps from observations. The CHIANTI database version 9.0 was used together with the \texttt{ChiantiPy} Python package (version 0.8.5), although we note that the data for the coronal  ions is the same as in version 8.  To perform the optimisations we used contribution functions computed with the same density as the previous analysis ($2 \times 10^9 ~\mathrm{cm}^{-3}$), and we computed the relative FIP biases using the radiances previously obtained. We used three sets of spectral lines that were observable in all regions and a fourth set that was observable only in regions CO1 and CO2; they are listed in Table~\ref{tab:lcr_lines}. We determine either the iron to sulfur \ion{Fe}{}/\ion{S}{} relative abundance, a mix between silicon and iron to sulfur (Fe \& Si)/S, or a mix between calcium and iron to argon (Ca \& Fe)/Ar.

\newcommand{\mltc}[1]{\multicolumn{1}{|c|}{#1}}
\begin{table*}[htbp]
\caption[Sets of spectral lines used to measure the relative FIP bias as various temperatures.]{Sets of spectral lines used to measure the relative FIP bias as various temperatures {using the LCR method on dataset B}. The table shows four different sets of spectral lines, their wavelengths, the temperature of maximum abundance $T_\textrm{max}$ and the FIP value.}

\begin{center}
\begin{tabular}{|l|l|S[table-format=3.3]|S|l|} 
\hline
Set \# &    Ion                 & \mltc{Wavelength}     & \mltc{$\log T_\textrm{max}$}  & \mltc{FIP}\\
    &				            & \mltc{(\AA)} 		    & \mltc{(K)} 					& \mltc{(eV)}\\ \hline
1   & \ion{S}{viii}         	&  198.553          	& 5.9     						& high (10.36)\\
    & \ion{Si}{vii}             & 275.361    		    & 5.8  	    					& low (8.15) \\
    & \ion{Si}{viii}            & 276.85    			& 5.9               			&  low\\
    & \ion{Fe}{x}               & 184.537       		& 6.0  				    		& low (7.9)\\
    & \ion{Fe}{xi}              & 180.401       		& 6.1  					    	& low\\ \hline
    
2   & \ion{S}{x}                & 264.23            	& 6.2  						    &high \\
    & \ion{Si}{x}               & 258.374   			& 6.1						    &low\\
    & \ion{Fe}{xiii}            & 202.044   			& 6.2						    &low\\
    & \ion{Fe}{xiv}             & 264.788       		& 6.3  						    &low\\ \hline

3   & \ion{S}{xii}          	& 288.434           	& 6.3  		    				&high \\
    & \ion{S}{xiii}         	& 256.685           	& 6.4  	    					&high \\
    & \ion{Fe}{xv}              & 284.163   			& 6.3    						&low\\
    & \ion{Fe}{xiv}             & 211.317   			& 6.3    						&low\\
    & \ion{Fe}{xvi}             & 262.976       		& 6.4     						&low\\ \hline

4   & \ion{Ar}{xiv}         	&  194.401          	& 6.5  		   					&high  (15.8)\\
    & \ion{Ca}{xiv}             & 193.866   			& 6.5    						& low (6.11)\\
    & \ion{Fe}{xv}              & 284.163   			& 6.3    						& low\\
\hline
 \end{tabular}
 \end{center}
 \label{tab:lcr_lines}
\end{table*}

The results from the FIP bias measurements at different temperatures are given in Table~\ref{tab:FIP_resultsLCR}. We assumed photospheric abundances from \cite{asplund09}. The relative FIP bias value obtained is listed for every selected region. The uncertainties are computed using a statistical approach based on the Bayes theorem\footnote{We use Monte-Carlo simulations to simulate noisy radiances by adding Gaussian random perturbations with a 20\% standard deviation assuming that this would be a typical error bar one would obtain for the radiances of EIS observations. From these noisy observations computed assuming different relative FIP biases we obtain $P(f_p \mid f_i)$, the probability distribution of the real relative FIP bias of the plasma $f_p$ knowing that we have obtained a measured (inferred) relative FIP bias $f_i$.}. It allows us to define the likelihood of the plasma's real relative FIP bias given the inferred relative FIP bias measured with spectroscopic observations. Using this probability distribution we compute the credibility interval at 75\% which is also listed for every measurement in Table.~\ref{tab:FIP_resultsLCR}.

Sulfur is an intermediate FIP element which can be subject to a slight enhancement of its abundance. It is therefore no surprise that we obtain a higher value when comparing iron and calcium to argon in set \#4, argon having a much higher FIP than sulfur. In the CO1 and CO2 regions we observe an enhancement of low FIP elements as compared to sulfur of a factor 2 and at least a factor 3 when compared to argon. We should however note that the argon and calcium lines are very weak. As we have used Gaussian noise with a standard deviation of 20\% to simulate noise for all lines, the uncertainties for the relative FIP bias obtained are most likely underestimated as these very weak lines (at least an order of magnitude lower than the iron line we also have in this set) would drag along more noise. The hotter loop regions HOT and HOT2 show photospheric abundances at lower temperatures (results of set \# 1). Given that the very hot loops we wanted to analyse within this region barely exist at those temperatures and that coronal plasma is optically thin, what we are seeing in this temperature range is probably the underlying quiet sun material which does generally have abundances closer to photospheric. Higher in temperature we see an enhancement in the abundance of low FIP elements closer to a factor 2.
We find similar results to the DEM analysis (Section \ref{sec:eis_full}) for regions CO1 and CO2. Concerning the hot loops, the LCR method yields lower abundance biases, still consistent with the DEM analysis. 

\begin{table*}[ht]
\begin{center}
\caption{First four lines: relative FIP bias values obtained using the LCR method in all selected regions indicated in Fig.~\ref{fig:eis_full}. We list the relative FIP bias values measured using the radiances observed and the confidence interval at 75$\%$. For example, for set \# 2 and region CL1, we obtained $f_i = 2.2$ and the confidence interval at 75\% is [1.5, 3.0]. The \ion{S}{xi} 285.822 \AA\ and \ion{S}{xii} 288.434 \AA\ lines were too weak in regions CL2 and CL3 and resulted in inaccurate results, which is why we do not include these measurements. Last line: FIP bias results derived using the DEM method described in Sect. \ref{sec:eis_full} on dataset B.}
\label{tab:FIP_resultsLCR}
\begin{tabular}{|c|c|c|c|c|c|c|c|}
\hline
    Set  & CL1                           & CL2                           & CL3                           & CO1                          & CO2                           & HOT                                   & HOT2          \\\hline     
    1    & $2.2\substack{+0.8 \\ -0.7}$  & $2.7\substack{+0.9 \\ -0.8}$  & $2.7\substack{+0.9 \\ -0.8}$  & $2.0\substack{+0.7 \\ -0.6}$ & $1.8\substack{+0.6 \\ -0.6}$  & $1.0\substack{+0.4 \\ -0.3}$    & $0.9\substack{+0.3 \\ -0.3}$  \\[3pt]
    2    & $1.9\substack{+0.6 \\ -0.5}$  & $1.6\substack{+0.5 \\ -0.4}$  & $2.0\substack{+0.6 \\ -0.5}$  & $2.0\substack{+0.6 \\ -0.5}$ & $2.0\substack{+0.6 \\ -0.5}$  & $1.7\substack{+0.5 \\ -0.5}$    & $1.9\substack{+0.6 \\ -0.5}$  \\[3pt]
    3    & $1.3\substack{+0.7 \\ -0.1}$  &       &   & $2.5\substack{+1.5 \\ -0.0}$ & $1.8\substack{+1.0 \\ -0.0}$  & $1.8\substack{+1.0 \\ -0.0}$    & $1.4\substack{+0.9 \\ -0.0}$  \\[3pt]
    4    &                               &                               &                               & $3.1\substack{+1.6 \\ -0.3}$     & $3.3\substack{+1.5 \\ -0.2}$      &                                      &                               \\[1pt]
\hline
    B  & $1-1.6$            &      $1 - 1.6$              &     $1 - 1.6$          & $ 3.2 $                          & $3.2$  & $3.2 $       & $3.2 $  \\[1pt] 
\hline

 \end{tabular}
 \end{center}
\end{table*}

\subsubsection{Evolution of the AR between 1 and 3 August}

\begin{figure*}[!thb]
 \begin{tabular}{lllll}
\includegraphics[scale=.28,trim = 165 80 160 0, clip ]{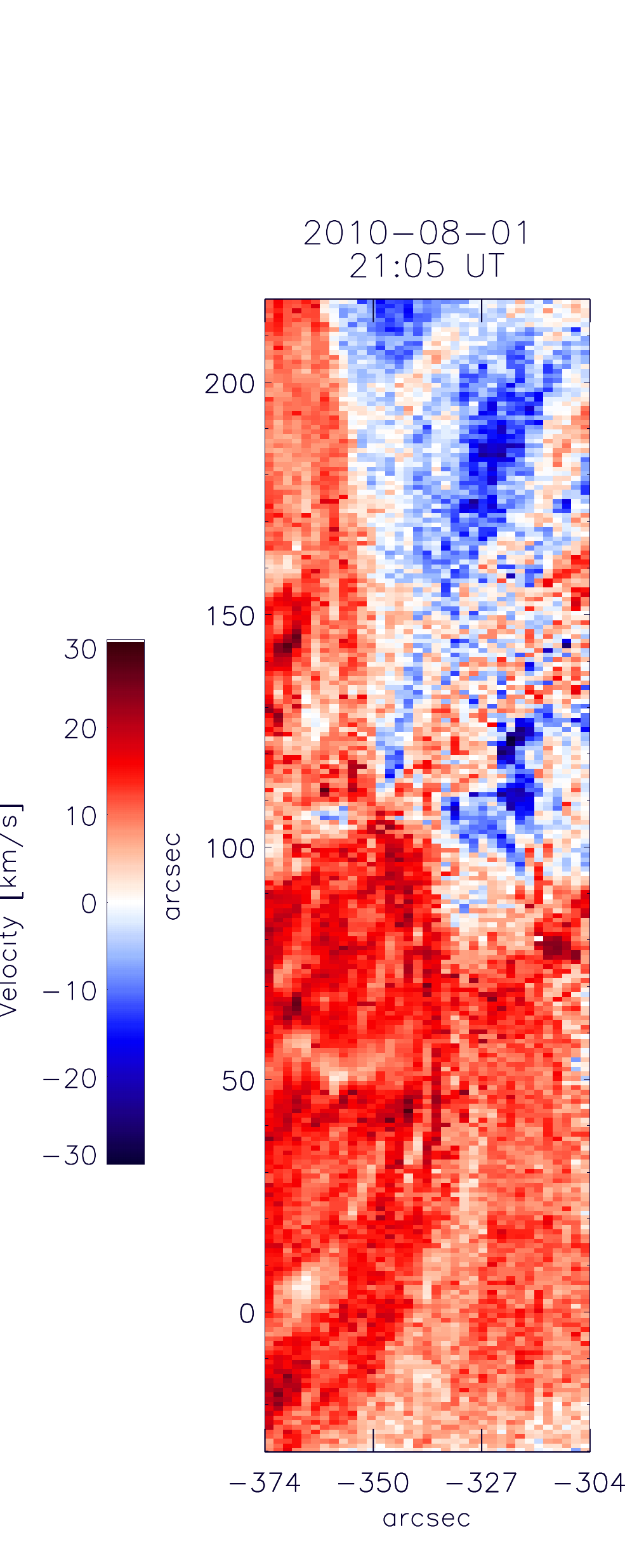}
\includegraphics[scale=0.28, trim = 30 125 5 0, clip]{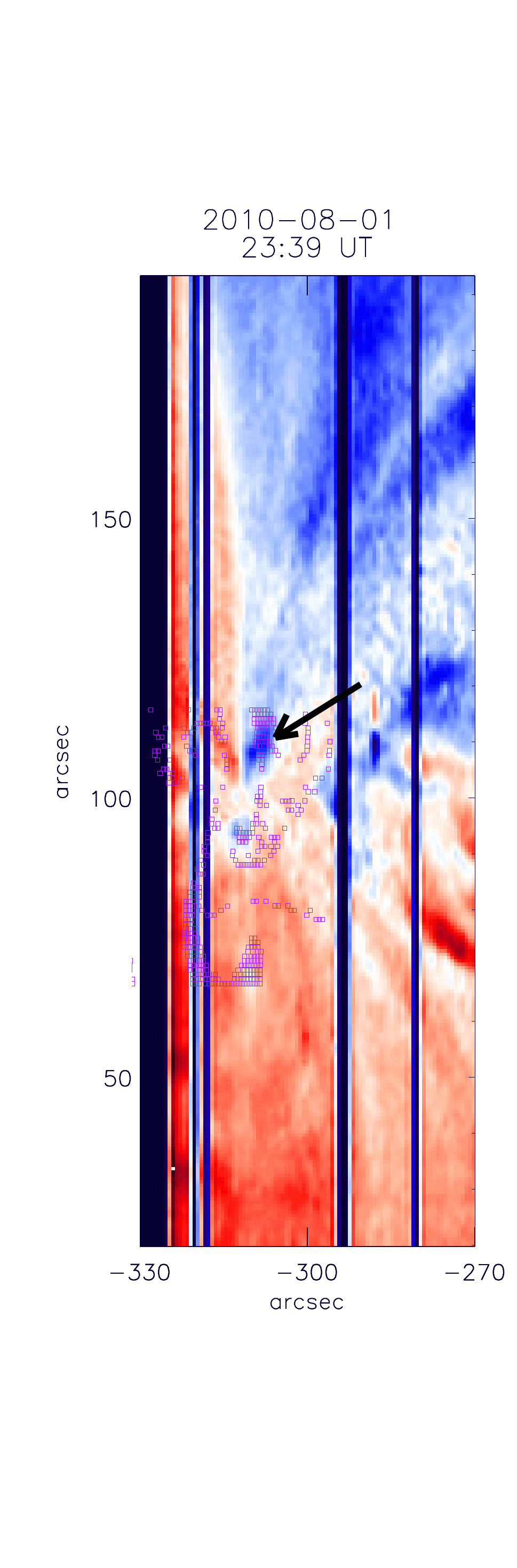}
\includegraphics[scale=.35, trim = 50 50 90 0, clip]{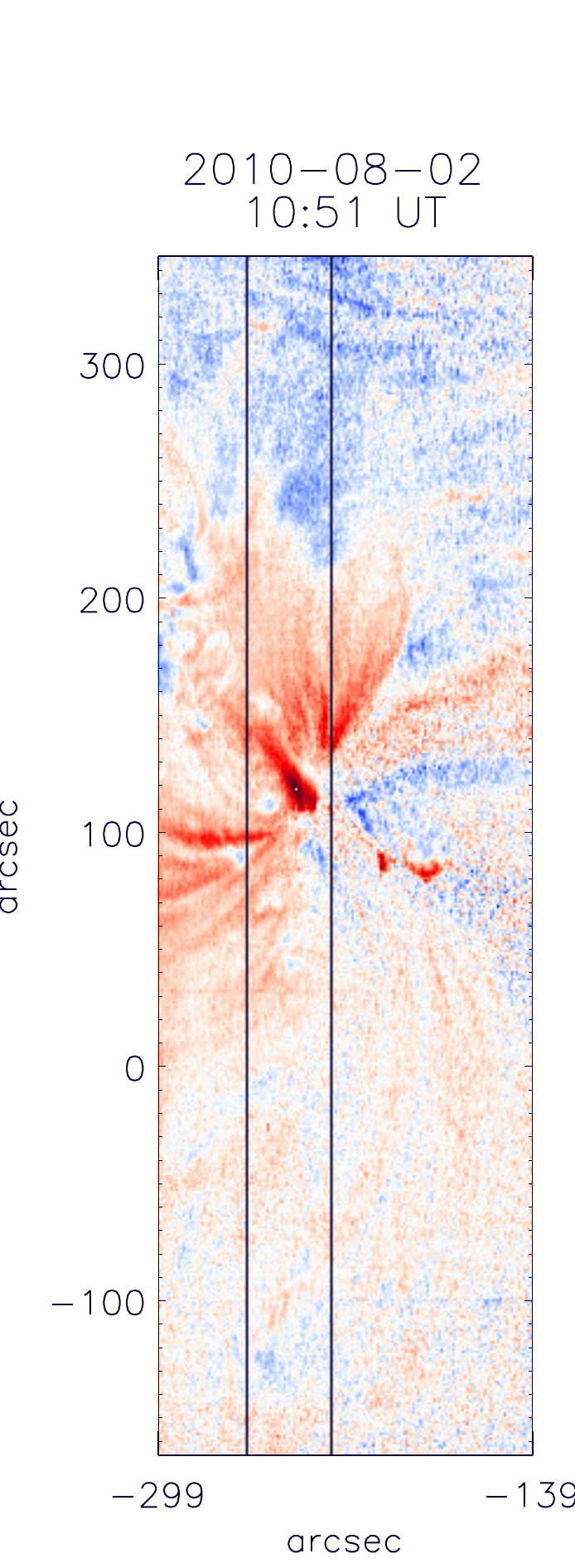}
\includegraphics[scale=.35,trim = 50 50 90 0, clip]{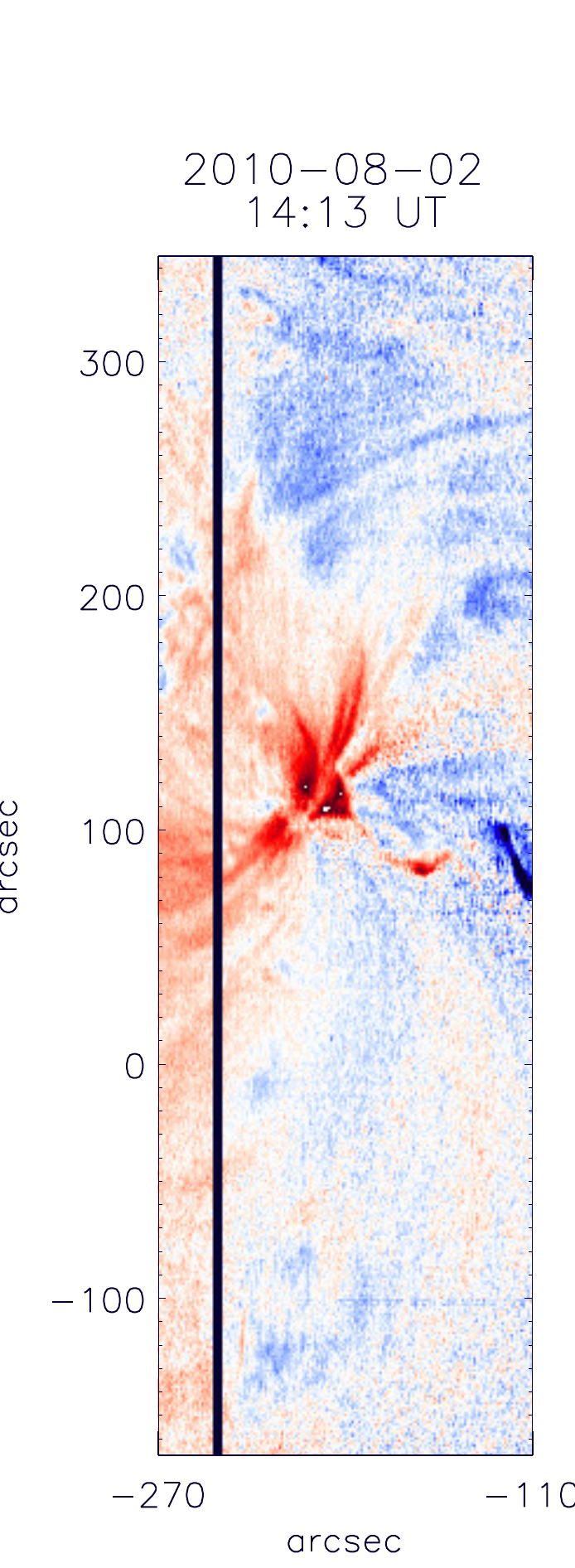}
\includegraphics[scale=.35, trim = 90 50 0 0, clip]{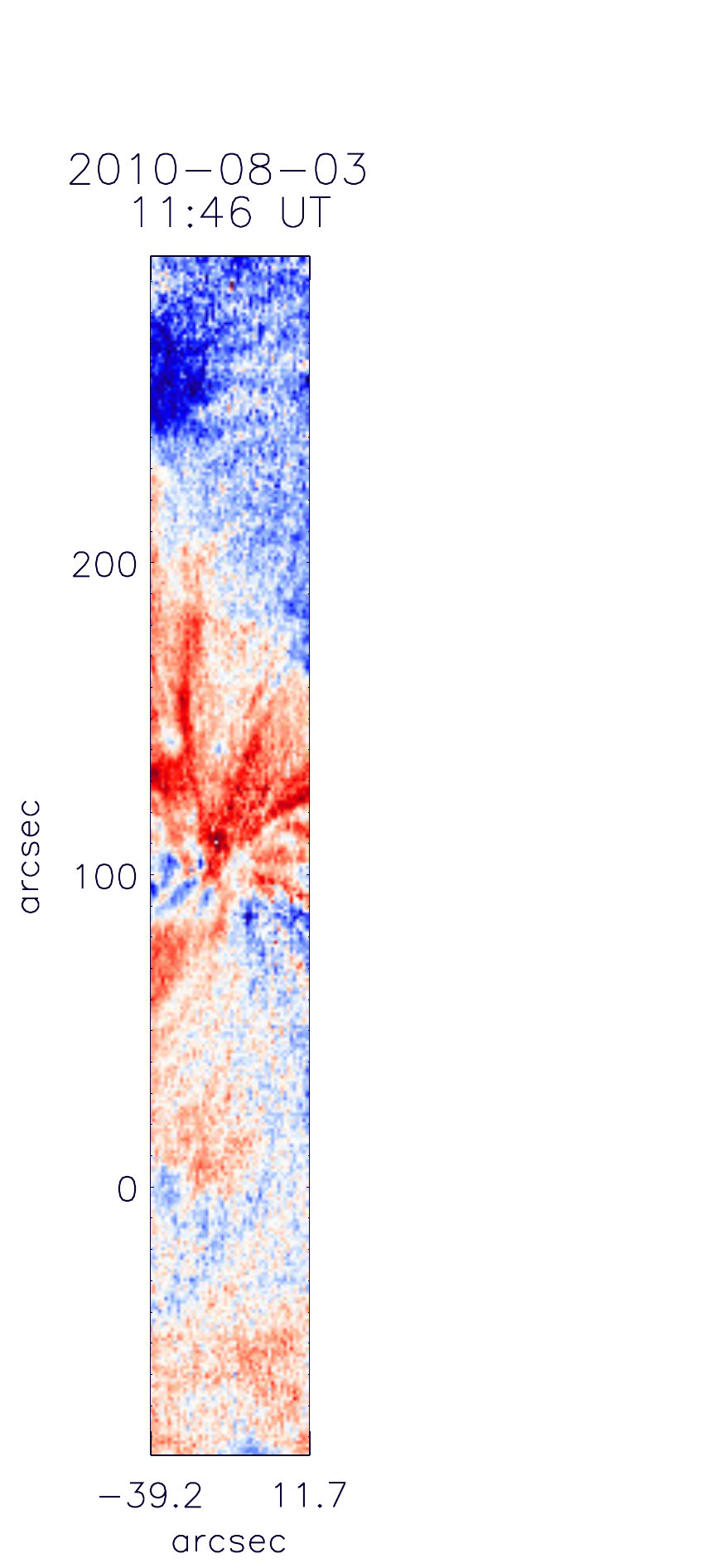}
\end{tabular}
\caption{Velocity maps from  \ion{Fe}{xii} 195.2 \AA. From the left to the right: 1 August (first and second panels, dataset A and a zoom on dataset B), 2 August at 10:51, 14:13 UT, and 3rd August. The black arrow points at the outflow region CO1. The second panel also shows, in violet crosses, the location of the footpoints of the open magnetic field. Note that the plotted FOV is different in each panel.}
 \label{fig:fe12_v}     
\end{figure*}

{Figure \ref{fig:fe12_v}  second panel shows a zoom of the velocity map of Figure \ref{fig:eis_full}. The other panels shows the maps of Doppler shift in the range $\mathrm{\pm 30 ~km/s}$ for the A, C and D rasters (1st, 2nd and 3rd August) derived from \ion{Fe}{xii}. The reference position of the line was obtained as follow.} For each spatial pixel the line profile was fitted using a double-Gaussian as suggested by \cite{young09}. The position of the brightest component of the  \ion{Fe}{xii} doublet was averaged  (in the post-processed data, that is applying the correction for orbital variation and variations along the slit) in a QS region within the raster. We do not have an appropriate QS area, however, we did this for the datasets C and D, selecting a box of 20 pixels in the bottom of the raster. For the dataset A we adopted as reference the averaged line position within the whole raster \citep{peter99}.

\begin{figure*}[!htb]
\begin{tabular}{cccc}
\centering
 \includegraphics[scale=.3, trim = 50 0 50 0, clip]{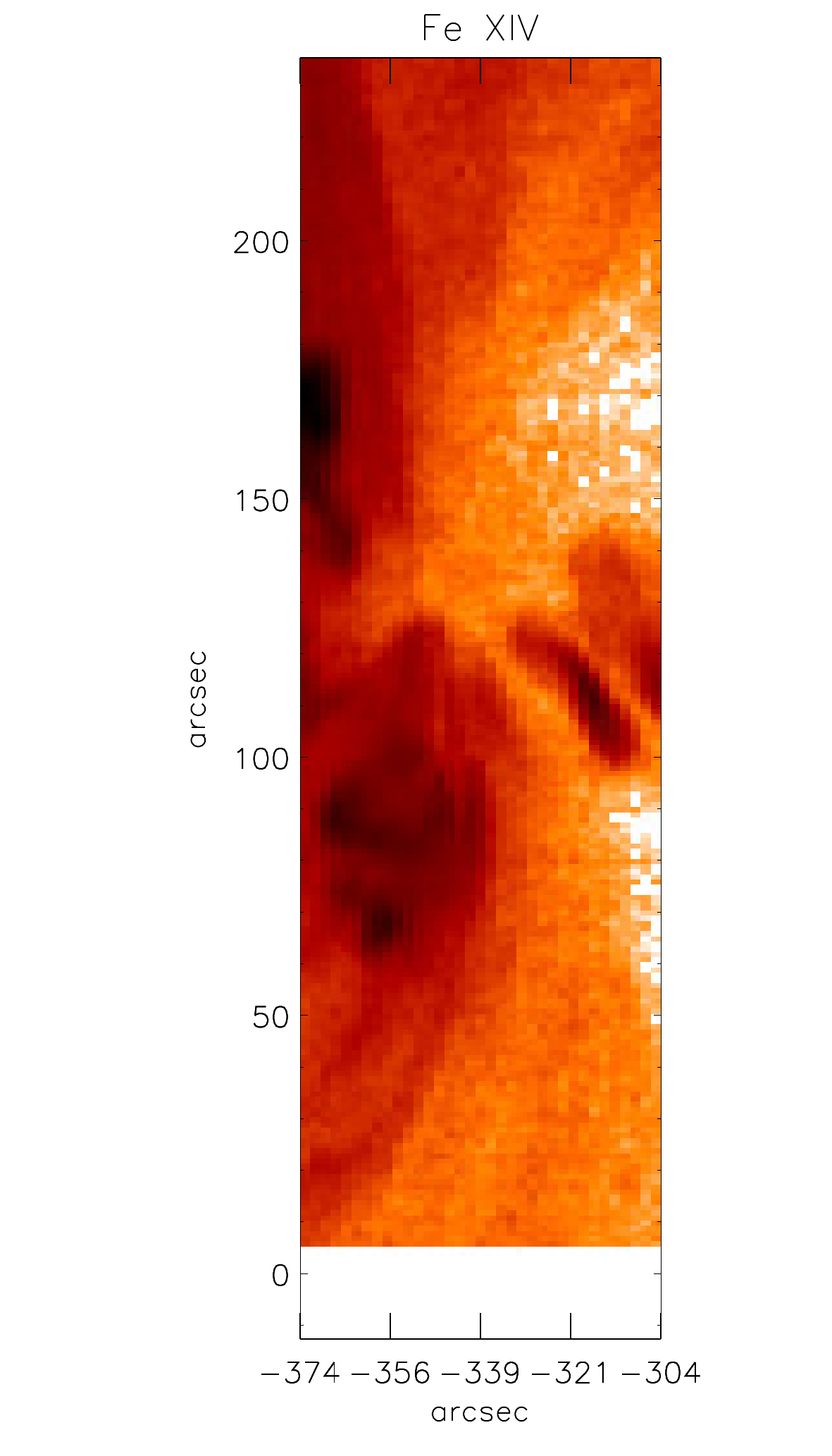}
 \includegraphics[scale=.3, trim = 0 0 50 0, clip]{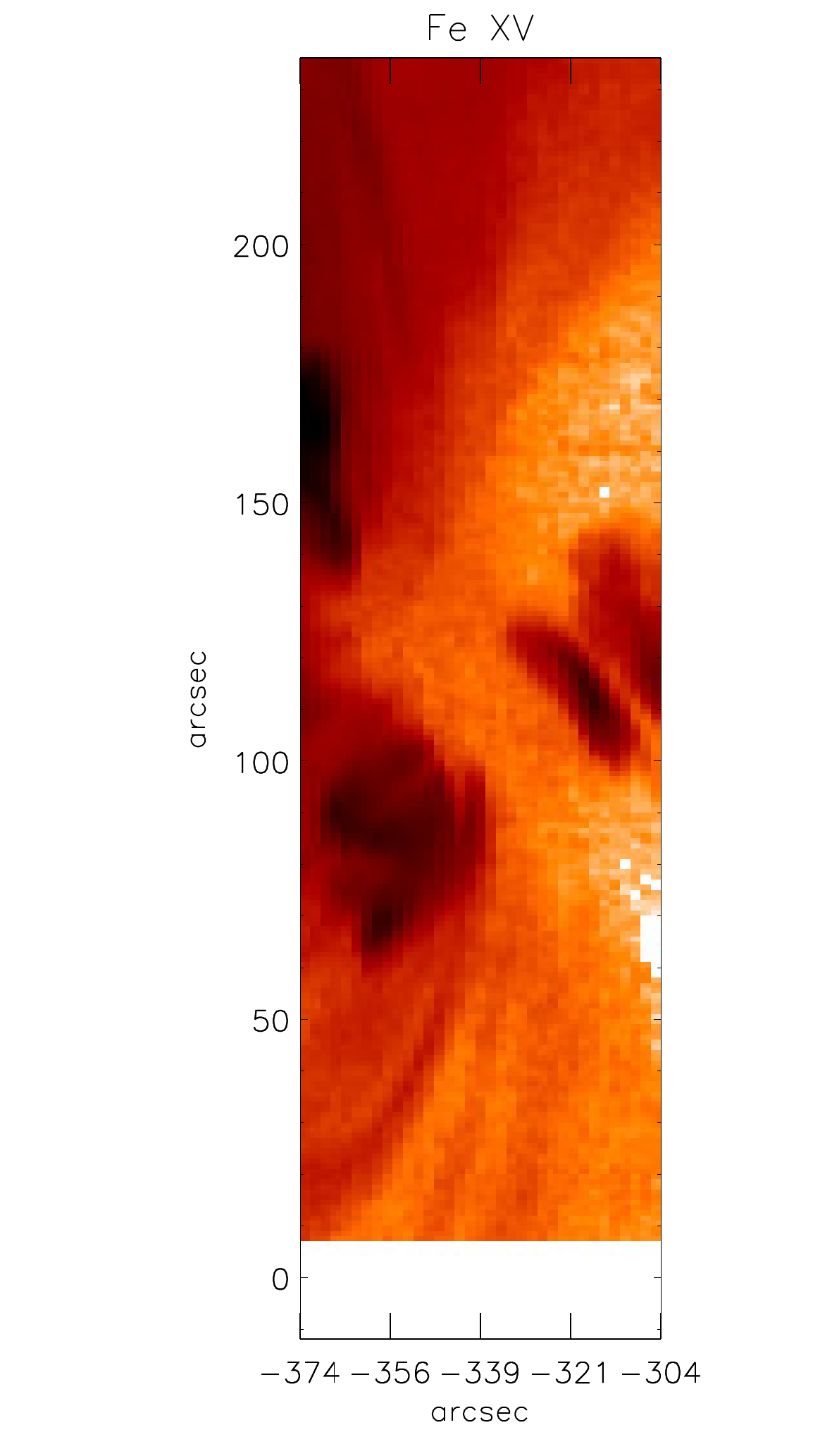}
 \includegraphics[scale=.3, trim = 0 0 50 0, clip]{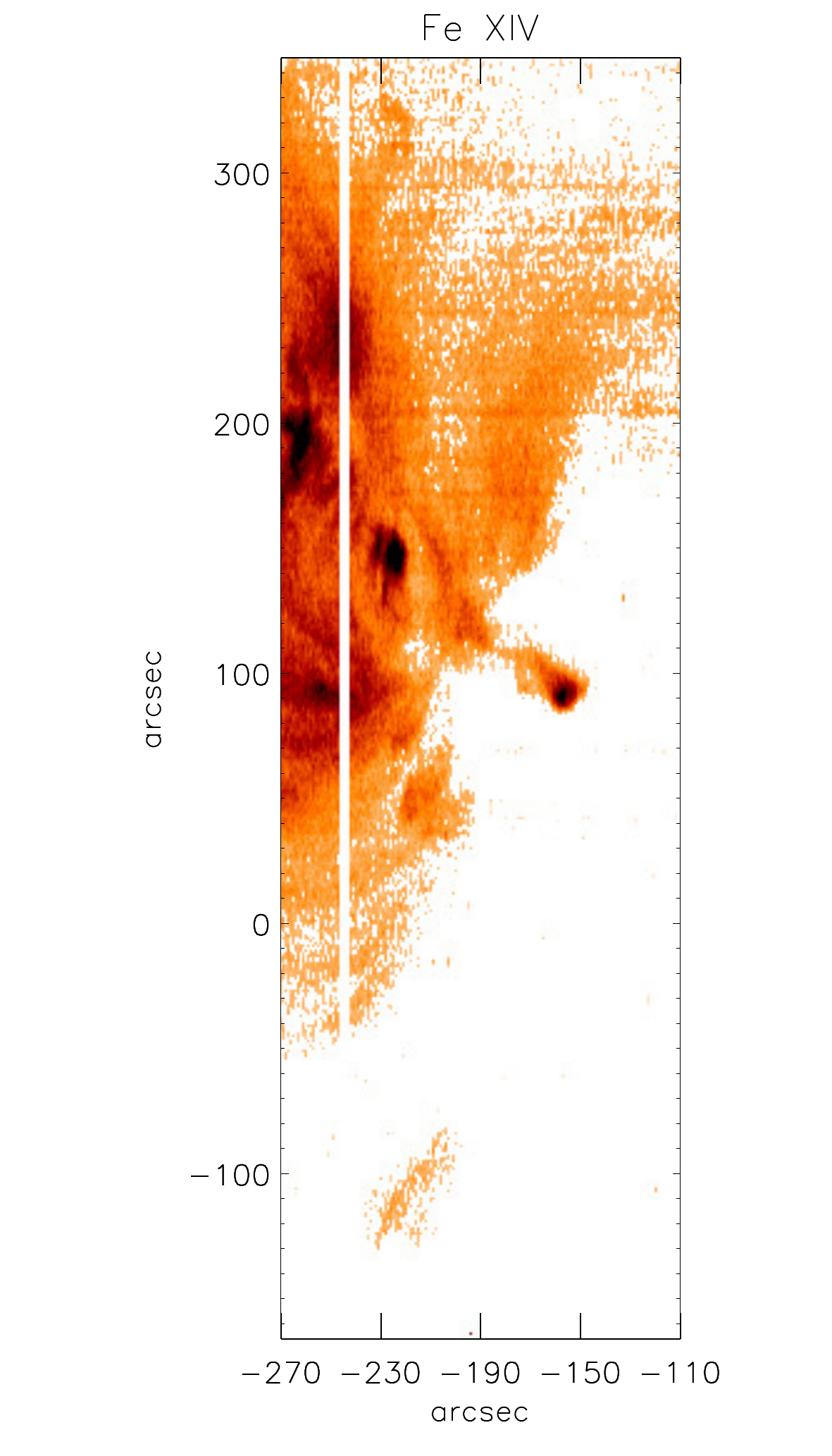}
 \includegraphics[scale=.3, trim = 0 0 50 0, clip]{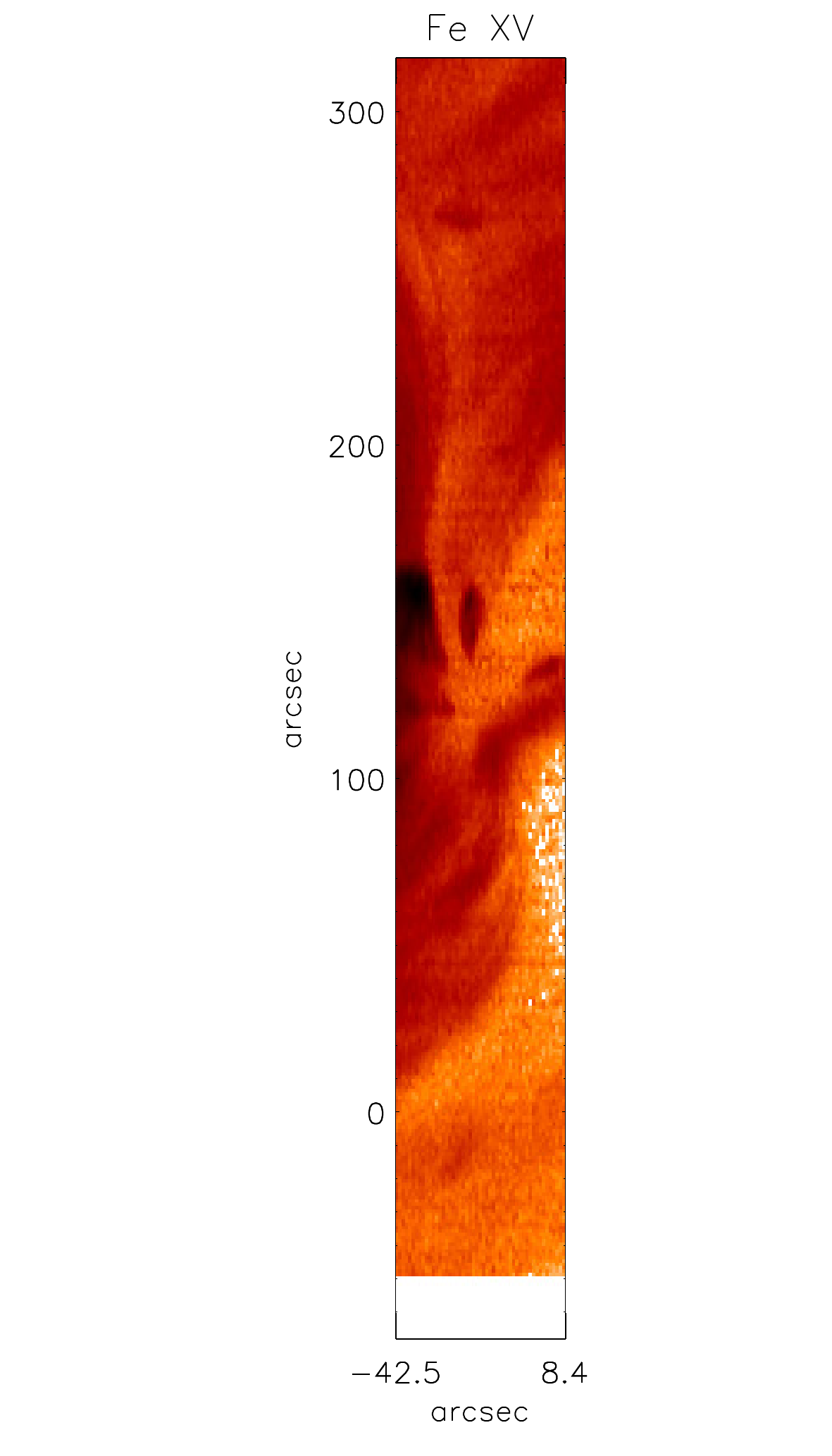}
\end{tabular}  
\caption{Negative intensity maps for hot iron lines. From left to right, \ion{Fe}{xiv} 274.2 \AA~ and \ion{Fe}{xv} 284.1 \AA~ for August 1 at 10:51 UT, \ion{Fe}{xiv} 211.3 \AA~for August 2 at 14:13 UT and \ion{Fe}{xv} 284.1 \AA~for August 3 at 11:28 UT.}
 \label{fig:fe_hot}      
\end{figure*}

{In the second velocity map in Figure \ref{fig:fe12_v} have overplotted in violet the open magnetic field lines footpoints obtained from the NLFFF extrapolation. We see that the outflow 'CO1' region is consistent with the open flux located close to the spot. The first panel in the figure is the map obtained from the data taken two hours earlier than the second panel} (and which has a slightly different FOV). The two maps are similar even though the lower exposure time of the first dataset makes the map noisier. The small outflow region (CO1 is part of it) on the north part of the spot is still present. The larger upflow areas on the top of the raster are possibly associated with the quasi-horizontal loops (as projection effect) that are better visible in the third and fourth panels (see also Figure \ref{fig:fe12_i} and Figure \ref{fig:sdo_context}). The third and fourth panels (2nd August) show some red-shift spots associated with the jet precursor activity along the jet channel. These may be the signatures of flows due to reconnection events. In this respect we also plot in Figure \ref{fig:fe_hot} intensity maps for the hotter {\ion{Fe}{xiv}--\ion{}{xv}} (depending on the available lines in the data) during the three days. We see the hot core of the AR with short loops on the 1st of August (first two panels). On the 2 August we clearly see an hot blob in the jet channel in correspondence of the red-shifted blob of Figure \ref{fig:fe12_v}, panels 3 and 4. We remind that these data are taken few hours before the jet. On the day of the jet and afterward, the north part of the spot becomes completely red-shifted and loop-shaped. 

\subsection{Linking the results from EUV data with the modelling}
\label{sec:uv_mhd}

\begin{figure*}[!htb]
    \centerline{\includegraphics[width=0.99\textwidth]{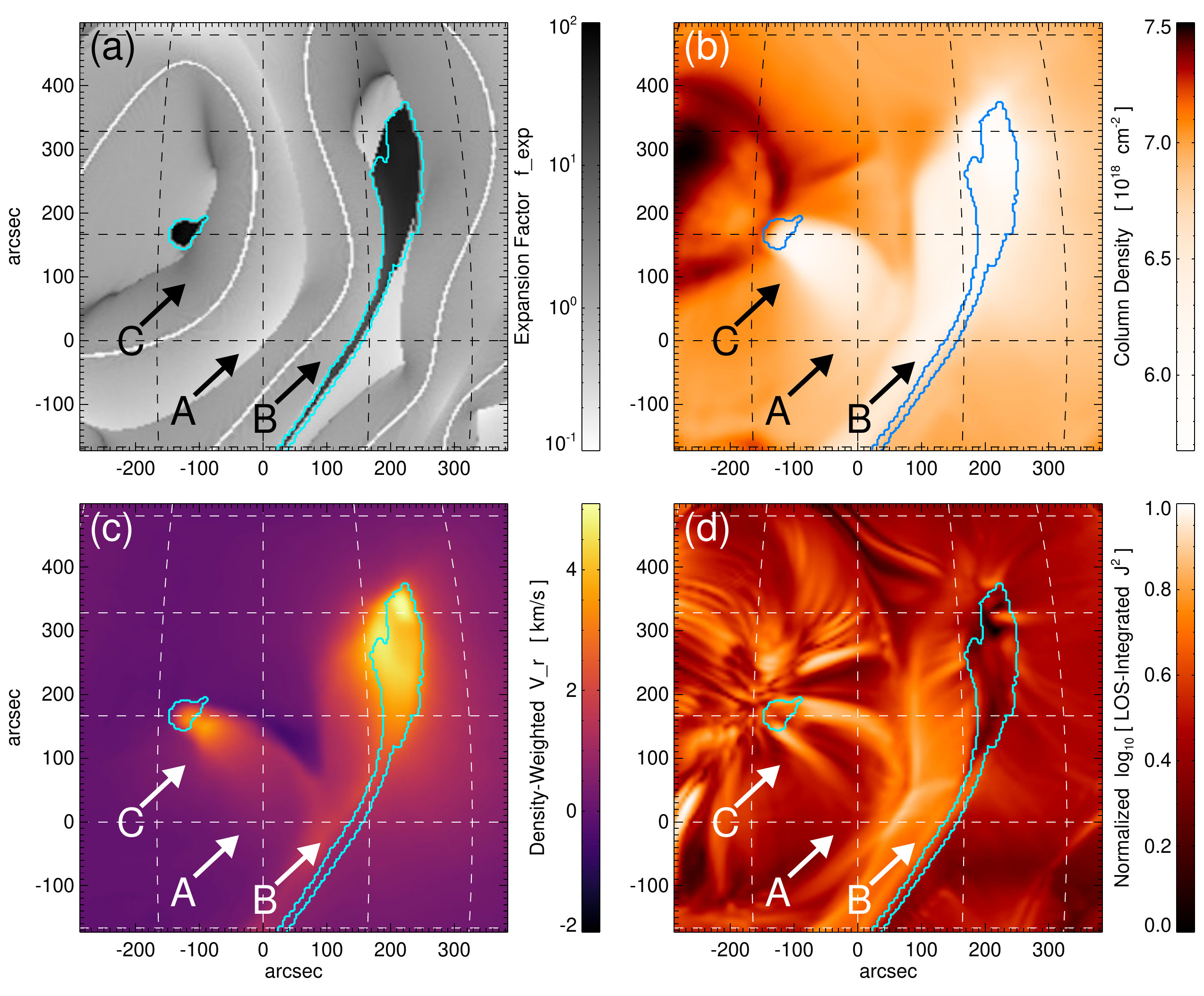}}
    \caption{Magnetic field and plasma properties in the sub-domain of the MHD results shown in Figure~\ref{fig:local_mag}. (a) Magnetic flux tube expansion factor. {The polarity inversion lines of magnetic flux distribution are shown as the white lines.} (b) Line-of-sight integrated number density. (c) Line-of-sight integrated density-weighted radial velocity. (d) Synthetic emission from line-of-sight integrated $J^2$. {In each panel the cyan/blue contours represent the open magnetic field regions.} See text for details. The arrows labeled 'A', 'B' and 'C' indicate the location of coronal features identified in Figure \ref{fig:aia_context}. {Specifically, 'C' is the observed location of the jet, 'A' is a loop foot point that originally connects to the central spot and participates in the jet eruption, and 'B' points to the elongated separatrix dome adjacent to the active region.}   
    }
    \label{fig:arms_zoom}     
\end{figure*}

In this section we examine the MHD simulation results in the context of the EUV observations and discuss the quantities from the global modelling that are provided as inputs to the more detailed FIP modelling in Sect.~\ref{sec:res_fip_charge}. Figure~\ref{fig:arms_zoom} displays several magnetic field and plasma properties of the MHD solution in the vicinity of the AR jet, i.e. in the same viewpoint shown in Figure~\ref{fig:local_mag}. 

In Figure~\ref{fig:arms_zoom}(a), we show the flux tube expansion factor \citep{WangYM1990}, calculated as $f_{\rm exp} = \left( {r_0}/{r_1} \right)^2 \left( |B_r(\mathbf{r}_0)|/|B_r(\mathbf{r}_1)| \right)$, where the starting foot point of each magnetic field line (flux tube) is given by $\mathbf{r}_0$. We construct a uniform $256 \times 256$ grid in $(\theta,\phi)$ for the field line foot points on the $r_0 = R_\odot$ lower boundary, with $\theta \in [-13^{\circ}, 35^{\circ}]$ and $\phi \in [69^{\circ}, 122^{\circ}]$. For the field line geometric quantities in this section, we use the near-Sun portion of the global domain, $r \in [ 1, 2.5 ] \, R_{\odot}$, i.e., if the field line passes through the $r=2.5R_\odot$ surface, we consider it ``open.'' {The expansion factors for closed field lines are calculated in exactly the same way. However, since $(r_0/r_1)^2 = 1$ for closed field lines, only the ratio $|B_r(\mathbf{r}_0)|/|B_r(\mathbf{r}_1)|$ contributes to $f_{\rm exp}$.} The geometry of the open-field magnetic flux tubes are a key parameter in both semi-empirical models of the solar wind \citep[e.g. the Wang-Sheeley-Arge model;][]{Arge2004, WangYM2012} and the more sophisticated, first-principles 1D calculations \citep[e.g.][]{hansteen12,Cranmer2012,Pinto2017}. Likewise, the field line length of closed flux tubes constrain the low-frequency input (and available resonances) for waves and turbulence in these loops and therefore is an important component of the FIP fractionation modeling based on the wave interactions at the corona-chromosphere transition layer.

The {open magnetic field regions are indicated with the cyan/blue contours in each panel}. In Figure~\ref{fig:arms_zoom}(a), we see that the largest expansion factors ($f_{\rm exp} \gtrsim 100$) are clearly associated with the regions of open field. The presence of open field around the spot area is confirmed by the results from the PFSS and NLFFF models shown in Figures~\ref{fig:local_mag}(a),(b), \ref{fig:OpenField}, which use HMI magnetograms or synoptic maps, as well as the MHD model using the NSO/GONG synoptic map (Figures~\ref{fig:local_mag}(c)).

Figure~\ref{fig:arms_zoom}(b) plots the column number density, $N_{\rm col}(\theta,\phi) = \int_{\rm los}{ n_e(r,\theta,\phi) \, d\ell}$, where we approximate the line-of-sight integral as along the radial direction and take $n_e = n_p \equiv \rho/m_p$. For panels~\ref{fig:arms_zoom}(b) and (c) the range of the radial line-of-sight was $1.02R_\odot \le r \le 2.50R_\odot$ in order to highlight the low-coronal structure. For panel~\ref{fig:arms_zoom}(d) the radial line-of-sight was $1.05R_\odot \le r \le 2.50R_\odot$ to minimize the contribution of the lower boundary current structure. Figure~\ref{fig:arms_zoom}(c) shows the line-of-sight integrated, density-weighted radial velocity, given by $\langle V_r \rangle = \int{ w \; V_r \, dr}$, where the weights are defined as the normalized density distribution with height $w(r,\theta,\phi) = n(r,\theta,\phi) / \max{[ N_{\rm col} ]}$.

{The density map of panel (b) can be qualitatively compared to the AR density around the spot shown in the first panel of the second row of Fig. \ref{fig:fe12_i}}, obtained from about 1MK plasma emission data of the 1st of August. The east denser structures are there, while the less dense jet channel in the model is replaced in the data by a much more dense short loops structure. We know that the latter will be modified and reduced prior and after the jet (see also Figure \ref{fig:aia_context}) leaving the space to the more 'empty' jet channel.

The velocity map in figure (c) cannot be quantitatively compared with the velocity maps from the data, as the latter are LOS averaged values. {However, it is interesting to see that the model identifies an outflow area in the jet channel similarly} to the data (see Fig. \ref{fig:eis_full} panel 5 and Fig. \ref{fig:fe12_v}).

Figure~\ref{fig:arms_zoom}(d) shows a synthetic EUV emission formulation that is a simplified version of the \citet{Cheung2012} electric current density-based emission proxy. Here we use  $I_J(\theta,\phi) = \int J^2(r,\theta,\phi) \, dr$, which avoids averaging over a field-line for every point along the radial line of sight. The \citet{Cheung2012} current-based emission proxy has been favorably compared to the morphology of AIA 131\AA\, \citep[see also][]{Hoeksema2020}, and here it shows more qualitative agreement with our hottest lines, specifically \ion{Fe}{xiv}--\ion{}{xv} (formed around 1.5 -- 2.5 MK, see Fig.
\ref{fig:fe_hot}).

\subsection{Modelling the FIP bias}
\label{sec:res_fip_charge}

The modelling described above allows us to identify magnetic field strengths associated with the various regions singled out for spectroscopic analysis, and use this information to calculate expected FIP fractionations. We first consider the closed field regions, CL1, CL2, and CL3 {on Figure 9.} We assume that the FIP fractionation in closed loops is caused by Alfv\'en waves generated in the corona {\citep{laming17}}, presumably by nanoflare reconnection events. This provides a rationale for having waves resonant with the loop, where the wave travel time from one loop footpoint to the other is equal to an integral number of wave half periods. In such a case, fractionation by the ponderomotive force is restricted to the top of the chromosphere where H is becoming ionized. Moving low FIP ions through a background of protons requires a strong ponderomotive force, and even a small neutral fraction can diffuse back down the concentration gradient eliminating any fractions. Consequently elements have to be almost completely ionized to fractionate, and intermediate elements like \ion{S}{} behave much more as high FIP elements. This leads to significant fractionation in e.g. the \ion{Fe}{}/\ion{S}{} abundance ratio. 

A further consideration for waves of coronal origin from nanoflares is that the Alfv\'en speed be sufficiently high. This means that waves from a reconnection event can reach the chromosphere and cause fractionation before the heat conduction front arrives to evaporate the chromospheric plasma up into the corona. The simplest (possibly naive) criterion for FIP fractionation should then be that the coronal Alfv\'en speed being higher than the electron thermal speed. The closed region CL1, CL2 and CL3 all have one footpoint close to the open field jet origin region, and the other footpoint on the other side of the neutral line. In all cases the Alfv\'en speed at the loop footpoints is quite high. The magnetic field here is typically $\pm 25$G giving an Alfv\'en speed of over 1000 km s$^{-1}$. Higher up in the corona the Alfv\'en speed is  100 - 200 km s$^{-1}$. For comparison the electron thermal speed is $\sim 2000$ km s$^{-1}$.  We therefore do not predict significant FIP fractionation in CL1, CL2, or CL3, and any prediction we did make would not include this dynamical effect when the heat conduction speed is greater than the Alfv\'en speed. The hot regions have much higher coronal Alfv\'en speeds, typically above 500 km s$^{-1}$ at the lowest value, and higher than that for significant portions of the loop.

In Table \ref{tab:fip} we give predicted FIP fractionations for various elements relative to O for both the HOT and HOT2 regions. The loop lengths and magnetic fields are taken from the MHD/GONG reconstruction, and {in the absence of other information on MHD wave fields,} the coronal wave amplitudes are chosen to give the Fe/O fractionation of about 3. {These wave amplitudes give Poynting fluxes commensurate with those required by coronal heating \citep{laming15,dahlburg16}, and could in principle (but not in our case) be determined from reconnection parameters or boundary conditions.} The wave angular frequency is taken to be the resonant frequency for each loop, assuming that the Alfv\'en waves have a coronal origin. For each model we give two fractionations, one arising solely from the ponderomotive force as would be revealed by spectroscopy of emissions from close to the solar surface, and a second incorporating also the effects on abundances of the conservation of the first adiabatic invariant in the expanding magnetic field lines \citep{laming17b,laming19,kuroda20}, as would be detected {\it in--situ}. We assume (as before) that the plasma becomes collisionless at a density of $\sim 10^6$ cm$^{-3}$, which in this open field region occurs at a radius of about 2.5 R$_{\odot}$. The magnetic field expansion of a factor of 625 is taken from {the data plotted in} Figure \ref{fig:arms_zoom}, panel (a), {for open field regions only}.

\begin{table*}[tbh]
\centering
\begin{tabular}{|c|cc|cc|cc|}
  \hline
  & \multicolumn{2}{c}{HOT}  & \multicolumn{2}{c}{HOT2}   & \multicolumn{2}{c|}{CO1}  \\
  \hline
loop length (km)& \multicolumn{2}{c}{25,000}  & \multicolumn{2}{c}{75,000}   &  \multicolumn{2}{c|}{open}\\
loop B(G)& \multicolumn{2}{c}{10}  & \multicolumn{2}{c}{15}   & \multicolumn{2}{c|}{25}  \\
wave ang. freq.(rad s$^{-1}$) & \multicolumn{2}{c}{0.1241}  & \multicolumn{2}{c}{0.0527}   & \multicolumn{2}{c|}{0.021}\\
coronal wave amp. (km s$^{-1}$)& \multicolumn{2}{c}{68}  & \multicolumn{2}{c}{88}   & \multicolumn{2}{c|}{150}\\
  \hline
He/H&  0.64& 0.53& 0.69& 0.57& 0.73& 0.60\\
He/O& 0.61& 0.74& 0.56& 0.68& 0.86& 1.04\\
C/O& 0.78& 0.83& 0.79& 0.84& 0.93& 0.99\\
N/O& 0.66& 0.68& 0.67& 0.69& 0.88& 0.91\\
Ne/O& 0.69& 0.65& 0.69& 0.65& 0.88& 0.83\\
Mg/O& 2.30& 2.05& 2.27& 2.02& 2.31& 2.05\\
Si/O& 2.18& 1.83& 2.19& 1.84& 2.24& 1.88\\
S/O& 1.11& 0.88& 1.02& 0.81& 1.06& 0.84\\
Ar/O& 0.84& 0.60& 0.85& 0.59& 0.93& 0.66\\
Ca/O& 3.63& 2.60& 3.90& 2.79& 3.14& 2.25\\
Fe/O& 3.35& 1.96& 3.58& 2.08& 2.98& 1.75\\
\hline
\end{tabular}
\caption{FIP fractionations for HOT, HOT2, and CO1. The first fractionation is due to the ponderomotive force only, the second includes conservation of the 
first adiabatic invariant.}
\label{tab:fip}
\end{table*}

The magnetic field strength also affects the FIP fractionation in open field regions, both by the adiabatic invariant {\it in--situ}, and lower down by a different mechanism. Here we consider the fractionation to occur as waves coming up from the photosphere are mode converted to fast mode and Alfv\'en waves at the layer where sound and Alfv\'en speeds are equal, with all fractionation occurring above this layer \citep{laming21}. In stronger magnetic field regions, this mode conversion layer lies deeper in the chromosphere, allowing fractionation to occur over a greater depth, with consequently stronger results. In Table \ref{tab:fip} we also give model fractionations for the CO1 region, for the same two cases as before. Due to the excitation of slow mode wave by the Alfv\'en driver in the open field, the fractionation is much less sensitive to the Alfv\'en wave amplitude than is the case for closed fields, as seen in \citet{laming19}. While similar FIP fractionation can be predicted for open field as for closed field  if the magnetic field is sufficiently strong, there is an important difference in the fractionation of the \ion{S}{}/\ion{O}{} abundance ratio. In the open field, \ion{S}{} behaves more like a low FIP element, while in closed field it more closely resembles a high FIP element. The reasons for this have been previously discussed \citep{laming19,kuroda20}, relating to the different layer is the chromosphere over which the ponderomotive force develops in open and closed field, and validated with different solar wind and SEP observations \citep{reames18}. It is curious then that CO1 appears to exhibit abundances more appropriate to a closed loop than to open field (see Fig. \ref{fig:eis_full} and Table 4),  but this appears to be related to its magnetic field strength of $\sim 25$G. Following revision of the S chromospheric ionization balance \citep{laming21}, magnetic fields of order 100 G are required to enhance the \ion{S}{} abundance in this way.

\subsection{Heliospheric Back-mapping and {\it In--situ} Observational Results} 
\label{res:insitu}

\begin{figure*}[!htb]
    \includegraphics[width=1.0\textwidth]{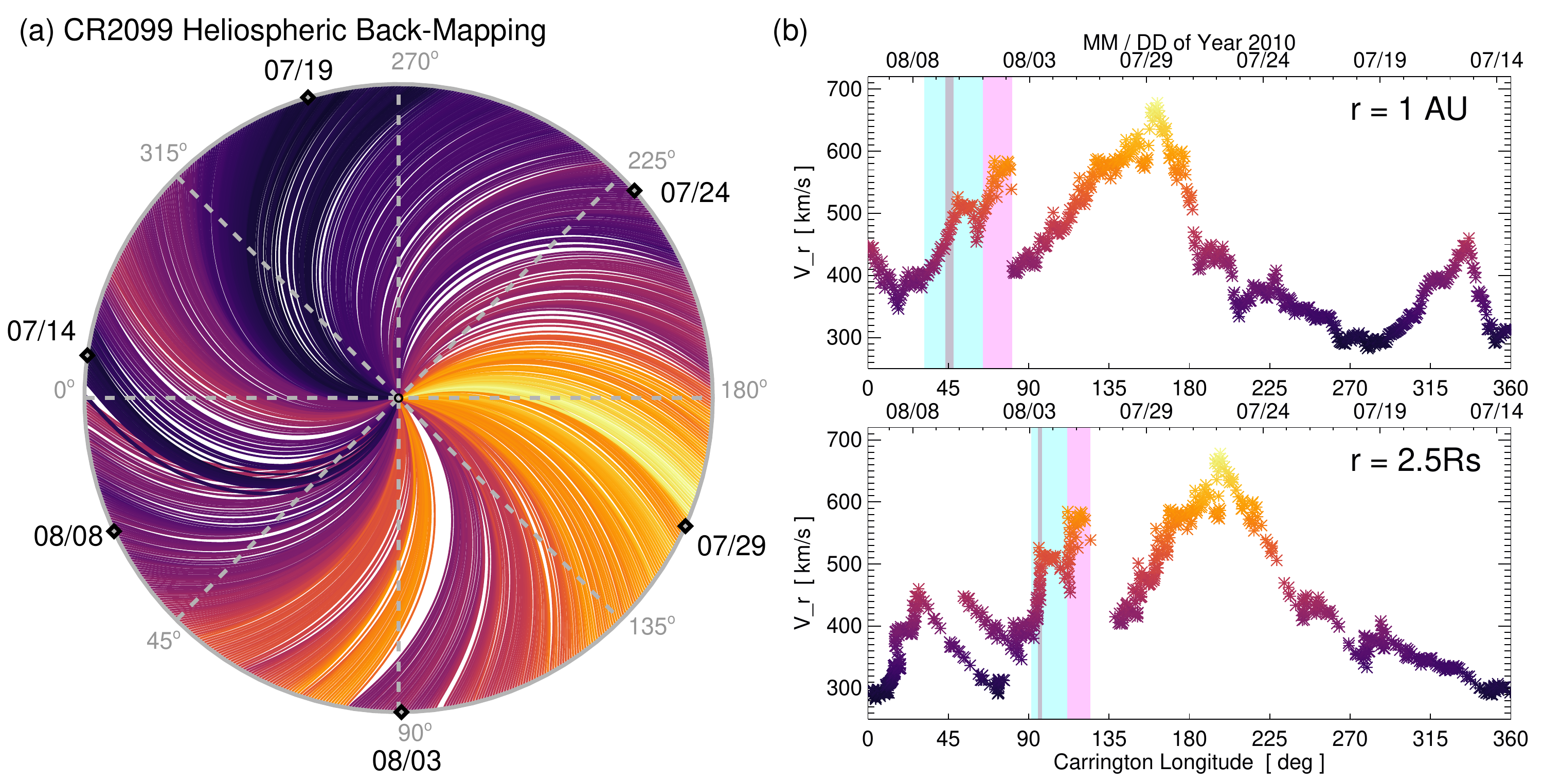}
    \caption{Constant $v_r$, ballistic back-mapping during CR2099. Panel (a) Parker spiral streamlines color-coded by the 1~AU radial proton velocity. Panel (b) Conversion of the uniform temporal sampling in Carrington longitude at 1~AU into the Carrington longitude distribution at the $r=2.5R_\odot$ source surface and origin time estimate. {The vertical cyan bar represents the heliospheric back-mapping of the {\it in-situ} solar wind interval potentially associated with the August~02 coronal jet, including the period of low \ion{S}{}/\ion{O}{} in red/orange. The earlier magenta region corresponds to the ICME interval indicated in Figure~\ref{fig:Jet2010insitu}.}}
    \label{fig:CRmap}       
\end{figure*}

\begin{figure*}[!th]
    \includegraphics[width=\textwidth]{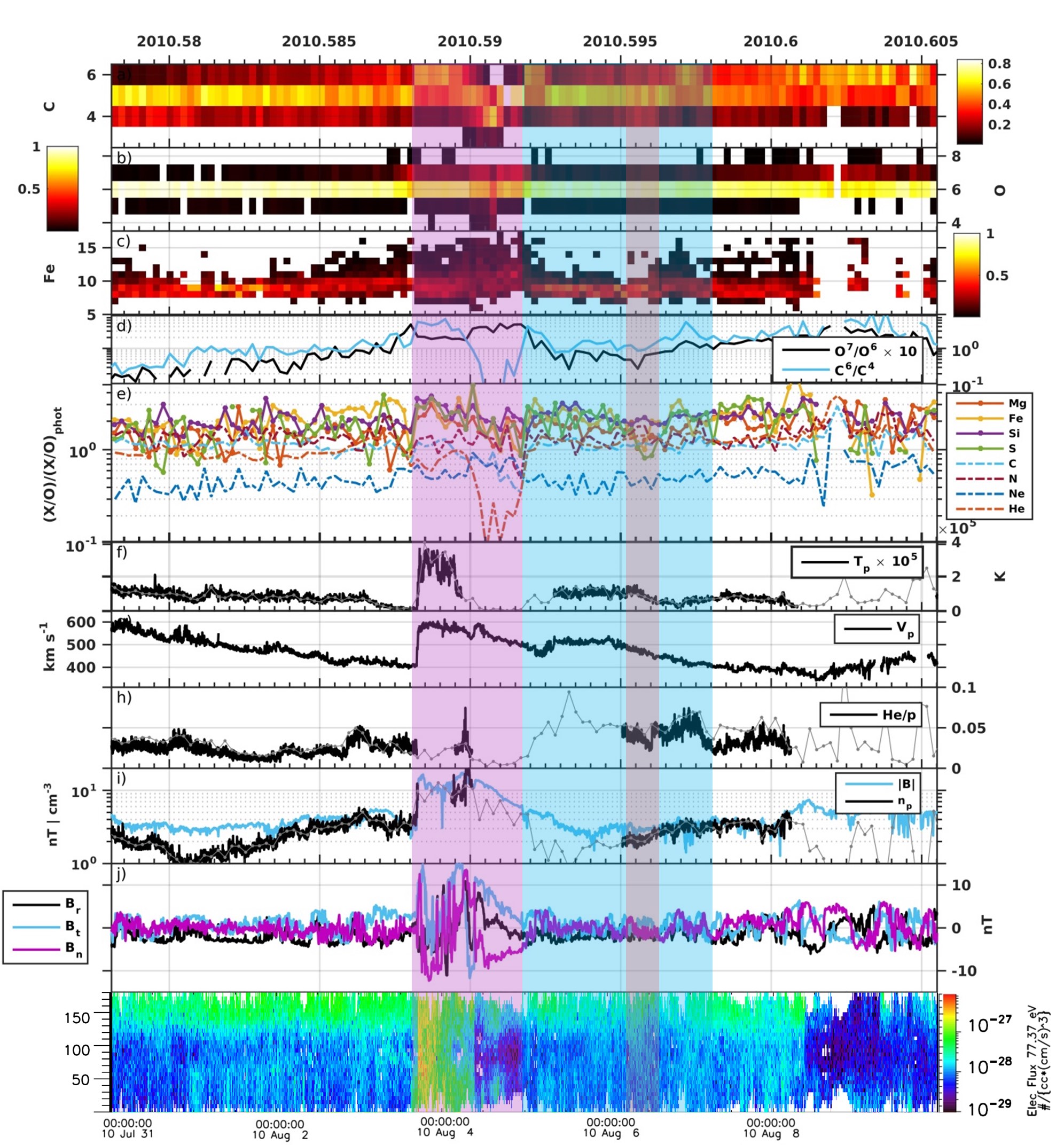}
    \caption{Multi-panel plot within the timeframe of the 2010 jet arrival period at L1 between day $212-222$. The purple shaded region denotes the ICME period and the shaded blue region are the possible boundaries of the jet based on Figure \ref{fig:CRmap}. Panels (a)--(c) are the normalized charge state distributions for C, O, and Fe. Panel (d) shows O$^{7+}$/O$^{6+}$ $\times10$ and C$^{6+}$/C$^{4+}$. Panel (e) shows the elemental composition compared to O divided by its photospheric ratio, X/O/X/O$_{phot}$. Panel (f)--(i) shows the proton temperature, velocity, He/p, and density and magnetic field magnitude, respectively at high resolution ($<$1 hour) in black and 2 hour resolution in gray. Panel (j) shows the radial, tangential, and normal magnetic field components. Panel (k) shows the pitch angle electron flux at 77.37 eV with vertical axis in degrees.}
    \label{fig:Jet2010insitu}   
\end{figure*}

Figure~\ref{fig:CRmap} illustrates the back-mapping procedure for CR2099 which starts at 2010-07-13 09:38UT and goes through 2010-08-09 14:48UT.  Figure~\ref{fig:CRmap}(a) plots the Parker spiral streamlines, color-coded by radial velocity. Figure~\ref{fig:CRmap}(b) plots the bulk proton radial velocity in 1-hour averages measured by ACE at 1~AU in Carrington longitude (here time runs from right-to-left, as indicated with the top $x$-axis labels). {The vertical bars identify three {\it in--situ} intervals that are highlighted in Figure~\ref{fig:Jet2010insitu} and are discussed in detail below: magenta indicates an ICME interval, cyan indicates the estimated time of arrival of the jet material, and the red/orange region within the cyan interval has a distinct elemental composition.} The estimated arrival time of the 2010-08-02 coronal jet material is taken to be between DOY 217.0 (August~05) to DOY 219.5 (August~07 12:00UT) corresponding to the range of Carrington longitudes 31.6--64.3$^\circ$ at 1~AU. This range maps back to Carrington longitudes 91.3--111.7$^\circ$ at $r=2.5R_\odot$, or equivalently, comfortably within the $\pm$10$^\circ$ range of the central meridian (Carrington longitude $\phi=94^\circ$) on August~02, which is indeed the location of the spot (see Figure \ref{fig:global_mag}).

{\it In--situ} measurements of {the solar wind and our jet interval of interest} were investigated using MAG \citep{Smith1998}, Solar Wind Electron Proton Alpha Monitor (SWEPAM) \citep{McComas1998}, and Solar Wind Ion Composition Spectrometer (SWICS) \citep{Gloeckler1998} instruments on the Advanced Composition Explorer \citep[ACE;][]{Stone1998}, along with the 3DP instrument \citep{lin1995} on the Wind spacecraft showing the pitch angle distribution of electron flux. Figure \ref{fig:Jet2010insitu} is a multipanel plot showing solar wind properties at the ACE instruments for panels \ref{fig:Jet2010insitu}(a)--(j) and at Wind for panel \ref{fig:Jet2010insitu}(k).

The purple box across the data shows {an ICME interval} with boundaries taken from \cite{Richardson2010}. {The ICME-driven shock arrives on August 3 at $\sim$17:00UT followed by a large sheath region and what appears to be two coherent magnetic flux rope ejecta \citep[e.g., see][]{Harrison2012,Moestl2012}. The ICME interval exhibits distinct {\it in--situ} ionic and elemental composition signatures that have been interpreted as filament material \citep{Sharma2012}, which may be expected from previous analyses \citep[e.g.][]{Lepri2010,Lepri2021}. This ICME period has been linked to a series of three sequential CME eruptions of large, quiescent filaments on August 1st at $\sim$06:30UT, $\sim$16:30UT, and $\sim$21:00UT \citep[e.g.][]{Schrijver2011,Torok2011,Temmer2012}. The filament channel source region of the third eruption is approximately 30$^{\circ}$ west and 15--20$^{\circ}$ north of AR 11092 and its quiescent filament erupts $\sim$20~hr earlier than our jet event on August 2nd.} 

The blue shaded region is the predicted arrival period of the jet using the ballistic back-mapping from Figure \ref{fig:CRmap}. The blue shaded region is characterized by a period of steady $\sim500$~km~s$^{-1}$ wind followed by monotonically decreasing speed towards the second half as shown in panel \ref{fig:Jet2010insitu}(g). The proton temperature remains between $0.5-1\times10^5$~K, and the density $<3$~cm$^{-3}$, both typical properties of fast solar wind.

The relative ionization fractions of \ion{C}{}, \ion{O}{}, and \ion{Fe}{}, respectively, shown in panels \ref{fig:Jet2010insitu}(a)--(c) are observed to shift to higher ionized states, with elevated C$^{6+}$ and O$^{7+}$ along with the appearance of higher ionized Fe charge states, Fe$^{>12+}$, towards the second half of the blue region. This is also seen in panel \ref{fig:Jet2010insitu}(d) showing elevated values of O$^{7+}$/O$^{6+}$ and C$^{6+}$/C$^{4+}$ indicating higher coronal temperature or higher coronal densities. Panel \ref{fig:Jet2010insitu}(e) shows the FIP bias as X/O/X/O$_{phot}$ for each element, X, listed. Within the blue region we see a gradual decrease in the abundance of low FIP elements \ion{Mg}{}, \ion{Fe}{}, \ion{Si}{} and high FIP S relative to \ion{O}{} that initially range between $2-3$ begin to dip below 2 within the red box.  Other ratios, \ion{Ne}{}/\ion{O}{}, \ion{N}{}/\ion{O}{}, \ion{C}{}/\ion{O}{} and \ion{He}{}/\ion{O}{} hardly change. The fact that \ion{S}{}/\ion{O}{} is high in the blue region suggests an origin in strong open magnetic fields, with its decrease in the red region, more pronounced than the decrease in any other element, implies either a weaker magnetic field open region, or a closed loop as the origin for this plasma. Other abundance ratios are not so conclusive. \ion{Ne}{}/\ion{O}{} strongly favours a closed loop origin, \ion{C}{}/\ion{O}{} and \ion{N}{}/\ion{O}{} favour open field, while \ion{He}{}/\ion{O}{} here measured above 1 is at variance with expectations from \ion{He}{}/\ion{H}{} in panel \ref{fig:Jet2010insitu}(h), typically about 0.5 for an absolute abundance ratio of about 0.04. Although not purely photospheric, this transition to near photospheric abundances in \ion{Mg}{}/\ion{O}{}, \ion{Fe}{}/\ion{O}{}, \ion{Si}{}/\ion{O}{}, and \ion{S}{}/\ion{O}{} along with a reduction in the ionization stage of charge states, may also be an indication that its source lies deeper in the chromosphere. These properties may also suggest that the ejection site was already at open field lines with plasma flowing out with coronal composition as we inferred with our EUV and MHD analysis while the shift to lower FIP values within the jet structure may indicate the plasma ejected also contained some mixture of coronal and chromospheric composition.

Moreover, prior to the red region within the blue box, the temperature and velocity increase slightly then remain steady before reaching the red box as shown in panel \ref{fig:Jet2010insitu}(f)--(g). This period also shows enriched He/p, $\sim 0.5-1.0$ values as seen in ICMEs \citep{Hirshberg1972, Borrini1983}. The red box marks a transition to cooler proton temperature and monotonically decreasing velocity values indicating a rapid expansion in the plasma. Furthermore, the period within the blue box is characterized by low magnetic field magnitude, as shown in panel \ref{fig:Jet2010insitu}(i), where B gradually decreases and remains below 5 nT while the magnetic field components in panel \ref{fig:Jet2010insitu}(j) show small fluctuations in the B$_N$ and B$_T$ and a steady B$_R$ component. During this period in blue, the IMF points back towards the Sun, where B$_{R} < 0$, B$_{T} \sim \pm 3$, and B$_{N} \sim \pm3$. This can also be observed with the electron pitch angle distributions in panel \ref{fig:Jet2010insitu}(k) which observe an uninterrupted concentrated beam of electrons directed towards the Sun, anti parallel (135--180$^{\circ}$) to the interplanetary magnetic field (IMF), that are likely rooted at the Sun.

\section{Discussion and Conclusions}
\label{sec:discussion}

In this work we have combined data and modelling work in order to reconstruct the solar and heliospheric magnetic and plasma properties for the days around the appearance of a coronal jet. The goal was to show that it is possible to reconcile data and modeling results, and that they should be used together to provide all the elements to reconstruct the history of this ejection from its solar source to its possible travel into the interplanetary medium. This event was taken as test case in preparation of the Solar Orbiter data exploitation.
The main results are summarized and discussed in the following. 

The large scale {coronal} magnetic field distribution for the day of the jet was obtained applying different methods and initial conditions (Sections \ref{sec:GLB_res}). The results are consistent in finding open flux nearby the spot at the jet location that is, a channel through which active region plasma can flow and reach the interplanetary medium.

A more detailed analysis in the spot and nearby areas, which takes account of the local temporal variation of the magnetic field, also indicates the presence of a pseudostreamer-like closed-flux system west of the spot. This configuration can easily produce reconnection followed by plasma ejection in the adjacent created open field lines (Sections \ref{sec:loc02} and \ref{sec:loc1-3}). The footpoints of the open area adjacent to the spot are present  before, during and after the ejection, even though with some morphological change. These together with the signatures of reconnection (HXR emission) and flux emergence found by \cite{mulay16}, are consistent with reconnection (possibly interchange type) and re-organization of the magnetic field. 

The additional information provided by the EUV coronal observations is the presence of upflows in these open regions before, during and after the jet, as well as small scale dynamics activity in the jet channel. This reinforces the picture where plasma from the spot flows out in the open corona. As discussed in Section \ref{sec:uv_mhd} and shown in Figure \ref{fig:arms_zoom} our MHD model reproduces consistently the plasma outflow from EUV observations and the magnetic field environment.  

Our magnetic field analysis highlights one of the known critical key points affecting the quality of the results, that is the important role of the magnetic map built from observations and used to extrapolate the initial field. We have seen in Section \ref{methods:field} that low spatial resolution and/or insufficient updates of the map can negatively affect the extrapolation. Having a second magnetometer, such as the Solar Orbiter/PHI, observing at different angles than the Earth view, will allow to have a much frequent update of such maps, including information like the backside face of the Sun. This is a major improvement which will reduce uncertainties on the magnetic field temporal variation. Additionally, the reduced solar distance together with the high spatial resolution of the PHI instrument, will provide high quality data for local magnetic field analysis. 

Solar Orbiter also will leave the ecliptic plane and reach about 30$^\circ$ in latitude, allowing for the first time a view of the solar poles. This will be a major input for the study of the solar magnetic field with the consequent improvement of the global magnetic field modelling. 

Solar Orbiter carries the UV spectrometer SPICE which, for the first time, will provide  data ({even though not on a continuous observation basis)} to perform spectro-stereoscopy \citep[e.g.][]{Podladchikova21} with other Earth view observatories such as IRIS and Hinode/EIS. One of the major potential advantages of these multi-point of view measurements is the reconstruction of the 3D vector velocity through Doppler measurements. These measures will allow a better identification and characterization of the outflow regions even in magnetically complex areas as active regions, and they will permit a more direct comparison with the magnetic field extrapolation. Furthermore, having the real amplitude of the plasma outflow will provide stronger constraints to the wind and FIP bias models, as well as back mapping diagnostics and {\it in--situ} data comparison. {Unfortunately the remote sensing instruments on board Solar Orbiter cannot observe on daily basis, as done by the other existing EUV-UV spectrometers or imagers positioned at Earth distance, so that these unique multi-point of view configurations remain occasional.}

In this work we use the plasma FIP bias as a tool for the connectivity between the heliosphere and its source regions on the Sun. Using two independent methods applied to the EUV data we found the outflow open areas at the side of the spot to be the source of coronal FIP biased plasma (Sections \ref{sec:eis_full} and \ref{se:LCR}). Using the results from the coronal MHD model we simulated the strength of the FIP fractionation in the magnetically open and closed selected areas (Sections \ref{sec:uv_mhd} and \ref{sec:res_fip_charge}). In some cases the results show a clear difference between these two regions, with the closed  areas affected by a more important fractionation, consistently with the results from the EUV data. However, one of the closed areas (CO1) shows a less pronounced FIP bias. This is a case where further investigation would be needed. In fact, such result may be consistent with a low FIP bias coming from closed regions which have been opened up as consequence of local magnetic reconnection. Fractionation in both open and closed field regions increases with increasing magnetic field strength, but in different ways, and also depends on the degree to which Alfv\'en waves dominate over compressive fluctuations (the Alfv\'enicity). In closed field it is also sensitive to how close the Alfv\'en waves are to resonance with the loop (i.e. when the wave travel time from one footpoint to the other is to an integral number of wave half-periods). Such constraints provide significant distinct FIP fractionation imprints depending on these parameters, providing a potential tool for selecting the most appropriate source regions of the plasma measured {\it in--situ}. As a consequence we emphasize the importance of having, additionally to a high quality magnetic field instruments, {\it in--situ} measures of the  Alfve\'nicity  as close as possible to the Sun, and in conjunction with solar FIP bias at candidate solar locations. This exactly what Solar Orbiter provides. These {\it in--situ} measures will be less affected by modification resulting from the interaction with the local  magnetic field and plasma as the wind travels through the heliosphere.

From {\it in--situ} observations within the period where the jet is predicted through the backmapping method, measurements show some changes in both the chemical composition and ionization level of charge states along with temperature and velocity that may suggest some part of the jet may have been observed. {In particular, the closed field with resonant Alfv\'en waves does not fractionate S, but open field can, depending on magnetic field strength. The observations in Table 4 and models in Table 5 do not see or predict significant fractionation in S, and this matches almost exclusively with the latter part of the putative ``jet'' in Figure 15 denoted by the red shaded column beginning approximately at 2010 August 6, 05:00:00, which illustrates the  utility of abundances in identifying solar wind sources regions. However, the {\it in--situ} measurements show no other relevant signatures of the jet structure.

We do not think the jet signature is intertwined or overwhelmed by the ICME passage. 
The beginning of the ICME has a moderately strong FIP effect, but not so large as to be remarkable in remote sensing observations. The strong depletion of He towards the end of the ICME however is remarkable. It has not been possible to model depletion of He down to He/O $\sim$ 1\% with the ponderomotive force model. This degree of depletion presumably indicates gravitational settling in a closed loop prior to eruption \citep[e.g][]{laming19}.

Furthermore, jet may have evolved and expanded such that it is indistinguishable from the surrounding solar wind.
} 
Another possibility could be that the instruments at 1AU do not measure the main jet structure missing the majority of the ejected jet material. {This demonstrates the need for additional {\it in--situ} measurements made closer to the Sun, such as those that are made by Solar Orbiter and Parker Solar Probe where smaller structures, the jets in this case, are easier to distinguish from solar wind and can be more effectively traced back to the Sun. This understanding of the spatial and temporal evolution of magnetic and plasma structures as they travel through the heliosphere is a major
challenge, and multi-measurement observations at different distances is the ideal thing to have to resolve this aspect. The monitoring of the heliosphere in the neighbour of the Earth remains, however, fundamental in the frame of the Space Weather forecasting, as well as to understand the physical processes dominating the diluted plasma and magnetic field of the heliopshere.}

Another advantage of {additional \it in-situ} measurements below 1 AU, is that measurements of heavy ions, such as those from the Heavy Ion Sensor (HIS) \citep{owen-etal-2020} on Solar Orbiter, can be taken at higher resolution compared to 1AU that will allow for more detailed characterization of jet substructure to study the plasma origin, release, and evolution from the Sun. 

In conclusion, with this work we have shown that collaborative data and modelling efforts are essential to build a comprehensive picture which allows to trace and follow the plasma travelling through the heliosphere. This is needed to identify the heliosphere and solar wind sources regions. We have also discussed the difficulties of our work which was limited, beside others, by the available UV and {\it in--situ} data. With the Solar Orbiter entering the nominal mission phase in 2022 we expect most of these limitations to be solved.

\section*{Declarations}

\begin{itemize}
\item[$\bullet$] \emph{Funding:} This work was supported by the International Space Science Institute through its International Team program.

\item[$\bullet$] \emph{Availability of data and materials:} The Hinode/EIS data are freely available through the European Hinode Data Cente: \url{http://sdc.uio.no/sdc/}. 

STEREO and SDO/AIA data are available through the MEDOC datacenter:  \url{https://idoc-medoc.ias.u-psud.fr/sitools/client-user/index.html?project=Medoc-Solar-Portal}.

GONG synoptic maps: \url{ftp://gong2.nso.edu/oQR/zqs/201008/mrzqs100802/mrzqs100802t1154c2099_034.fits.gz.}

SDO/HMI radial field synoptic map: \url{http://jsoc.stanford.edu/data/hmi/synoptic/hmi.Synoptic_Mr_small.2099.fits}.

\item[$\bullet$] \emph{Code availability:} 

The CHIANTI and ChiantiPy atomic database and software for the EUV diagnostic analysis\footnote{http://www.chiantidatabase.org/} are freely available through SolarSoft: \url{https://www.lmsal.com/solarsoft/}.

The PFSS extrapolations are based on an IDL implementation of the current-free case of the formula presented in \citet{ZhaoXP1993}. Our results are comparable to those obtained with the PFSS package provided in SolarSoft: \url{https://www.lmsal.com/~derosa/pfsspack/}.

The Adaptively Refined MHD Solver \citep[ARMS;][]{DeVore2008} is a proprietary research code designed for massively parallel high-performance computing clusters such as the NASA Center for Climate Simulation. For queries about the ARMS simulation data herein or the IDL post-processing analysis routines, please contact \url{blynch@berkeley.edu}. For queries about ARMS, including obtaining the source code, please contact \url{c.richard.devore@nasa.gov}.  

FIPLCR finds the optimal linear combination of EUV spectral lines for calculating the FIP bias using the LCR method: 
\url{https://git.ias.u-psud.fr/nzambran/fiplcr}

\end{itemize}

\begin{acknowledgements}
This collaborative work was made with the support of the International Space Science Institute through its International Team program (Team 418). S.P. acknowledges the funding by CNES through the MEDOC data and operations center. S.P. is grateful to Dr. Hans-Peter Doerr for the discussion on images co-registration. B.J.L. was supported in part by NASA grants 80NSSC18K0645, 80NSSC18K1553, and 80NSSC20K1448. J.M.L. was supported by the NASA Heliophysics Guest Investigator (80HQTR19T0029) and Supporting Research Programs (80HQTR20T0076), and by Basic Research Funds of the Office of Naval Research. I.C. acknowledges DFG-grant WI 3211/5-1. G.D.Z. acknowledges support from STFC (UK) via the consolidated grants to the atomic astrophysics group at DAMTP, University of Cambridge (ST/P000665/1. and ST/T000481/1). T.W. acknowledges financial support by DLR-grants 50 OC 1701  and 50 OC 2101 and DFG-grant WI 3211/5-1. Y.J.R. acknowledges support from the Rackham Merit Fellowship at the University of Michigan, the Newkirk Fellowship at the High Altitude Observatory, and Future Faculty Leaders Fellowship at the Center for Astrophysics. S.T.L. acknowledges support from NASA grants: NNX16AP03H, NNX16AH01G, and 80NSSC20K1063;  NSF Grants: 1460170 and AGS 1358268; and DOD Grant: N00173-14-1-G904. R.F.W-S. acknowledges DLR grant 50OT1702 and DFG grant WI 2139/11-1. G.P. has received funding from the European Research Council (ERC) under the European Union’s Horizon 2020 research and innovation program (grant agreement No. 724326). CHIANTI is a collaborative project involving George Ma-son University, the University of Michigan, the NASA Goddard Space Flight Centre (USA) and the University of Cambridge (UK). Hinode is a Japanese mission developed and launched by ISAS/JAXA, with NAOJ as domestic partner and NASA and UKSA as international partners. It is operated by these agencies in co-operation with ESA and NSC (Norway). "Courtesy of NASA/SDO and the AIA, EVE, and HMI science teams." The EUVI images are supplied courtesy of the STEREO Sun Earth Connection Coronal and Heliospheric Investigation (SECCHI) team.
\end{acknowledgements}

\bibliographystyle{spbasic}      
\bibliography{FIP,NLFFF}   

\begin{thebibliography}{175}
\providecommand{\natexlab}[1]{#1}
\providecommand{\url}[1]{{#1}}
\providecommand{\urlprefix}{URL }
\expandafter\ifx\csname urlstyle\endcsname\relax
  \providecommand{\doi}[1]{DOI~\discretionary{}{}{}#1}\else
  \providecommand{\doi}{DOI~\discretionary{}{}{}\begingroup
  \urlstyle{rm}\Url}\fi
\providecommand{\eprint}[2][]{\url{#2}}

\bibitem[{{Abbo} et~al.(2016){Abbo}, {Ofman}, {Antiochos}, {Hansteen}, {Harra},
  {Ko}, {Lapenta}, {Li}, {Riley}, {Strachan}, {von Steiger}, and
  {Wang}}]{abbo16}
{Abbo} L, {Ofman} L, {Antiochos} SK, {Hansteen} VH, {Harra} L, {Ko} YK,
  {Lapenta} G, {Li} B, {Riley} P, {Strachan} L, {von Steiger} R, {Wang} YM
  (2016) {Slow Solar Wind: Observations and Modeling}. \ssr, 201(1-4):55--108,
  \doi{10.1007/s11214-016-0264-1}

\bibitem[{{Altschuler} and {Newkirk}(1969)}]{Altschuler1969}
{Altschuler} MD, {Newkirk} G (1969) {Magnetic Fields and the Structure of the
  Solar Corona. I: Methods of Calculating Coronal Fields}. \solphys,
  9:131--149, \doi{10.1007/BF00145734}

\bibitem[{{Antiochos} et~al.(2011){Antiochos}, {Miki{\'c}}, {Titov},
  {Lionello}, and {Linker}}]{Antiochos2011}
{Antiochos} SK, {Miki{\'c}} Z, {Titov} VS, {Lionello} R, {Linker} JA (2011) {A
  Model for the Sources of the Slow Solar Wind}. \apj, 731:112,
  \doi{10.1088/0004-637X/731/2/112}, \eprint{1102.3704}

\bibitem[{{Arge} et~al.(2004){Arge}, {Luhmann}, {Odstrcil}, {Schrijver}, and
  {Li}}]{Arge2004}
{Arge} CN, {Luhmann} JG, {Odstrcil} D, {Schrijver} CJ, {Li} Y (2004) {Stream
  structure and coronal sources of the solar wind during the May 12th, 1997
  CME}. Journal of Atmospheric and Solar-Terrestrial Physics,
  66(15-16):1295--1309, \doi{10.1016/j.jastp.2004.03.018}

\bibitem[{{Asplund} et~al.(2009){Asplund}, {Grevesse}, {Sauval}, and
  {Scott}}]{asplund09}
{Asplund} M, {Grevesse} N, {Sauval} AJ, {Scott} P (2009) {The Chemical
  Composition of the Sun}. \araa, 47:481--522,
  \doi{10.1146/annurev.astro.46.060407.145222}, \eprint{0909.0948}

\bibitem[{{Badman} et~al.(2020){Badman}, {Bale}, {Mart{\'\i}nez Oliveros},
  {Panasenco}, {Velli}, {Stansby}, {Buitrago-Casas}, {R{\'e}ville}, {Bonnell},
  {Case}, {Dudok de Wit}, {Goetz}, {Harvey}, {Kasper}, {Korreck}, {Larson},
  {Livi}, {MacDowall}, {Malaspina}, {Pulupa}, {Stevens}, and
  {Whittlesey}}]{Badman2020}
{Badman} ST, {Bale} SD, {Mart{\'\i}nez Oliveros} JC, {Panasenco} O, {Velli} M,
  {Stansby} D, {Buitrago-Casas} JC, {R{\'e}ville} V, {Bonnell} JW, {Case} AW,
  {Dudok de Wit} T, {Goetz} K, {Harvey} PR, {Kasper} JC, {Korreck} KE, {Larson}
  DE, {Livi} R, {MacDowall} RJ, {Malaspina} DM, {Pulupa} M, {Stevens} ML,
  {Whittlesey} PL (2020) {Magnetic Connectivity of the Ecliptic Plane within
  0.5 au: Potential Field Source Surface Modeling of the First Parker Solar
  Probe Encounter}. \apjs, 246(2):23, \doi{10.3847/1538-4365/ab4da7}

\bibitem[{{Baker} et~al.(2018){Baker}, {Brooks}, {van Driel-Gesztelyi},
  {James}, {D{\'e}moulin}, {Long}, {Warren}, and {Williams}}]{baker18}
{Baker} D, {Brooks} DH, {van Driel-Gesztelyi} L, {James} AW, {D{\'e}moulin} P,
  {Long} DM, {Warren} HP, {Williams} DR (2018) {Coronal Elemental Abundances in
  Solar Emerging Flux Regions}. \apj, 856:71, \doi{10.3847/1538-4357/aaadb0},
  \eprint{1801.08424}

\bibitem[{{Bemporad} et~al.(2003){Bemporad}, {Poletto}, {Suess}, {Ko},
  {Parenti}, {Riley}, {Romoli}, and {Zurbuchen}}]{bemporad03}
{Bemporad} A, {Poletto} G, {Suess} ST, {Ko} YK, {Parenti} S, {Riley} P,
  {Romoli} M, {Zurbuchen} TZ (2003) {Temporal Evolution of a Streamer Complex:
  Coronal and in Situ Plasma Parameters}. \apj, 593(2):1146--1163,
  \doi{10.1086/376605}

\bibitem[{Bochsler(2007)}]{bochsler07}
Bochsler P (2007) Minor ions in the solar wind. Astron Astrophys Rev, 14:1--40,
  \doi{10.1007/s00159-006-0002-x}

\bibitem[{{Borrini} et~al.(1983){Borrini}, {Gosling}, {Bame}, and
  {Feldman}}]{Borrini1983}
{Borrini} G, {Gosling} JT, {Bame} SJ, {Feldman} WD (1983) {Helium Abundance
  Variations in the Solar Wind}. \solphys, 83(2):367--378,
  \doi{10.1007/BF00148286}

\bibitem[{{B\"{u}rgi} and {Geiss}(1986)}]{Burgi1986}
{B\"{u}rgi} A, {Geiss} J (1986) {Helium and minor ions in the corona and solar
  wind - Dynamics and charge states}. \solphys, 103:347--383,
  \doi{10.1007/BF00147835}

\bibitem[{{Bu{\v{c}}{\'\i}k}(2020)}]{Bucik2020}
{Bu{\v{c}}{\'\i}k} R (2020) {$^{3}$He-Rich Solar Energetic Particles: Solar
  Sources}. \ssr, 216(2):24, \doi{10.1007/s11214-020-00650-5},
  \eprint{2002.09442}

\bibitem[{{Cheung} and {DeRosa}(2012)}]{Cheung2012}
{Cheung} MCM, {DeRosa} ML (2012) {A Method for Data-driven Simulations of
  Evolving Solar Active Regions}. \apj, 757(2):147,
  \doi{10.1088/0004-637X/757/2/147}, \eprint{1208.2954}

\bibitem[{{Corti} et~al.(2007){Corti}, {Poletto}, {Suess}, {Moore}, and
  {Sterling}}]{corti07}
{Corti} G, {Poletto} G, {Suess} ST, {Moore} RL, {Sterling} AC (2007)
  {Cool-Plasma Jets that Escape into the Outer Corona}. \apj,
  659(2):1702--1712, \doi{10.1086/512233}

\bibitem[{{Cranmer}(2012)}]{Cranmer2012}
{Cranmer} SR (2012) {Self-Consistent Models of the Solar Wind}. \ssr,
  172(1-4):145--156, \doi{10.1007/s11214-010-9674-7}, \eprint{1007.0954}

\bibitem[{Cranmer et~al.(1999)Cranmer, Kohl, Noci, Antonucci, Tondello, Huber,
  Strachan, Panasyuk, Gardner, Romoli, Fineschi, Dobrzycka, Raymond, Nicolosi,
  Siegmund, Spadaro, Benna, Ciaravella, Giordano, and Habbal}]{cranmer99}
Cranmer SR, Kohl JL, Noci G, Antonucci E, Tondello G, Huber MCE, Strachan L,
  Panasyuk AV, Gardner LD, Romoli M, Fineschi S, Dobrzycka D, Raymond JC,
  Nicolosi P, Siegmund OHW, Spadaro S, Benna C, Ciaravella A, Giordano S,
  Habbal SR (1999) An empirical model of a polar coronal hole at solar minimum.
  Astrophys J, 511:481--501, \doi{0.1086/306675}

\bibitem[{Cranmer et~al.(2008)Cranmer, Panasyuk, and Kohl}]{cranmer08}
Cranmer SR, Panasyuk AV, Kohl JL (2008) Improved constraints on the
  preferential heating and acceleration of oxygen ions in the extended solar
  corona. Astrophys J, 678:1480--1497, \doi{10.1086/586890},
  \eprint{arXiv:0802.0144}

\bibitem[{{Culhane} et~al.(2007){Culhane}, {Harra}, {James}, {Al-Janabi},
  {Bradley}, {Chaudry}, {Rees}, {Tandy}, {Thomas}, {Whillock}, {Winter},
  {Doschek}, {Korendyke}, {Brown}, {Myers}, {Mariska}, {Seely}, {Lang}, {Kent},
  {Shaughnessy}, {Young}, {Simnett}, {Castelli}, {Mahmoud}, {Mapson-Menard},
  {Probyn}, {Thomas}, {Davila}, {Dere}, {Windt}, {Shea}, {Hagood}, {Moye},
  {Hara}, {Watanabe}, {Matsuzaki}, {Kosugi}, {Hansteen}, and
  {Wikstol}}]{culhane07}
{Culhane} JL, {Harra} LK, {James} AM, {Al-Janabi} K, {Bradley} LJ, {Chaudry}
  RA, {Rees} K, {Tandy} JA, {Thomas} P, {Whillock} MCR, {Winter} B, {Doschek}
  GA, {Korendyke} CM, {Brown} CM, {Myers} S, {Mariska} J, {Seely} J, {Lang} J,
  {Kent} BJ, {Shaughnessy} BM, {Young} PR, {Simnett} GM, {Castelli} CM,
  {Mahmoud} S, {Mapson-Menard} H, {Probyn} BJ, {Thomas} RJ, {Davila} J, {Dere}
  K, {Windt} D, {Shea} J, {Hagood} R, {Moye} R, {Hara} H, {Watanabe} T,
  {Matsuzaki} K, {Kosugi} T, {Hansteen} V, {Wikstol} {\O} (2007) {The EUV
  Imaging Spectrometer for Hinode}. \solphys, 243:19--61

\bibitem[{{Dahlburg} et~al.(2016){Dahlburg}, {Laming}, {Taylor}, and
  {Obenschain}}]{dahlburg16}
{Dahlburg} RB, {Laming} JM, {Taylor} BD, {Obenschain} K (2016) {Ponderomotive
  Acceleration in Coronal Loops}. \apj, 831(2):160,
  \doi{10.3847/0004-637X/831/2/160}, \eprint{1608.04372}

\bibitem[{{Dahlin} et~al.(2019){Dahlin}, {Antiochos}, and
  {DeVore}}]{Dahlin2019}
{Dahlin} JT, {Antiochos} SK, {DeVore} CR (2019) {A Model for Energy Buildup and
  Eruption Onset in Coronal Mass Ejections}. \apj, 879(2):96,
  \doi{10.3847/1538-4357/ab262a}, \eprint{1905.13218}

\bibitem[{{de Pablos} et~al.(2021){de Pablos}, {Long}, {Owen}, {Valori},
  {Nicolaou}, and {Harra}}]{depablos2021}
{de Pablos} D, {Long} DM, {Owen} CJ, {Valori} G, {Nicolaou} G, {Harra} LK
  (2021) {Matching Temporal Signatures of Solar Features to Their Corresponding
  Solar-Wind Outflows}. \solphys, 296(4):68, \doi{10.1007/s11207-021-01813-5},
  \eprint{2103.09077}

\bibitem[{{Del Zanna}(1999)}]{delzanna_thesis:1999}
{Del Zanna} G (1999) {Extreme ultraviolet spectroscopy of the solar corona}.
  PhD thesis, University of Central Lancashire, UK,
  \urlprefix\url{http://www.damtp.cam.ac.uk/user/astro/gd232/research/thesis/gdz\_phd\_thesis.pdf}

\bibitem[{{Del Zanna}(2013)}]{delzanna13}
{Del Zanna} G (2013) {A revised radiometric calibration for the Hinode/EIS
  instrument}. \aap, 555:A47, \doi{10.1051/0004-6361/201220810},
  \eprint{1211.6771}

\bibitem[{{Del Zanna}(2019)}]{delzanna19}
{Del Zanna} G (2019) {The EUV spectrum of the Sun: Quiet- and active-Sun
  irradiances and chemical composition}. \aap, 624:A36,
  \doi{10.1051/0004-6361/201834842}, \eprint{1901.08841}

\bibitem[{{Del Zanna} and {Mason}(2018)}]{delzanna18}
{Del Zanna} G, {Mason} HE (2018) {Solar UV and X-ray spectral diagnostics}.
  \lrsp, 15:5, \doi{10.1007/s41116-018-0015-3}

\bibitem[{{Del Zanna} et~al.(2015){Del Zanna}, {Dere}, {Young}, {Landi}, and
  {Mason}}]{delzanna15}
{Del Zanna} G, {Dere} KP, {Young} PR, {Landi} E, {Mason} HE (2015) {CHIANTI -
  An atomic database for emission lines. Version 8}. \aap, 582:A56,
  \doi{10.1051/0004-6361/201526827}, \eprint{1508.07631}

\bibitem[{{Dere} et~al.(1997){Dere}, {Landi}, {Mason}, {Monsignori Fossi}, and
  {Young}}]{dere97}
{Dere} KP, {Landi} E, {Mason} HE, {Monsignori Fossi} BC, {Young} PR (1997)
  Chianti - an atomic database for emission lines. \aaps, 125:149--173

\bibitem[{{Dere} et~al.(2019){Dere}, {Del Zanna}, {Young}, {Landi}, and
  {Sutherland}}]{dere19}
{Dere} KP, {Del Zanna} G, {Young} PR, {Landi} E, {Sutherland} RS (2019)
  {dere19{\textemdash}An Atomic Database for Emission Lines. XV. Version 9,
  Improvements for the X-Ray Satellite Lines}. \apjs, 241:22,
  \doi{10.3847/1538-4365/ab05cf}, \eprint{1902.05019}

\bibitem[{{DeRosa} and {Barnes}(2018)}]{DeRosa2018}
{DeRosa} ML, {Barnes} G (2018) {Does Nearby Open Flux Affect the Eruptivity of
  Solar Active Regions?} \apj, 861(2):131, \doi{10.3847/1538-4357/aac77a},
  \eprint{1802.01199}

\bibitem[{{DeVore}(1991)}]{DeVore1991}
{DeVore} CR (1991) {Flux-corrected transport techniques for multidimensional
  compressible magnetohydrodynamics}. \jcoph, 92:142--160,
  \doi{10.1016/0021-9991(91)90295-V}

\bibitem[{{DeVore} and {Antiochos}(2008)}]{DeVore2008}
{DeVore} CR, {Antiochos} SK (2008) {Homologous Confined Filament Eruptions via
  Magnetic Breakout}. \apj, 680:740--756, \doi{10.1086/588011}

\bibitem[{{Domingo} et~al.(1995){Domingo}, {Fleck}, and {Poland}}]{domingo95}
{Domingo} V, {Fleck} B, {Poland} AI (1995) {The SOHO Mission: an Overview}.
  \solphys, 162(1-2):1--37, \doi{10.1007/BF00733425}

\bibitem[{Drake et~al.(1997)Drake, Laming, and Widing}]{drake97}
Drake JJ, Laming JM, Widing KG (1997) Stellar coronal abundances. v. evidence
  for the first ionization potential effect in alpha centauri. Astrophys J,
  478:403--416, \doi{10.1086/303755}

\bibitem[{{Dufresne} and {Del Zanna}(2019)}]{dufresne19}
{Dufresne} RP, {Del Zanna} G (2019) {Modelling ion populations in astrophysical
  plasmas: carbon in the solar transition region}. \aap, 626:A123,
  \doi{10.1051/0004-6361/201935133}, \eprint{1901.08992}

\bibitem[{{Dufresne} et~al.(2020){Dufresne}, {Del Zanna}, and
  {Badnell}}]{dufresne20}
{Dufresne} RP, {Del Zanna} G, {Badnell} NR (2020) {Effects of density on the
  oxygen ionization equilibrium in collisional plasmas}. \mnras,
  497(2):1443--1456, \doi{10.1093/mnras/staa2005}, \eprint{2007.00465}

\bibitem[{{Edmondson} and {Lynch}(2017)}]{Edmondson2017}
{Edmondson} JK, {Lynch} BJ (2017) {Formation and Reconnection of
  Three-dimensional Current Sheets with a Guide Field in the Solar Corona}.
  \apj, 849:28, \doi{10.3847/1538-4357/aa83ba}

\bibitem[{{Edmondson} et~al.(2010){Edmondson}, {Antiochos}, {DeVore}, {Lynch},
  and {Zurbuchen}}]{Edmondson2010a}
{Edmondson} JK, {Antiochos} SK, {DeVore} CR, {Lynch} BJ, {Zurbuchen} TH (2010)
  {Interchange Reconnection and Coronal Hole Dynamics}. \apj, 714:517--531,
  \doi{10.1088/0004-637X/714/1/517}

\bibitem[{{Feldman}(1992)}]{feldman92b}
{Feldman} U (1992) Elemental abundances in the upper solar atmosphere.
  \physscr, 46:202--220

\bibitem[{{Fisk} and {Schwadron}(2001)}]{Fisk2001}
{Fisk} LA, {Schwadron} NA (2001) {The Behavior of the Open Magnetic Field of
  the Sun}. \apj, 560(1):425--438, \doi{10.1086/322503}

\bibitem[{Frazin et~al.(1999)Frazin, Ciaravella, Dennis, Fineschi, Gardner,
  Michels, O'Neal, Raymond, Wu, Kohl, Modigliani, and Noci}]{frazin99}
Frazin RA, Ciaravella A, Dennis E, Fineschi S, Gardner LD, Michels DJ, O'Neal
  R, Raymond JC, Wu CR, Kohl JL, Modigliani A, Noci G (1999) Uvcs/soho ion
  kinetics in coronal streamers. Space Science Rev, 87:189--192,
  \doi{10.1023/A:1005184014654}

\bibitem[{{Geiss} et~al.(1995){Geiss}, {Gloeckler}, and {von
  Steiger}}]{geiss95b}
{Geiss} J, {Gloeckler} G, {von Steiger} R (1995) {Origin of the Solar Wind From
  Composition Data}. \ssr, 72(1-2):49--60, \doi{10.1007/BF00768753}

\bibitem[{{Gibson} et~al.(2011){Gibson}, {de Toma}, {Emery}, {Riley}, {Zhao},
  {Elsworth}, {Leamon}, {Lei}, {McIntosh}, {Mewaldt}, {Thompson}, and
  {Webb}}]{Gibson2011}
{Gibson} SE, {de Toma} G, {Emery} B, {Riley} P, {Zhao} L, {Elsworth} Y,
  {Leamon} RJ, {Lei} J, {McIntosh} S, {Mewaldt} RA, {Thompson} BJ, {Webb} D
  (2011) {The Whole Heliosphere Interval in the Context of a Long and
  Structured Solar Minimum: An Overview from Sun to Earth}. \solphys,
  274:5--27, \doi{10.1007/s11207-011-9921-4}

\bibitem[{{Giunta} et~al.(2015){Giunta}, {Fludra}, {Lanzafame}, {O'Mullane},
  {Summers}, and {Curdt}}]{Giunta2015}
{Giunta} AS, {Fludra} A, {Lanzafame} AC, {O'Mullane} MG, {Summers} HP, {Curdt}
  W (2015) {On Extreme-ultraviolet Helium Line Intensity Enhancement Factors on
  the Sun}. \apj, 803(2):66, \doi{10.1088/0004-637X/803/2/66}

\bibitem[{{Gloeckler} et~al.(1998){Gloeckler}, {Cain}, {Ipavich}, {Tums},
  {Bedini}, {Fisk}, {Zurbuchen}, {Bochsler}, {Fischer}, {Wimmer-Schweingruber},
  {Geiss}, and {Kallenbach}}]{Gloeckler1998}
{Gloeckler} G, {Cain} J, {Ipavich} FM, {Tums} EO, {Bedini} P, {Fisk} LA,
  {Zurbuchen} TH, {Bochsler} P, {Fischer} J, {Wimmer-Schweingruber} RF, {Geiss}
  J, {Kallenbach} R (1998) {Investigation of the composition of solar and
  interstellar matter using solar wind and pickup ion measurements with SWICS
  and SWIMS on the ACE spacecraft}. \ssr, 86:497--539,
  \doi{10.1023/A:1005036131689}

\bibitem[{{Gruesbeck} et~al.(2011){Gruesbeck}, {Lepri}, {Zurbuchen}, and
  {Antiochos}}]{Gruesbeck2011}
{Gruesbeck} JR, {Lepri} ST, {Zurbuchen} TH, {Antiochos} SK (2011) {Constraints
  on Coronal Mass Ejection Evolution from in Situ Observations of Ionic Charge
  States}. \apj, 730:103, \doi{10.1088/0004-637X/730/2/103}

\bibitem[{{Hansteen} and {Velli}(2012)}]{hansteen12}
{Hansteen} VH, {Velli} M (2012) {Solar Wind Models from the Chromosphere to 1
  AU}. \ssr, 172(1-4):89--121, \doi{10.1007/s11214-012-9887-z}

\bibitem[{{Harra} et~al.(2021){Harra}, {Brooks}, {Bale}, {Mandrini},
  {Barczynski}, {Sharma}, {Badman}, {Vargas Dom{\'\i}nguez}, and
  {Pulupa}}]{harra2021}
{Harra} L, {Brooks} DH, {Bale} SD, {Mandrini} CH, {Barczynski} K, {Sharma} R,
  {Badman} ST, {Vargas Dom{\'\i}nguez} S, {Pulupa} M (2021) {The active region
  source of a type III radio storm observed by Parker Solar Probe during
  encounter 2}. \aap, 650:A7, \doi{10.1051/0004-6361/202039514},
  \eprint{2102.04964}

\bibitem[{{Harrison} et~al.(2012){Harrison}, {Davies}, {M{\"o}stl}, {Liu},
  {Temmer}, {Bisi}, {Eastwood}, {de Koning}, {Nitta}, {Rollett}, {Farrugia},
  {Forsyth}, {Jackson}, {Jensen}, {Kilpua}, {Odstrcil}, and
  {Webb}}]{Harrison2012}
{Harrison} RA, {Davies} JA, {M{\"o}stl} C, {Liu} Y, {Temmer} M, {Bisi} MM,
  {Eastwood} JP, {de Koning} CA, {Nitta} N, {Rollett} T, {Farrugia} CJ,
  {Forsyth} RJ, {Jackson} BV, {Jensen} EA, {Kilpua} EKJ, {Odstrcil} D, {Webb}
  DF (2012) {An Analysis of the Origin and Propagation of the Multiple Coronal
  Mass Ejections of 2010 August 1}. \apj, 750(1):45,
  \doi{10.1088/0004-637X/750/1/45}

\bibitem[{Heber et~al.(2021)Heber, Burnett, McKeegan, Steele, Jurewicz, Rieck,
  Guan, Wieler, and Burnett}]{heber21}
Heber VS, Burnett DS, McKeegan KD, Steele RCJ, Jurewicz AJG, Rieck KD, Guan Y,
  Wieler R, Burnett DS (2021) Elemental abundances of major elements in the
  solar wind as measured in genesis targets and implications on solar wind
  fractionation. Astrophys J, 907:15, \doi{10.3847/1538-4357/abc94a}

\bibitem[{{Higginson} and {Lynch}(2018)}]{Higginson2018}
{Higginson} AK, {Lynch} BJ (2018) {Structured Slow Solar Wind Variability:
  Streamer-blob Flux Ropes and Torsional Alfv\'{e}n Waves}. \apj, 859:6,
  \doi{10.3847/1538-4357/aabc08}

\bibitem[{{Higginson} et~al.(2017{\natexlab{a}}){Higginson}, {Antiochos},
  {DeVore}, {Wyper}, and {Zurbuchen}}]{Higginson2017a}
{Higginson} AK, {Antiochos} SK, {DeVore} CR, {Wyper} PF, {Zurbuchen} TH
  (2017{\natexlab{a}}) {Dynamics of Coronal Hole Boundaries}. \apj, 837:113,
  \doi{10.3847/1538-4357/837/2/113}, \eprint{1611.04968}

\bibitem[{{Higginson} et~al.(2017{\natexlab{b}}){Higginson}, {Antiochos},
  {DeVore}, {Wyper}, and {Zurbuchen}}]{Higginson2017b}
{Higginson} AK, {Antiochos} SK, {DeVore} CR, {Wyper} PF, {Zurbuchen} TH
  (2017{\natexlab{b}}) {The Formation of Heliospheric Arcs of Slow Solar Wind}.
  \apjl, 840(1):L10, \doi{10.3847/2041-8213/aa6d72}, \eprint{1701.08797}

\bibitem[{{Hill}(2018)}]{Hill2018}
{Hill} F (2018) {The Global Oscillation Network Group Facility{\textemdash}An
  Example of Research to Operations in Space Weather}. \spwea,
  16(10):1488--1497, \doi{10.1029/2018SW002001}

\bibitem[{{Hirshberg} et~al.(1972){Hirshberg}, {Bame}, and
  {Robbins}}]{Hirshberg1972}
{Hirshberg} J, {Bame} SJ, {Robbins} DE (1972) {Solar flares and solar wind
  helium enrichments: July 1965 July 1967}. \solphys, 23(2):467--486,
  \doi{10.1007/BF00148109}

\bibitem[{{Hoeksema}(1991)}]{Hoeksema1991}
{Hoeksema} JT (1991) {Large-scale solar and heliospheric magnetic fields.}
  \adv, 11(1):15--24

\bibitem[{{Hoeksema} et~al.(1983){Hoeksema}, {Wilcox}, and
  {Scherrer}}]{Hoeksema1983}
{Hoeksema} JT, {Wilcox} JM, {Scherrer} PH (1983) {The structure of the
  heliospheric current sheet: 1978-1982}. \jgr, 88(A12):9910--9918,
  \doi{10.1029/JA088iA12p09910}

\bibitem[{{Hoeksema} et~al.(2020){Hoeksema}, {Abbett}, {Bercik}, {Cheung},
  {DeRosa}, {Fisher}, {Hayashi}, {Kazachenko}, {Liu}, {Lumme}, {Lynch}, {Sun},
  and {Welsch}}]{Hoeksema2020}
{Hoeksema} JT, {Abbett} WP, {Bercik} DJ, {Cheung} MCM, {DeRosa} ML, {Fisher}
  GH, {Hayashi} K, {Kazachenko} MD, {Liu} Y, {Lumme} E, {Lynch} BJ, {Sun} X,
  {Welsch} BT (2020) {The Coronal Global Evolutionary Model: Using HMI Vector
  Magnetogram and Doppler Data to Determine Coronal Magnetic Field Evolution}.
  \apjs, 250(2):28, \doi{10.3847/1538-4365/abb3fb}, \eprint{2006.14579}

\bibitem[{{Hundhausen}(1972)}]{hundhausen72}
{Hundhausen} AJ (1972) {Coronal Expansion and Solar Wind}. Springer-Verlag,
  Berlin

\bibitem[{{Hundhausen} et~al.(1968){Hundhausen}, {Gilbert}, and
  {Bame}}]{Hundhausen1968}
{Hundhausen} AJ, {Gilbert} HE, {Bame} SJ (1968) {The State of Ionization of
  Oxygen in the Solar Wind}. \apjl, 152:L3, \doi{10.1086/180165}

\bibitem[{{Innes} et~al.(2011){Innes}, {Cameron}, and {Solanki}}]{innes11}
{Innes} DE, {Cameron} RH, {Solanki} SK (2011) {EUV jets, type III radio bursts
  and sunspot waves investigated using SDO/AIA observations}. \aap, 531:L13,
  \doi{10.1051/0004-6361/201117255}, \eprint{1106.3417}

\bibitem[{{Karpen} et~al.(2012){Karpen}, {Antiochos}, and
  {DeVore}}]{Karpen2012}
{Karpen} JT, {Antiochos} SK, {DeVore} CR (2012) {The Mechanisms for the Onset
  and Explosive Eruption of Coronal Mass Ejections and Eruptive Flares}. \apj,
  760:81, \doi{10.1088/0004-637X/760/1/81}

\bibitem[{{Karpen} et~al.(2017){Karpen}, {DeVore}, {Antiochos}, and
  {Pariat}}]{Karpen2017}
{Karpen} JT, {DeVore} CR, {Antiochos} SK, {Pariat} E (2017)
  {Reconnection-Driven Coronal-Hole Jets with Gravity and Solar Wind}. \apj,
  834:62, \doi{10.3847/1538-4357/834/1/62}, \eprint{1606.09201}

\bibitem[{{Ko} et~al.(1997){Ko}, {Fisk}, {Geiss}, {Gloeckler}, and
  {Guhathakurta}}]{Ko1997}
{Ko} YK, {Fisk} LA, {Geiss} J, {Gloeckler} G, {Guhathakurta} M (1997) {An
  Empirical Study of the Electron Temperature and Heavy Ion Velocities in the
  South Polar Coronal Hole}. \solphys, 171:345--361

\bibitem[{{Ko} et~al.(2016){Ko}, {Young}, {Muglach}, {Warren}, and
  {Ugarte-Urra}}]{ko16}
{Ko} YK, {Young} PR, {Muglach} K, {Warren} HP, {Ugarte-Urra} I (2016)
  {Correlation of Coronal Plasma Properties and Solar Magnetic Field in a
  Decaying Active Region}. \apj, 826:126, \doi{10.3847/0004-637X/826/2/126}

\bibitem[{{Ko} et~al.(2018){Ko}, {Roberts}, and {Lepri}}]{ko18}
{Ko} YK, {Roberts} DA, {Lepri} ST (2018) Boundary of the slow solar wind.
  Astrophys J, 864:139, \doi{10.3847/1538-4357/aad69e}

\bibitem[{Kohl et~al.(1997)Kohl, Noci, Antonucci, Huber, Gardner, Nicolosi,
  Strachan, Fineschi, Raymond, Romoli, Spadaro, Panasyuk, Siegmund, Benna,
  Ciaravella, Cranmer, Giordano, Karovska, and Martin}]{kohl97}
Kohl JL, Noci G, Antonucci E, Huber MCE, Gardner LD, Nicolosi P, Strachan L,
  Fineschi S, Raymond JC, Romoli M, Spadaro D, Panasyuk AV, Siegmund OHW, Benna
  C, Ciaravella A, Cranmer SR, Giordano S, Karovska M, Martin R (1997) First
  results from the soho ultraviolet coronagraph spectrometer. Solar Phys,
  175:613--644, \doi{10.1023/A:1004903206467}

\bibitem[{Kohl et~al.(1998)Kohl, Noci, Antonucci, Tondello, Huber, Strachan,
  Panasyuk, Gardner, Romoli, Fineschi, Dobrzycka, Raymond, Nicolosi, Siegmund,
  Spadaro, Benna, Ciaravella, Giordano, and Habbal}]{kohl98}
Kohl JL, Noci G, Antonucci E, Tondello G, Huber SR M C Eand~Cranmer, Strachan
  L, Panasyuk AV, Gardner LD, Romoli M, Fineschi S, Dobrzycka D, Raymond JC,
  Nicolosi P, Siegmund OH, Spadaro D, Benna C, Ciaravella A, Giordano S, Habbal
  SR (1998) Uvcs/soho empirical determinations of anisotropic velocity
  distributions in the solar corona. Astrophys J Lett, 501:L127--L131,
  \doi{10.1086/311434}

\bibitem[{Kohl et~al.(2006)Kohl, Noci, Cranmer, and Raymond}]{kohl06}
Kohl JL, Noci G, Cranmer SR, Raymond JC (2006) Ultraviolet spectroscopy of the
  extended solar corona. Astron Astrophys Rev, 13:31--157, \doi{10.1086/510710}

\bibitem[{{Krucker} et~al.(2020){Krucker}, {Hurford}, {Grimm}, {K{\"o}gl},
  {Gr{\"o}belbauer}, {Etesi}, {Casadei}, {Csillaghy}, {Benz}, {Arnold},
  {Molendini}, {Orleanski}, {Schori}, {Xiao}, {Kuhar}, {Hochmuth}, {Felix},
  {Schramka}, {Marcin}, {Kobler}, {Iseli}, {Dreier}, {Wiehl}, {Kleint},
  {Battaglia}, {Lastufka}, {Sathiapal}, {Lapadula}, {Bednarzik}, {Birrer},
  {Stutz}, {Wild}, {Marone}, {Skup}, {Cichocki}, {Ber}, {Rutkowski}, {Bujwan},
  {Juchnikowski}, {Winkler}, {Darmetko}, {Michalska}, {Seweryn}, {Bia{\l}ek},
  {Osica}, {Sylwester}, {Kowalinski}, {{\'S}cis{\l}owski}, {Siarkowski},
  {St{\k{e}}{\'s}licki}, {Mrozek}, {Podg{\'o}rski}, {Meuris}, {Limousin},
  {Gevin}, {Le Mer}, {Brun}, {Strugarek}, {Vilmer}, {Musset}, {Maksimovi{\'c}},
  {F{\'a}rn{\'\i}k}, {Koz{\'a}{\v{c}}ek}, {Ka{\v{s}}parov{\'a}}, {Mann},
  {{\"O}nel}, {Warmuth}, {Rendtel}, {Anderson}, {Bauer}, {Dionies}, {Paschke},
  {Pl{\"u}schke}, {Woche}, {Schuller}, {Veronig}, {Dickson}, {Gallagher},
  {Maloney}, {Bloomfield}, {Piana}, {Massone}, {Benvenuto}, {Massa},
  {Schwartz}, {Dennis}, {van Beek}, {Rodr{\'\i}guez-Pacheco}, and
  {Lin}}]{krucker2020}
{Krucker} S, {Hurford} GJ, {Grimm} O, {K{\"o}gl} S, {Gr{\"o}belbauer} HP,
  {Etesi} L, {Casadei} D, {Csillaghy} A, {Benz} AO, {Arnold} NG, {Molendini} F,
  {Orleanski} P, {Schori} D, {Xiao} H, {Kuhar} M, {Hochmuth} N, {Felix} S,
  {Schramka} F, {Marcin} S, {Kobler} S, {Iseli} L, {Dreier} M, {Wiehl} HJ,
  {Kleint} L, {Battaglia} M, {Lastufka} E, {Sathiapal} H, {Lapadula} K,
  {Bednarzik} M, {Birrer} G, {Stutz} S, {Wild} C, {Marone} F, {Skup} KR,
  {Cichocki} A, {Ber} K, {Rutkowski} K, {Bujwan} W, {Juchnikowski} G, {Winkler}
  M, {Darmetko} M, {Michalska} M, {Seweryn} K, {Bia{\l}ek} A, {Osica} P,
  {Sylwester} J, {Kowalinski} M, {{\'S}cis{\l}owski} D, {Siarkowski} M,
  {St{\k{e}}{\'s}licki} M, {Mrozek} T, {Podg{\'o}rski} P, {Meuris} A,
  {Limousin} O, {Gevin} O, {Le Mer} I, {Brun} S, {Strugarek} A, {Vilmer} N,
  {Musset} S, {Maksimovi{\'c}} M, {F{\'a}rn{\'\i}k} F, {Koz{\'a}{\v{c}}ek} Z,
  {Ka{\v{s}}parov{\'a}} J, {Mann} G, {{\"O}nel} H, {Warmuth} A, {Rendtel} J,
  {Anderson} J, {Bauer} S, {Dionies} F, {Paschke} J, {Pl{\"u}schke} D, {Woche}
  M, {Schuller} F, {Veronig} AM, {Dickson} ECM, {Gallagher} PT, {Maloney} SA,
  {Bloomfield} DS, {Piana} M, {Massone} AM, {Benvenuto} F, {Massa} P,
  {Schwartz} RA, {Dennis} BR, {van Beek} HF, {Rodr{\'\i}guez-Pacheco} J, {Lin}
  RP (2020) {The Spectrometer/Telescope for Imaging X-rays (STIX)}. \aap,
  642:A15, \doi{10.1051/0004-6361/201937362}

\bibitem[{{Kruse} et~al.(2021){Kruse}, {Heidrich-Meisner}, and
  {Wimmer-Schweingruber}}]{kruse-etal-2021}
{Kruse} M, {Heidrich-Meisner} V, {Wimmer-Schweingruber} RF (2021) {Evaluation
  of a potential field source surface model with elliptical source surfaces via
  ballistic back mapping of in situ spacecraft data}. \aap, 645:A83,
  \doi{10.1051/0004-6361/202039120}

\bibitem[{Kuroda and Laming(2020)}]{kuroda20}
Kuroda N, Laming JM (2020) Magnetic field geometry and composition variation in
  slow solar winds: The case of sulfur. Astrophys J, 895:36,
  \doi{10.3847/1538-4357/ab8870}, \eprint{arXiv:2005.10842}

\bibitem[{Laming(2004a)}]{laming04b}
Laming JM (2004a) On collisionless electron-ion temperature equilibration in
  the fast solar wind. Astrophys J, 604:874--883, \doi{10.1086/382066},
  \eprint{arXiv:astro-ph/0312387}

\bibitem[{Laming(2004b)}]{laming04}
Laming JM (2004b) A unified picture of the first ionization potential and
  inverse first ionization potential effects. Astrophys J, 614:1063--1072,
  \doi{10.1086/423780}, \eprint{arXiv:astro-ph/0405230}

\bibitem[{Laming(2009)}]{laming09}
Laming JM (2009) Non-wkb models of the first ionization potential effect:
  Implications for solar coronal heating and the coronal helium and neon
  abundances. Astrophys J, 695:954--969, \doi{10.1088/0004-637X/695/2/954},
  \eprint{arXiv:0901.3350}

\bibitem[{Laming(2012)}]{laming12}
Laming JM (2012) Non-wkb models of the first ionization potential effect: The
  role of slow mode waves. Astrophys J, 744:115,
  \doi{10.1088/0004-637X/744/2/115}, \eprint{arXiv:1110.4357}

\bibitem[{{Laming}(2015)}]{laming15}
{Laming} JM (2015) {The FIP and Inverse FIP Effects in Solar and Stellar
  Coronae}. \lrsp, 12:2, \doi{10.1007/lrsp-2015-2}, \eprint{1504.08325}

\bibitem[{Laming(2017)}]{laming17}
Laming JM (2017) The first ionization potential effect from the ponderomotive
  force: On the polarization and coronal origin of alfvén waves. Astrophys J,
  844:153, \doi{10.3847/1538-4357/aa7cf1}, \eprint{arXiv:1707.05378}

\bibitem[{Laming(2021)}]{laming21}
Laming JM (2021) The fip and inverse fip effects in solar flares. Astrophys J,
  \eprint{arXiv:2101.03038}

\bibitem[{{Laming} and {Lepri}(2007)}]{Laming2007}
{Laming} JM, {Lepri} ST (2007) {Ion Charge States in the Fast Solar Wind: New
  Data Analysis and Theoretical Refinements}. \apj, 660:1642--1652,
  \doi{10.1086/513505}, \eprint{astro-ph/0702131}

\bibitem[{Laming and Lepri(2007)}]{laming07}
Laming JM, Lepri ST (2007) Ion charge states in the fast solar wind: New data
  analysis and theoretical refinements. Astrophys J, 660:1642--1652,
  \doi{10.1086/513505}, \eprint{arXiv:astro-ph/0702131}

\bibitem[{Laming et~al.(2017)Laming, Heber, Burnett, Guan, Hervig, Huss,
  Jurewicz, Koeman-Shields, McKeegan, Nittler, Reisenfeld, Rieck, Wang, Wiens,
  and Woolum}]{laming17b}
Laming JM, Heber VS, Burnett DS, Guan Y, Hervig R, Huss GR, Jurewicz AJG,
  Koeman-Shields EC, McKeegan KD, Nittler LR, Reisenfeld DB, Rieck KD, Wang J,
  Wiens RC, Woolum DS (2017) Determining the elemental and isotopic composition
  of the pre-solar nebula from genesis data analysis: The case of oxygen.
  Astrophys J Lett, 851:L12, \doi{10.3847/2041-8213/aa9bf0},
  \eprint{arXiv:1711.07503}

\bibitem[{Laming et~al.(2019)Laming, Vourlidas, Korendyke, Chua, Cranmer, Ko,
  Kuroda, Provornikova, Raymond, Raouafi, Strachan, Tun-Beltran, Weberg, and
  Wood}]{laming19}
Laming JM, Vourlidas A, Korendyke C, Chua D, Cranmer SR, Ko YK, Kuroda N,
  Provornikova E, Raymond JC, Raouafi NE, Strachan L, Tun-Beltran S, Weberg M,
  Wood BE (2019) Element abundances: A new diagnostic for the solar wind.
  Astrophys J, 879:124, \doi{10.3847/1538-4357/ab23f1},
  \eprint{arXiv:1905.09319}

\bibitem[{{Landi} et~al.(2012){Landi}, {Gruesbeck}, {Lepri}, {Zurbuchen}, and
  {Fisk}}]{Landi2012b}
{Landi} E, {Gruesbeck} JR, {Lepri} ST, {Zurbuchen} TH, {Fisk} LA (2012) {Charge
  State Evolution in the Solar Wind. II. Plasma Charge State Composition in the
  Inner Corona and Accelerating Fast Solar Wind}. \apj, 761:48,
  \doi{10.1088/0004-637X/761/1/48}

\bibitem[{{Leamon} and {McIntosh}(2009)}]{Leamon2009}
{Leamon} RJ, {McIntosh} SW (2009) {How the Solar Wind Ties to its Photospheric
  Origins}. \apjl, 697:L28--L32, \doi{10.1088/0004-637X/697/1/L28},
  \eprint{0904.0614}

\bibitem[{{Lepri} and {Rivera}(2021)}]{Lepri2021}
{Lepri} ST, {Rivera} YJ (2021) {Elemental Abundances of Prominence Material
  inside ICMEs}. \apj, 912(1):51, \doi{10.3847/1538-4357/abea9f}

\bibitem[{{Lepri} and {Zurbuchen}(2010)}]{Lepri2010}
{Lepri} ST, {Zurbuchen} TH (2010) {Direct Observational Evidence of Filament
  Material Within Interplanetary Coronal Mass Ejections}. \apjl, 723:L22--L27,
  \doi{10.1088/2041-8205/723/1/L22}

\bibitem[{{Lepri} et~al.(2013){Lepri}, {Landi}, and {Zurbuchen}}]{Lepri2013}
{Lepri} ST, {Landi} E, {Zurbuchen} TH (2013) {Solar Wind Heavy Ions over Solar
  Cycle 23: ACE/SWICS Measurements}. \apj, 768:94,
  \doi{10.1088/0004-637X/768/1/94}

\bibitem[{{Lin}(1970)}]{Lin1970}
{Lin} RP (1970) {The Emission and Propagation of 40 keV Solar Flare Electrons.
  I: The Relationship of 40 keV Electron to Energetic Proton and Relativistic
  Electron Emission by the Sun}. \solphys, 12(2):266--303,
  \doi{10.1007/BF00227122}

\bibitem[{{Lin} et~al.(1995){Lin}, {Anderson}, {Ashford}, {Carlson}, {Curtis},
  {Ergun}, {Larson}, {McFadden}, {McCarthy}, {Parks}, {R{\`e}me}, {Bosqued},
  {Coutelier}, {Cotin}, {D'Uston}, {Wenzel}, {Sanderson}, {Henrion}, {Ronnet},
  and {Paschmann}}]{lin1995}
{Lin} RP, {Anderson} KA, {Ashford} S, {Carlson} C, {Curtis} D, {Ergun} R,
  {Larson} D, {McFadden} J, {McCarthy} M, {Parks} GK, {R{\`e}me} H, {Bosqued}
  JM, {Coutelier} J, {Cotin} F, {D'Uston} C, {Wenzel} KP, {Sanderson} TR,
  {Henrion} J, {Ronnet} JC, {Paschmann} G (1995) {A Three-Dimensional Plasma
  and Energetic Particle Investigation for the Wind Spacecraft}. \ssr,
  71(1-4):125--153, \doi{10.1007/BF00751328}

\bibitem[{{Linker} et~al.(2017){Linker}, {Caplan}, {Downs}, {Riley}, {Mikic},
  {Lionello}, {Henney}, {Arge}, {Liu}, {Derosa}, {Yeates}, and
  {Owens}}]{Linker2017}
{Linker} JA, {Caplan} RM, {Downs} C, {Riley} P, {Mikic} Z, {Lionello} R,
  {Henney} CJ, {Arge} CN, {Liu} Y, {Derosa} ML, {Yeates} A, {Owens} MJ (2017)
  {The Open Flux Problem}. \apj, 848(1):70, \doi{10.3847/1538-4357/aa8a70},
  \eprint{1708.02342}

\bibitem[{{Lionello} et~al.(2014){Lionello}, {Velli}, {Downs}, {Linker},
  {Miki{\'c}}, and {Verdini}}]{Lionello2014}
{Lionello} R, {Velli} M, {Downs} C, {Linker} JA, {Miki{\'c}} Z, {Verdini} A
  (2014) {Validating a Time-dependent Turbulence-driven Model of the Solar
  Wind}. \apj, 784(2):120, \doi{10.1088/0004-637X/784/2/120},
  \eprint{1402.4188}

\bibitem[{{Lynch} et~al.(2014){Lynch}, {Edmondson}, and {Li}}]{Lynch2014}
{Lynch} BJ, {Edmondson} JK, {Li} Y (2014) {Interchange Reconnection Alfv{\'e}n
  Wave Generation}. \solphys, 289:3043--3058, \doi{10.1007/s11207-014-0506-x},
  \eprint{1401.7965}

\bibitem[{{Lynch} et~al.(2016){Lynch}, {Masson}, {Li}, {Devore}, {Luhmann},
  {Antiochos}, and {Fisher}}]{Lynch2016b}
{Lynch} BJ, {Masson} S, {Li} Y, {Devore} CR, {Luhmann} JG, {Antiochos} SK,
  {Fisher} GH (2016) {A model for stealth coronal mass ejections}. \jgr,
  121:10677, \doi{10.1002/2016JA023432}

\bibitem[{{MacNeice} et~al.(2000){MacNeice}, {Olson}, {Mobarry}, {de
  Fainchtein}, and {Packer}}]{MacNeice2000}
{MacNeice} P, {Olson} KM, {Mobarry} C, {de Fainchtein} R, {Packer} C (2000)
  {PARAMESH: A parallel adaptive mesh refinement community toolkit}. \cophc,
  126:330--354, \doi{10.1016/S0010-4655(99)00501-9}

\bibitem[{{Masson} et~al.(2019){Masson}, {Antiochos}, and
  {DeVore}}]{Masson2019}
{Masson} S, {Antiochos} SK, {DeVore} CR (2019) {Escape of Flare-accelerated
  Particles in Solar Eruptive Events}. arXiv e-prints, arXiv:1909.13578,
  \eprint{1909.13578}

\bibitem[{{McComas} et~al.(1998){McComas}, {Bame}, {Barker}, {Feldman},
  {Phillips}, {Riley}, and {Griffee}}]{McComas1998}
{McComas} DJ, {Bame} SJ, {Barker} P, {Feldman} WC, {Phillips} JL, {Riley} P,
  {Griffee} JW (1998) {Solar Wind Electron Proton Alpha Monitor (SWEPAM) for
  the Advanced Composition Explorer}. \ssr, 86:563--612,
  \doi{10.1023/A:1005040232597}

\bibitem[{{McComas} et~al.(2003){McComas}, {Elliott}, {Schwadron}, {Gosling},
  {Skoug}, and {Goldstein}}]{mccomas03}
{McComas} DJ, {Elliott} HA, {Schwadron} NA, {Gosling} JT, {Skoug} RM,
  {Goldstein} BE (2003) The three-dimensional solar wind around solar maximum.
  Geophys Res Lett, 30:1517, \doi{https://doi.org/10.1029/2003GL017136}

\bibitem[{{M{\"o}stl} et~al.(2012){M{\"o}stl}, {Farrugia}, {Kilpua}, {Jian},
  {Liu}, {Eastwood}, {Harrison}, {Webb}, {Temmer}, {Odstrcil}, {Davies},
  {Rollett}, {Luhmann}, {Nitta}, {Mulligan}, {Jensen}, {Forsyth}, {Lavraud},
  {de Koning}, {Veronig}, {Galvin}, {Zhang}, and {Anderson}}]{Moestl2012}
{M{\"o}stl} C, {Farrugia} CJ, {Kilpua} EKJ, {Jian} LK, {Liu} Y, {Eastwood} JP,
  {Harrison} RA, {Webb} DF, {Temmer} M, {Odstrcil} D, {Davies} JA, {Rollett} T,
  {Luhmann} JG, {Nitta} N, {Mulligan} T, {Jensen} EA, {Forsyth} R, {Lavraud} B,
  {de Koning} CA, {Veronig} AM, {Galvin} AB, {Zhang} TL, {Anderson} BJ (2012)
  {Multi-point Shock and Flux Rope Analysis of Multiple Interplanetary Coronal
  Mass Ejections around 2010 August 1 in the Inner Heliosphere}. \apj,
  758(1):10, \doi{10.1088/0004-637X/758/1/10}, \eprint{1209.2866}

\bibitem[{{Mulay} et~al.(2016){Mulay}, {Tripathi}, {Del Zanna}, and
  {Mason}}]{mulay16}
{Mulay} SM, {Tripathi} D, {Del Zanna} G, {Mason} H (2016) {Multiwavelength
  study of 20 jets that emanate from the periphery of active regions}. \aap,
  589:A79, \doi{10.1051/0004-6361/201527473}, \eprint{1602.00151}

\bibitem[{{M{\"u}ller} et~al.(2020){M{\"u}ller}, {St. Cyr}, {Zouganelis},
  {Gilbert}, {Marsden}, {Nieves-Chinchilla}, {Antonucci}, {Auch{\`e}re},
  {Berghmans}, {Horbury}, {Howard}, {Krucker}, {Maksimovic}, {Owen}, {Rochus},
  {Rodriguez-Pacheco}, {Romoli}, {Solanki}, {Bruno}, {Carlsson}, {Fludra},
  {Harra}, {Hassler}, {Livi}, {Louarn}, {Peter}, {Sch{\"u}hle}, {Teriaca}, {del
  Toro Iniesta}, {Wimmer-Schweingruber}, {Marsch}, {Velli}, {De Groof},
  {Walsh}, and {Williams}}]{muller2020}
{M{\"u}ller} D, {St Cyr} OC, {Zouganelis} I, {Gilbert} HR, {Marsden} R,
  {Nieves-Chinchilla} T, {Antonucci} E, {Auch{\`e}re} F, {Berghmans} D,
  {Horbury} TS, {Howard} RA, {Krucker} S, {Maksimovic} M, {Owen} CJ, {Rochus}
  P, {Rodriguez-Pacheco} J, {Romoli} M, {Solanki} SK, {Bruno} R, {Carlsson} M,
  {Fludra} A, {Harra} L, {Hassler} DM, {Livi} S, {Louarn} P, {Peter} H,
  {Sch{\"u}hle} U, {Teriaca} L, {del Toro Iniesta} JC, {Wimmer-Schweingruber}
  RF, {Marsch} E, {Velli} M, {De Groof} A, {Walsh} A, {Williams} D (2020) {The
  Solar Orbiter mission. Science overview}. \aap, 642:A1,
  \doi{10.1051/0004-6361/202038467}, \eprint{2009.00861}

\bibitem[{{Neugebauer} et~al.(1998){Neugebauer}, {Forsyth}, {Galvin}, {Harvey},
  {Hoeksema}, {Lazarus}, {Lepping}, {Linker}, {Mikic}, {Steinberg}, {von
  Steiger}, {Wang}, and {Wimmer-Schweingruber}}]{neugebauer98}
{Neugebauer} M, {Forsyth} RJ, {Galvin} AB, {Harvey} KL, {Hoeksema} JT,
  {Lazarus} AJ, {Lepping} RP, {Linker} JA, {Mikic} Z, {Steinberg} JT, {von
  Steiger} R, {Wang} YM, {Wimmer-Schweingruber} RF (1998) {Spatial structure of
  the solar wind and comparisons with solar data and models}. \jgr,
  103(A7):14587--14600, \doi{10.1029/98JA00798}

\bibitem[{{Neugebauer} et~al.(2002){Neugebauer}, {Liewer}, {Smith}, {Skoug},
  and {Zurbuchen}}]{Neugebauer2002}
{Neugebauer} M, {Liewer} PC, {Smith} EJ, {Skoug} RM, {Zurbuchen} TH (2002)
  {Sources of the solar wind at solar activity maximum}. Journal of Geophysical
  Research (Space Physics), 107:1488, \doi{10.1029/2001JA000306}

\bibitem[{{Neugebauer} et~al.(2004){Neugebauer}, {Liewer}, {Goldstein}, {Zhou},
  and {Steinberg}}]{Neugebauer2004}
{Neugebauer} M, {Liewer} PC, {Goldstein} BE, {Zhou} X, {Steinberg} JT (2004)
  {Solar wind stream interaction regions without sector boundaries}. Journal of
  Geophysical Research (Space Physics), 109(A18):A10102,
  \doi{10.1029/2004JA010456}

\bibitem[{{Odstrcil} et~al.(2004){Odstrcil}, {Riley}, and
  {Zhao}}]{Odstrcil2004}
{Odstrcil} D, {Riley} P, {Zhao} XP (2004) {Numerical simulation of the 12 May
  1997 interplanetary CME event}. \jgr, 109:A02116, \doi{10.1029/2003JA010135}

\bibitem[{{Oran} et~al.(2017){Oran}, {Landi}, {van der Holst}, {Sokolov}, and
  {Gombosi}}]{Oran2017}
{Oran} R, {Landi} E, {van der Holst} B, {Sokolov} IV, {Gombosi} TI (2017)
  {Alfv\'{e}n Wave Turbulence as a Coronal Heating Mechanism: Simultaneously
  Predicting the Heating Rate and the Wave-induced Emission Line Broadening}.
  \apj, 845(2):98, \doi{10.3847/1538-4357/aa7fec}, \eprint{1401.0565}

\bibitem[{{Owen} et~al.(2020){Owen}, {Bruno}, {Livi}, {Louarn}, {Al Janabi},
  {Allegrini}, {Amoros}, {Baruah}, {Barthe}, {Berthomier}, {Bordon},
  {Brockley-Blatt}, {Brysbaert}, {Capuano}, {Collier}, {DeMarco}, {Fedorov},
  {Ford}, {Fortunato}, {Fratter}, {Galvin}, {Hancock}, {Heirtzler}, {Kataria},
  {Kistler}, {Lepri}, {Lewis}, {Loeffler}, {Marty}, {Mathon}, {Mayall}, {Mele},
  {Ogasawara}, {Orlandi}, {Pacros}, {Penou}, {Persyn}, {Petiot}, {Phillips},
  {P{\v{r}}ech}, {Raines}, {Reden}, {Rouillard}, {Rousseau}, {Rubiella},
  {Seran}, {Spencer}, {Thomas}, {Trevino}, {Verscharen}, {Wurz}, {Alapide},
  {Amoruso}, {Andr{\'e}}, {Anekallu}, {Arciuli}, {Arnett}, {Ascolese},
  {Bancroft}, {Bland}, {Brysch}, {Calvanese}, {Castronuovo},
  {{\v{C}}erm{\'a}k}, {Chornay}, {Clemens}, {Coker}, {Collinson}, {D'Amicis},
  {Dandouras}, {Darnley}, {Davies}, {Davison}, {De Los Santos}, {Devoto},
  {Dirks}, {Edlund}, {Fazakerley}, {Ferris}, {Frost}, {Fruit}, {Garat},
  {G{\'e}not}, {Gibson}, {Gilbert}, {de Giosa}, {Gradone}, {Hailey}, {Horbury},
  {Hunt}, {Jacquey}, {Johnson}, {Lavraud}, {Lawrenson}, {Leblanc}, {Lockhart},
  {Maksimovic}, {Malpus}, {Marcucci}, {Mazelle}, {Monti}, {Myers}, {Nguyen},
  {Rodriguez-Pacheco}, {Phillips}, {Popecki}, {Rees}, {Rogacki}, {Ruane},
  {Rust}, {Salatti}, {Sauvaud}, {Stakhiv}, {Stange}, {Stubbs}, {Taylor},
  {Techer}, {Terrier}, {Thibodeaux}, {Urdiales}, {Varsani}, {Walsh}, {Watson},
  {Wheeler}, {Willis}, {Wimmer-Schweingruber}, {Winter}, {Yardley}, and
  {Zouganelis}}]{owen-etal-2020}
{Owen} CJ, {Bruno} R, {Livi} S, {Louarn} P, {Al Janabi} K, {Allegrini} F,
  {Amoros} C, {Baruah} R, {Barthe} A, {Berthomier} M, {Bordon} S,
  {Brockley-Blatt} C, {Brysbaert} C, {Capuano} G, {Collier} M, {DeMarco} R,
  {Fedorov} A, {Ford} J, {Fortunato} V, {Fratter} I, {Galvin} AB, {Hancock} B,
  {Heirtzler} D, {Kataria} D, {Kistler} L, {Lepri} ST, {Lewis} G, {Loeffler} C,
  {Marty} W, {Mathon} R, {Mayall} A, {Mele} G, {Ogasawara} K, {Orlandi} M,
  {Pacros} A, {Penou} E, {Persyn} S, {Petiot} M, {Phillips} M, {P{\v{r}}ech} L,
  {Raines} JM, {Reden} M, {Rouillard} AP, {Rousseau} A, {Rubiella} J, {Seran}
  H, {Spencer} A, {Thomas} JW, {Trevino} J, {Verscharen} D, {Wurz} P, {Alapide}
  A, {Amoruso} L, {Andr{\'e}} N, {Anekallu} C, {Arciuli} V, {Arnett} KL,
  {Ascolese} R, {Bancroft} C, {Bland} P, {Brysch} M, {Calvanese} R,
  {Castronuovo} M, {{\v{C}}erm{\'a}k} I, {Chornay} D, {Clemens} S, {Coker} J,
  {Collinson} G, {D'Amicis} R, {Dandouras} I, {Darnley} R, {Davies} D,
  {Davison} G, {De Los Santos} A, {Devoto} P, {Dirks} G, {Edlund} E,
  {Fazakerley} A, {Ferris} M, {Frost} C, {Fruit} G, {Garat} C, {G{\'e}not} V,
  {Gibson} W, {Gilbert} JA, {de Giosa} V, {Gradone} S, {Hailey} M, {Horbury}
  TS, {Hunt} T, {Jacquey} C, {Johnson} M, {Lavraud} B, {Lawrenson} A, {Leblanc}
  F, {Lockhart} W, {Maksimovic} M, {Malpus} A, {Marcucci} F, {Mazelle} C,
  {Monti} F, {Myers} S, {Nguyen} T, {Rodriguez-Pacheco} J, {Phillips} I,
  {Popecki} M, {Rees} K, {Rogacki} SA, {Ruane} K, {Rust} D, {Salatti} M,
  {Sauvaud} JA, {Stakhiv} MO, {Stange} J, {Stubbs} T, {Taylor} T, {Techer} JD,
  {Terrier} G, {Thibodeaux} R, {Urdiales} C, {Varsani} A, {Walsh} AP, {Watson}
  G, {Wheeler} P, {Willis} G, {Wimmer-Schweingruber} RF, {Winter} B, {Yardley}
  J, {Zouganelis} I (2020) {The Solar Orbiter Solar Wind Analyser (SWA) suite}.
  \aap, 642:A16, \doi{10.1051/0004-6361/201937259}

\bibitem[{{Owocki} et~al.(1983){Owocki}, {Holzer}, and
  {Hundhausen}}]{Owocki1983}
{Owocki} SP, {Holzer} TE, {Hundhausen} AJ (1983) {The solar wind ionization
  state as a coronal temperature diagnostic}. \apj, 275:354--366,
  \doi{10.1086/161538}

\bibitem[{{Paraschiv} and {Donea}(2019)}]{paraschiv19}
{Paraschiv} AR, {Donea} A (2019) {On Solar Recurrent Coronal Jets: Coronal
  Geysers as Sources of Electron Beams and Interplanetary Type-III Radio
  Bursts}. \apj, 873(2):110, \doi{10.3847/1538-4357/ab04a6},
  \eprint{1903.04682}

\bibitem[{{Parenti}(2015)}]{parenti15}
{Parenti} S (2015) {Spectral Diagnostics of Cool Prominence and PCTR Optically
  Thin Plasmas}. In: {Vial} JC, {Engvold} O (eds) Solar Prominences,
  Astrophysics and Space Science Library, vol 415, p~61,
  \doi{10.1007/978-3-319-10416-4\_3}

\bibitem[{{Parenti} et~al.(2000){Parenti}, {Bromage}, {Poletto}, {Noci},
  {Raymond}, and {Bromage}}]{parenti00}
{Parenti} S, {Bromage} BJI, {Poletto} G, {Noci} G, {Raymond} JC, {Bromage} GE
  (2000) Characteristics of solar coronal streamers. element abundance,
  temperature and density from coordinated cds and uvcs soho observations.
  \aap, 363:800--814

\bibitem[{{Parenti} et~al.(2003){Parenti}, {Landi}, and {Bromage}}]{parenti03}
{Parenti} S, {Landi} E, {Bromage} BJI (2003) {SOHO-Ulysses Spring 2000
  Quadrature: Coronal Diagnostic Spectrometer and SUMER Results}. \apj,
  590:519--532

\bibitem[{{Parenti} et~al.(2019){Parenti}, {Del Zanna}, and {Vial}}]{parenti19}
{Parenti} S, {Del Zanna} G, {Vial} JC (2019) {Elemental composition in
  quiescent prominences}. \aap, 625:A52, \doi{10.1051/0004-6361/201935147},
  \eprint{1905.00871}

\bibitem[{{Pariat} et~al.(2015){Pariat}, {Dalmasse}, {DeVore}, {Antiochos}, and
  {Karpen}}]{Pariat2015}
{Pariat} E, {Dalmasse} K, {DeVore} CR, {Antiochos} SK, {Karpen} JT (2015)
  {Model for straight and helical solar jets. I. Parametric studies of the
  magnetic field geometry}. \aap, 573:A130, \doi{10.1051/0004-6361/201424209}

\bibitem[{Parker(1958)}]{parker58}
Parker EN (1958) Dynamics of the interplanetary gas and magnetic fields.
  Astrophys J, 128:664, \doi{10.1086/146579}

\bibitem[{{Parker}(1958)}]{Parker1958}
{Parker} EN (1958) {Dynamics of the Interplanetary Gas and Magnetic Fields.}
  \apj, 128:664, \doi{10.1086/146579}

\bibitem[{{Peleikis} et~al.(2017){Peleikis}, {Kruse}, {Berger}, and
  {Wimmer-Schweingruber}}]{peleikis17}
{Peleikis} T, {Kruse} M, {Berger} L, {Wimmer-Schweingruber} R (2017) {Origin of
  the solar wind: A novel approach to link in situ and remote observations. A
  study for SPICE and SWA on the upcoming Solar Orbiter mission}. \aap,
  602:A24, \doi{10.1051/0004-6361/201629727}

\bibitem[{{Peter} and {Judge}(1999)}]{peter99}
{Peter} H, {Judge} PG (1999) {On the Doppler Shifts of Solar Ultraviolet
  Emission Lines}. \apj, 522(2):1148--1166, \doi{10.1086/307672}

\bibitem[{{Pinto} and {Rouillard}(2017)}]{Pinto2017}
{Pinto} RF, {Rouillard} AP (2017) {A Multiple Flux-tube Solar Wind Model}.
  \apj, 838(2):89, \doi{10.3847/1538-4357/aa6398}, \eprint{1611.08744}

\bibitem[{{Podladchikova}(2021)}]{Podladchikova21}
{Podladchikova} Oea (2021) {Stereoscopic Measurements of Coronal Doppler
  Velocities}. \solphys

\bibitem[{{Poletto} et~al.(1996){Poletto}, {Parenti}, {Noci}, {Livi}, {Suess},
  {Balogh}, and {McComas}}]{poletto96}
{Poletto} G, {Parenti} S, {Noci} G, {Livi} S, {Suess} ST, {Balogh} A, {McComas}
  DJ (1996) {Searching for coronal plumes in ULYSSES observations of the far
  solar wind.} \aap, 316:374--383

\bibitem[{{Pomoell} and {Poedts}(2018)}]{Pomoell2018}
{Pomoell} J, {Poedts} S (2018) {EUHFORIA: European heliospheric forecasting
  information asset}. \jswsc, 8(27):A35, \doi{10.1051/swsc/2018020}

\bibitem[{Pottasch(1963)}]{pottasch63}
Pottasch SR (1963) The lower solar corona: Intepretation of the ultraviolet
  spectrum. Astrophys J, 137:945--966, \doi{10.1086/147569}

\bibitem[{{Raouafi} et~al.(2016){Raouafi}, {Patsourakos}, {Pariat}, {Young},
  {Sterling}, {Savcheva}, {Shimojo}, {Moreno-Insertis}, {DeVore}, {Archontis},
  {T{\"o}r{\"o}k}, {Mason}, {Curdt}, {Meyer}, {Dalmasse}, and
  {Matsui}}]{raouafi16}
{Raouafi} NE, {Patsourakos} S, {Pariat} E, {Young} PR, {Sterling} AC,
  {Savcheva} A, {Shimojo} M, {Moreno-Insertis} F, {DeVore} CR, {Archontis} V,
  {T{\"o}r{\"o}k} T, {Mason} H, {Curdt} W, {Meyer} K, {Dalmasse} K, {Matsui} Y
  (2016) {Solar Coronal Jets: Observations, Theory, and Modeling}. \ssr,
  201(1-4):1--53, \doi{10.1007/s11214-016-0260-5}, \eprint{1607.02108}

\bibitem[{{Raymond} et~al.(1997){Raymond}, {Kohl}, {Noci}, {Antonucci},
  {Tondello}, {Huber}, {Gardner}, {Nicolosi}, {Fineschi}, {Romoli}, {Spadaro},
  {Siegmund}, {Benna}, {Ciaravella}, {Cranmer}, {Giordano}, {Karovska},
  {Martin}, {Michels}, {Modigliani}, {Naletto}, {Panasyuk}, {Pernechele},
  {Poletto}, {Smith}, {Suleiman}, and {Strachan}}]{raymond97}
{Raymond} JC, {Kohl} JL, {Noci} G, {Antonucci} E, {Tondello} G, {Huber} MCE,
  {Gardner} LD, {Nicolosi} P, {Fineschi} S, {Romoli} M, {Spadaro} D, {Siegmund}
  OHW, {Benna} C, {Ciaravella} A, {Cranmer} S, {Giordano} S, {Karovska} M,
  {Martin} R, {Michels} J, {Modigliani} A, {Naletto} G, {Panasyuk} A,
  {Pernechele} C, {Poletto} G, {Smith} PL, {Suleiman} RM, {Strachan} L (1997)
  Composition of coronal streamers from the soho ultraviolet coronagraph
  spectrometer. \solphys, 175:645--665

\bibitem[{Reames(2018)}]{reames18}
Reames DV (2018) The ``fip effect'' and the origins of solar energetic
  particles and the solar wind. Solar Phys, 293:47--56,
  \doi{10.1007/s11207-018-1267-8}

\bibitem[{{Reames} et~al.(1985){Reames}, {von Rosenvinge}, and
  {Lin}}]{Reames1985}
{Reames} DV, {von Rosenvinge} TT, {Lin} RP (1985) {Solar He-3-rich events and
  nonrelativistic electron events - A new association}. \apj, 292:716--724,
  \doi{10.1086/163203}

\bibitem[{{Richardson} and {Cane}(2010)}]{Richardson2010}
{Richardson} IG, {Cane} HV (2010) {Interplanetary Coronal Mass Ejections During
  Solar Cycle 23}. In: {Maksimovic} M, {Issautier} K, {Meyer-Vernet} N,
  {Moncuquet} M, {Pantellini} F (eds) Twelfth International Solar Wind
  Conference, American Institute of Physics Conference Series, vol 1216, pp
  683--686, \doi{10.1063/1.3395959}

\bibitem[{{Riley} et~al.(2019){Riley}, {Linker}, {Mikic}, {Caplan}, {Downs},
  and {Thumm}}]{Riley2019}
{Riley} P, {Linker} JA, {Mikic} Z, {Caplan} RM, {Downs} C, {Thumm} JL (2019)
  {Can an Unobserved Concentration of Magnetic Flux Above the Poles of the Sun
  Resolve the Open Flux Problem?} \apj, 884(1):18,
  \doi{10.3847/1538-4357/ab3a98}

\bibitem[{{Rivera} et~al.(2019{\natexlab{a}}){Rivera}, {Landi}, and
  {Lepri}}]{Rivera2019b}
{Rivera} YJ, {Landi} E, {Lepri} ST (2019{\natexlab{a}}) {Identifying Spectral
  Lines to Study Coronal Mass Ejection Evolution in the Lower Corona}. \apjs,
  243(2):34, \doi{10.3847/1538-4365/ab2bfe}

\bibitem[{{Rivera} et~al.(2019{\natexlab{b}}){Rivera}, {Landi}, {Lepri}, and
  {Gilbert}}]{Rivera2019a}
{Rivera} YJ, {Landi} E, {Lepri} ST, {Gilbert} JA (2019{\natexlab{b}})
  {Empirical Modeling of CME Evolution Constrained to ACE/SWICS Charge State
  Distributions}. \apj, 874(2):164, \doi{10.3847/1538-4357/ab0e11}

\bibitem[{{Rochus} et~al.(2020){Rochus}, {Auch{\`e}re}, {Berghmans}, {Harra},
  {Schmutz}, {Sch{\"u}hle}, {Addison}, {Appourchaux}, {Aznar Cuadrado},
  {Baker}, {Barbay}, {Bates}, {BenMoussa}, {Bergmann}, {Beurthe}, {Borgo},
  {Bonte}, {Bouzit}, {Bradley}, {B{\"u}chel}, {Buchlin}, {B{\"u}chner},
  {Cab{\'e}}, {Cadiergues}, {Chaigneau}, {Chares}, {Choque Cortez}, {Coker},
  {Condamin}, {Coumar}, {Curdt}, {Cutler}, {Davies}, {Davison}, {Defise}, {Del
  Zanna}, {Delmotte}, {Delouille}, {Dolla}, {Dumesnil}, {D{\"u}rig}, {Enge},
  {Fran{\c{c}}ois}, {Fourmond}, {Gillis}, {Giordanengo}, {Gissot}, {Green},
  {Guerreiro}, {Guilbaud}, {Gyo}, {Haberreiter}, {Hafiz}, {Hailey}, {Halain},
  {Hansotte}, {Hecquet}, {Heerlein}, {Hellin}, {Hemsley}, {Hermans}, {Hervier},
  {Hochedez}, {Houbrechts}, {Ihsan}, {Jacques}, {J{\'e}r{\^o}me}, {Jones},
  {Kahle}, {Kennedy}, {Klaproth}, {Kolleck}, {Koller}, {Kotsialos},
  {Kraaikamp}, {Langer}, {Lawrenson}, {Le Clech'}, {Lenaerts}, {Liebecq},
  {Linder}, {Long}, {Mampaey}, {Markiewicz-Innes}, {Marquet}, {Marsch},
  {Matthews}, {Mazy}, {Mazzoli}, {Meining}, {Meltchakov}, {Mercier}, {Meyer},
  {Monecke}, {Monfort}, {Morinaud}, {Moron}, {Mountney}, {M{\"u}ller},
  {Nicula}, {Parenti}, {Peter}, {Pfiffner}, {Philippon}, {Phillips},
  {Plesseria}, {Pylyser}, {Rabecki}, {Ravet-Krill}, {Rebellato}, {Renotte},
  {Rodriguez}, {Roose}, {Rosin}, {Rossi}, {Roth}, {Rouesnel}, {Roulliay},
  {Rousseau}, {Ruane}, {Scanlan}, {Schlatter}, {Seaton}, {Silliman}, {Smit},
  {Smith}, {Solanki}, {Spescha}, {Spencer}, {Stegen}, {Stockman}, {Szwec},
  {Tamiatto}, {Tandy}, {Teriaca}, {Theobald}, {Tychon}, {van Driel-Gesztelyi},
  {Verbeeck}, {Vial}, {Werner}, {West}, {Westwood}, {Wiegelmann}, {Willis},
  {Winter}, {Zerr}, {Zhang}, and {Zhukov}}]{rochus20}
{Rochus} P, {Auch{\`e}re} F, {Berghmans} D, {Harra} L, {Schmutz} W,
  {Sch{\"u}hle} U, {Addison} P, {Appourchaux} T, {Aznar Cuadrado} R, {Baker} D,
  {Barbay} J, {Bates} D, {BenMoussa} A, {Bergmann} M, {Beurthe} C, {Borgo} B,
  {Bonte} K, {Bouzit} M, {Bradley} L, {B{\"u}chel} V, {Buchlin} E,
  {B{\"u}chner} J, {Cab{\'e}} F, {Cadiergues} L, {Chaigneau} M, {Chares} B,
  {Choque Cortez} C, {Coker} P, {Condamin} M, {Coumar} S, {Curdt} W, {Cutler}
  J, {Davies} D, {Davison} G, {Defise} JM, {Del Zanna} G, {Delmotte} F,
  {Delouille} V, {Dolla} L, {Dumesnil} C, {D{\"u}rig} F, {Enge} R,
  {Fran{\c{c}}ois} S, {Fourmond} JJ, {Gillis} JM, {Giordanengo} B, {Gissot} S,
  {Green} LM, {Guerreiro} N, {Guilbaud} A, {Gyo} M, {Haberreiter} M, {Hafiz} A,
  {Hailey} M, {Halain} JP, {Hansotte} J, {Hecquet} C, {Heerlein} K, {Hellin}
  ML, {Hemsley} S, {Hermans} A, {Hervier} V, {Hochedez} JF, {Houbrechts} Y,
  {Ihsan} K, {Jacques} L, {J{\'e}r{\^o}me} A, {Jones} J, {Kahle} M, {Kennedy}
  T, {Klaproth} M, {Kolleck} M, {Koller} S, {Kotsialos} E, {Kraaikamp} E,
  {Langer} P, {Lawrenson} A, {Le Clech'} JC, {Lenaerts} C, {Liebecq} S,
  {Linder} D, {Long} DM, {Mampaey} B, {Markiewicz-Innes} D, {Marquet} B,
  {Marsch} E, {Matthews} S, {Mazy} E, {Mazzoli} A, {Meining} S, {Meltchakov} E,
  {Mercier} R, {Meyer} S, {Monecke} M, {Monfort} F, {Morinaud} G, {Moron} F,
  {Mountney} L, {M{\"u}ller} R, {Nicula} B, {Parenti} S, {Peter} H, {Pfiffner}
  D, {Philippon} A, {Phillips} I, {Plesseria} JY, {Pylyser} E, {Rabecki} F,
  {Ravet-Krill} MF, {Rebellato} J, {Renotte} E, {Rodriguez} L, {Roose} S,
  {Rosin} J, {Rossi} L, {Roth} P, {Rouesnel} F, {Roulliay} M, {Rousseau} A,
  {Ruane} K, {Scanlan} J, {Schlatter} P, {Seaton} DB, {Silliman} K, {Smit} S,
  {Smith} PJ, {Solanki} SK, {Spescha} M, {Spencer} A, {Stegen} K, {Stockman} Y,
  {Szwec} N, {Tamiatto} C, {Tandy} J, {Teriaca} L, {Theobald} C, {Tychon} I,
  {van Driel-Gesztelyi} L, {Verbeeck} C, {Vial} JC, {Werner} S, {West} MJ,
  {Westwood} D, {Wiegelmann} T, {Willis} G, {Winter} B, {Zerr} A, {Zhang} X,
  {Zhukov} AN (2020) {The Solar Orbiter EUI instrument: The Extreme Ultraviolet
  Imager}. \aap, 642:A8, \doi{10.1051/0004-6361/201936663}

\bibitem[{{Schatten} et~al.(1969){Schatten}, {Wilcox}, and
  {Ness}}]{Schatten1969}
{Schatten} KH, {Wilcox} JM, {Ness} NF (1969) {A model of interplanetary and
  coronal magnetic fields}. \solphys, 6:442--455, \doi{10.1007/BF00146478}

\bibitem[{{Schrijver} and {Title}(2011)}]{Schrijver2011}
{Schrijver} CJ, {Title} AM (2011) {Long-range magnetic couplings between solar
  flares and coronal mass ejections observed by SDO and STEREO}. Journal of
  Geophysical Research (Space Physics), 116(A4):A04108,
  \doi{10.1029/2010JA016224}

\bibitem[{{Scott} et~al.(2018){Scott}, {Pontin}, {Yeates}, {Wyper}, and
  {Higginson}}]{Scott2018}
{Scott} RB, {Pontin} DI, {Yeates} AR, {Wyper} PF, {Higginson} AK (2018)
  {Magnetic Structures at the Boundary of the Closed Corona: Interpretation of
  S-Web Arcs}. \apj, 869(1):60, \doi{10.3847/1538-4357/aaed2b},
  \eprint{1805.04459}

\bibitem[{{Sharma} and {Srivastava}(2012)}]{Sharma2012}
{Sharma} R, {Srivastava} N (2012) {Presence of solar filament plasma detected
  in interplanetary coronal mass ejections by in situ spacecraft}. Journal of
  Space Weather and Space Climate, 2:A10, \doi{10.1051/swsc/2012010}

\bibitem[{{Sheeley}(1995)}]{sheely95}
{Sheeley} J N~R (1995) {A Volcanic Origin for High-FIP Material in the Solar
  Atmosphere}. \apj, 440:884, \doi{10.1086/175326}

\bibitem[{{Shimojo} et~al.(1996){Shimojo}, {Hashimoto}, {Shibata}, {Hirayama},
  {Hudson}, and {Acton}}]{shimojo96}
{Shimojo} M, {Hashimoto} S, {Shibata} K, {Hirayama} T, {Hudson} HS, {Acton} LW
  (1996) {Statistical Study of Solar X-Ray Jets Observed with the YOHKOH Soft
  X-Ray Telescope}. \pasj, 48:123--136, \doi{10.1093/pasj/48.1.123}

\bibitem[{{Smith} et~al.(1998){Smith}, {L'Heureux}, {Ness}, {Acu{\~n}a},
  {Burlaga}, and {Scheifele}}]{Smith1998}
{Smith} CW, {L'Heureux} J, {Ness} NF, {Acu{\~n}a} MH, {Burlaga} LF, {Scheifele}
  J (1998) {The ACE Magnetic Fields Experiment}. \ssr, 86:613--632,
  \doi{10.1023/A:1005092216668}

\bibitem[{{Solanki} et~al.(2020){Solanki}, {del Toro Iniesta}, {Woch},
  {Gandorfer}, {Hirzberger}, {Alvarez-Herrero}, {Appourchaux}, {Mart{\'\i}nez
  Pillet}, {P{\'e}rez-Grande}, {Sanchis Kilders}, {Schmidt}, {G{\'o}mez Cama},
  {Michalik}, {Deutsch}, {Fernandez-Rico}, {Grauf}, {Gizon}, {Heerlein},
  {Kolleck}, {Lagg}, {Meller}, {M{\"u}ller}, {Sch{\"u}hle}, {Staub}, {Albert},
  {Alvarez Copano}, {Beckmann}, {Bischoff}, {Busse}, {Enge}, {Frahm},
  {Germerott}, {Guerrero}, {L{\"o}ptien}, {Meierdierks}, {Oberdorfer},
  {Papagiannaki}, {Ramanath}, {Schou}, {Werner}, {Yang}, {Zerr}, {Bergmann},
  {Bochmann}, {Heinrichs}, {Meyer}, {Monecke}, {M{\"u}ller}, {Sperling},
  {{\'A}lvarez Garc{\'\i}a}, {Aparicio}, {Balaguer Jim{\'e}nez}, {Bellot
  Rubio}, {Cobos Carracosa}, {Girela}, {Hern{\'a}ndez Exp{\'o}sito}, {Herranz},
  {Labrousse}, {L{\'o}pez Jim{\'e}nez}, {Orozco Su{\'a}rez}, {Ramos},
  {Barandiar{\'a}n}, {Bastide}, {Campuzano}, {Cebollero}, {D{\'a}vila},
  {Fern{\'a}ndez-Medina}, {Garc{\'\i}a Parejo}, {Garranzo-Garc{\'\i}a},
  {Laguna}, {Mart{\'\i}n}, {Navarro}, {N{\'u}{\~n}ez Peral}, {Royo},
  {S{\'a}nchez}, {Silva-L{\'o}pez}, {Vera}, {Villanueva}, {Fourmond}, {de
  Galarreta}, {Bouzit}, {Hervier}, {Le Clec'h}, {Szwec}, {Chaigneau},
  {Buttice}, {Dominguez-Tagle}, {Philippon}, {Boumier}, {Le Cocguen},
  {Baranjuk}, {Bell}, {Berkefeld}, {Baumgartner}, {Heidecke}, {Maue}, {Nakai},
  {Scheiffelen}, {Sigwarth}, {Soltau}, {Volkmer}, {Blanco Rodr{\'\i}guez},
  {Domingo}, {Ferreres Sabater}, {Gasent Blesa}, {Rodr{\'\i}guez
  Mart{\'\i}nez}, {Osorno Caudel}, {Bosch}, {Casas}, {Carmona}, {Herms},
  {Roma}, {Alonso}, {G{\'o}mez-Sanjuan}, {Piqueras}, {Torralbo}, {Fiethe},
  {Guan}, {Lange}, {Michel}, {Bonet}, {Fahmy}, {M{\"u}ller}, and
  {Zouganelis}}]{solanki20}
{Solanki} SK, {del Toro Iniesta} JC, {Woch} J, {Gandorfer} A, {Hirzberger} J,
  {Alvarez-Herrero} A, {Appourchaux} T, {Mart{\'\i}nez Pillet} V,
  {P{\'e}rez-Grande} I, {Sanchis Kilders} E, {Schmidt} W, {G{\'o}mez Cama} JM,
  {Michalik} H, {Deutsch} W, {Fernandez-Rico} G, {Grauf} B, {Gizon} L,
  {Heerlein} K, {Kolleck} M, {Lagg} A, {Meller} R, {M{\"u}ller} R,
  {Sch{\"u}hle} U, {Staub} J, {Albert} K, {Alvarez Copano} M, {Beckmann} U,
  {Bischoff} J, {Busse} D, {Enge} R, {Frahm} S, {Germerott} D, {Guerrero} L,
  {L{\"o}ptien} B, {Meierdierks} T, {Oberdorfer} D, {Papagiannaki} I,
  {Ramanath} S, {Schou} J, {Werner} S, {Yang} D, {Zerr} A, {Bergmann} M,
  {Bochmann} J, {Heinrichs} J, {Meyer} S, {Monecke} M, {M{\"u}ller} MF,
  {Sperling} M, {{\'A}lvarez Garc{\'\i}a} D, {Aparicio} B, {Balaguer
  Jim{\'e}nez} M, {Bellot Rubio} LR, {Cobos Carracosa} JP, {Girela} F,
  {Hern{\'a}ndez Exp{\'o}sito} D, {Herranz} M, {Labrousse} P, {L{\'o}pez
  Jim{\'e}nez} A, {Orozco Su{\'a}rez} D, {Ramos} JL, {Barandiar{\'a}n} J,
  {Bastide} L, {Campuzano} C, {Cebollero} M, {D{\'a}vila} B,
  {Fern{\'a}ndez-Medina} A, {Garc{\'\i}a Parejo} P, {Garranzo-Garc{\'\i}a} D,
  {Laguna} H, {Mart{\'\i}n} JA, {Navarro} R, {N{\'u}{\~n}ez Peral} A, {Royo} M,
  {S{\'a}nchez} A, {Silva-L{\'o}pez} M, {Vera} I, {Villanueva} J, {Fourmond}
  JJ, {de Galarreta} CR, {Bouzit} M, {Hervier} V, {Le Clec'h} JC, {Szwec} N,
  {Chaigneau} M, {Buttice} V, {Dominguez-Tagle} C, {Philippon} A, {Boumier} P,
  {Le Cocguen} R, {Baranjuk} G, {Bell} A, {Berkefeld} T, {Baumgartner} J,
  {Heidecke} F, {Maue} T, {Nakai} E, {Scheiffelen} T, {Sigwarth} M, {Soltau} D,
  {Volkmer} R, {Blanco Rodr{\'\i}guez} J, {Domingo} V, {Ferreres Sabater} A,
  {Gasent Blesa} JL, {Rodr{\'\i}guez Mart{\'\i}nez} P, {Osorno Caudel} D,
  {Bosch} J, {Casas} A, {Carmona} M, {Herms} A, {Roma} D, {Alonso} G,
  {G{\'o}mez-Sanjuan} A, {Piqueras} J, {Torralbo} I, {Fiethe} B, {Guan} Y,
  {Lange} T, {Michel} H, {Bonet} JA, {Fahmy} S, {M{\"u}ller} D, {Zouganelis} I
  (2020) {The Polarimetric and Helioseismic Imager on Solar Orbiter}. \aap,
  642:A11, \doi{10.1051/0004-6361/201935325}, \eprint{1903.11061}

\bibitem[{{Spice Consortium} et~al.(2020){Spice Consortium}, {Anderson},
  {Appourchaux}, {Auch{\`e}re}, {Aznar Cuadrado}, {Barbay}, {Baudin},
  {Beardsley}, {Bocchialini}, {Borgo}, {Bruzzi}, {Buchlin}, {Burton},
  {B{\"u}chel}, {Caldwell}, {Caminade}, {Carlsson}, {Curdt}, {Davenne},
  {Davila}, {Deforest}, {Del Zanna}, {Drummond}, {Dubau}, {Dumesnil}, {Dunn},
  {Eccleston}, {Fludra}, {Fredvik}, {Gabriel}, {Giunta}, {Gottwald}, {Griffin},
  {Grundy}, {Guest}, {Gyo}, {Haberreiter}, {Hansteen}, {Harrison}, {Hassler},
  {Haugan}, {Howe}, {Janvier}, {Klein}, {Koller}, {Kucera}, {Kouliche},
  {Marsch}, {Marshall}, {Marshall}, {Matthews}, {McQuirk}, {Meining},
  {Mercier}, {Morris}, {Morse}, {Munro}, {Parenti}, {Pastor-Santos}, {Peter},
  {Pfiffner}, {Phelan}, {Philippon}, {Richards}, {Rogers}, {Sawyer},
  {Schlatter}, {Schmutz}, {Sch{\"u}hle}, {Shaughnessy}, {Sidher}, {Solanki},
  {Speight}, {Spescha}, {Szwec}, {Tamiatto}, {Teriaca}, {Thompson}, {Tosh},
  {Tustain}, {Vial}, {Walls}, {Waltham}, {Wimmer-Schweingruber}, {Woodward},
  {Young}, {de Groof}, {Pacros}, {Williams}, and {M{\"u}ller}}]{spice2020}
{Spice Consortium}, {Anderson} M, {Appourchaux} T, {Auch{\`e}re} F, {Aznar
  Cuadrado} R, {Barbay} J, {Baudin} F, {Beardsley} S, {Bocchialini} K, {Borgo}
  B, {Bruzzi} D, {Buchlin} E, {Burton} G, {B{\"u}chel} V, {Caldwell} M,
  {Caminade} S, {Carlsson} M, {Curdt} W, {Davenne} J, {Davila} J, {Deforest}
  CE, {Del Zanna} G, {Drummond} D, {Dubau} J, {Dumesnil} C, {Dunn} G,
  {Eccleston} P, {Fludra} A, {Fredvik} T, {Gabriel} A, {Giunta} A, {Gottwald}
  A, {Griffin} D, {Grundy} T, {Guest} S, {Gyo} M, {Haberreiter} M, {Hansteen}
  V, {Harrison} R, {Hassler} DM, {Haugan} SVH, {Howe} C, {Janvier} M, {Klein}
  R, {Koller} S, {Kucera} TA, {Kouliche} D, {Marsch} E, {Marshall} A,
  {Marshall} G, {Matthews} SA, {McQuirk} C, {Meining} S, {Mercier} C, {Morris}
  N, {Morse} T, {Munro} G, {Parenti} S, {Pastor-Santos} C, {Peter} H,
  {Pfiffner} D, {Phelan} P, {Philippon} A, {Richards} A, {Rogers} K, {Sawyer}
  C, {Schlatter} P, {Schmutz} W, {Sch{\"u}hle} U, {Shaughnessy} B, {Sidher} S,
  {Solanki} SK, {Speight} R, {Spescha} M, {Szwec} N, {Tamiatto} C, {Teriaca} L,
  {Thompson} W, {Tosh} I, {Tustain} S, {Vial} JC, {Walls} B, {Waltham} N,
  {Wimmer-Schweingruber} R, {Woodward} S, {Young} P, {de Groof} A, {Pacros} A,
  {Williams} D, {M{\"u}ller} D (2020) {The Solar Orbiter SPICE instrument. An
  extreme UV imaging spectrometer}. \aap, 642:A14,
  \doi{10.1051/0004-6361/201935574}, \eprint{1909.01183}

\bibitem[{{Stansby} et~al.(2020){Stansby}, {Baker}, {Brooks}, and
  {Owen}}]{Stansby2020}
{Stansby} D, {Baker} D, {Brooks} DH, {Owen} CJ (2020) {Directly comparing
  coronal and solar wind elemental fractionation}. \aap, 640:A28,
  \doi{10.1051/0004-6361/202038319}, \eprint{2005.00371}

\bibitem[{{Stone} et~al.(1998){Stone}, {Frandsen}, {Mewaldt}, {Christian},
  {Margolies}, {Ormes}, and {Snow}}]{Stone1998}
{Stone} EC, {Frandsen} AM, {Mewaldt} RA, {Christian} ER, {Margolies} D, {Ormes}
  JF, {Snow} F (1998) {The Advanced Composition Explorer}. \ssr, 86:1--22,
  \doi{10.1023/A:1005082526237}

\bibitem[{{Suess} and {Poletto}(2001)}]{suess01}
{Suess} S, {Poletto} G (2001) {The Fall 2000 and Fall 2001 SOHO-Ulysses
  Quadratures}. \ssr, 97:59--62, \doi{10.1023/A:1011865825442}

\bibitem[{{Summers} et~al.(2006){Summers}, {Dickson}, {O'Mullane}, {Badnell},
  {Whiteford}, {Brooks}, {Lang}, {Loch}, and {Griffin}}]{Summers2006}
{Summers} HP, {Dickson} WJ, {O'Mullane} MG, {Badnell} NR, {Whiteford} AD,
  {Brooks} DH, {Lang} J, {Loch} SD, {Griffin} DC (2006) {Ionization state,
  excited populations and emission of impurities in dynamic finite density
  plasmas: I. The generalized collisional radiative model for light elements}.
  Plasma Physics and Controlled Fusion, 48(2):263--293,
  \doi{10.1088/0741-3335/48/2/007}, \eprint{astro-ph/0511561}

\bibitem[{{Tadesse} et~al.(2014){Tadesse}, {Wiegelmann}, {Gosain}, {MacNeice},
  and {Pevtsov}}]{2014A&A...562A.105T}
{Tadesse} T, {Wiegelmann} T, {Gosain} S, {MacNeice} P, {Pevtsov} AA (2014)
  {First use of synoptic vector magnetograms for global nonlinear, force-free
  coronal magnetic field models}. \aap, 562:A105,
  \doi{10.1051/0004-6361/201322418}, \eprint{1309.5853}

\bibitem[{{Temmer} et~al.(2012){Temmer}, {Vr{\v{s}}nak}, {Rollett}, {Bein}, {de
  Koning}, {Liu}, {Bosman}, {Davies}, {M{\"o}stl}, {{\v{Z}}ic}, {Veronig},
  {Bothmer}, {Harrison}, {Nitta}, {Bisi}, {Flor}, {Eastwood}, {Odstrcil}, and
  {Forsyth}}]{Temmer2012}
{Temmer} M, {Vr{\v{s}}nak} B, {Rollett} T, {Bein} B, {de Koning} CA, {Liu} Y,
  {Bosman} E, {Davies} JA, {M{\"o}stl} C, {{\v{Z}}ic} T, {Veronig} AM,
  {Bothmer} V, {Harrison} R, {Nitta} N, {Bisi} M, {Flor} O, {Eastwood} J,
  {Odstrcil} D, {Forsyth} R (2012) {Characteristics of Kinematics of a Coronal
  Mass Ejection during the 2010 August 1 CME-CME Interaction Event}. \apj,
  749(1):57, \doi{10.1088/0004-637X/749/1/57}, \eprint{1202.0629}

\bibitem[{{Titov} et~al.(2011){Titov}, {Miki{\'c}}, {Linker}, {Lionello}, and
  {Antiochos}}]{Titov2011}
{Titov} VS, {Miki{\'c}} Z, {Linker} JA, {Lionello} R, {Antiochos} SK (2011)
  {Magnetic Topology of Coronal Hole Linkages}. \apj, 731:111,
  \doi{10.1088/0004-637X/731/2/111}, \eprint{1011.0009}

\bibitem[{{T{\"o}r{\"o}k} et~al.(2011){T{\"o}r{\"o}k}, {Panasenco}, {Titov},
  {Miki{\'c}}, {Reeves}, {Velli}, {Linker}, and {De Toma}}]{Torok2011}
{T{\"o}r{\"o}k} T, {Panasenco} O, {Titov} VS, {Miki{\'c}} Z, {Reeves} KK,
  {Velli} M, {Linker} JA, {De Toma} G (2011) {A Model for Magnetically Coupled
  Sympathetic Eruptions}. \apjl, 739(2):L63, \doi{10.1088/2041-8205/739/2/L63},
  \eprint{1108.2069}

\bibitem[{{T{\'o}th} et~al.(2011){T{\'o}th}, {van der Holst}, and
  {Huang}}]{Toth2011}
{T{\'o}th} G, {van der Holst} B, {Huang} Z (2011) {Obtaining Potential Field
  Solutions with Spherical Harmonics and Finite Differences}. \apj, 732(2):102,
  \doi{10.1088/0004-637X/732/2/102}, \eprint{1104.5672}

\bibitem[{{van der Holst} et~al.(2014){van der Holst}, {Sokolov}, {Meng},
  {Jin}, {Manchester}, {T{\'o}th}, and {Gombosi}}]{vanderHolst2014}
{van der Holst} B, {Sokolov} IV, {Meng} X, {Jin} M, {Manchester} I W~B,
  {T{\'o}th} G, {Gombosi} TI (2014) {Alfv{\'e}n Wave Solar Model (AWSoM):
  Coronal Heating}. \apj, 782(2):81, \doi{10.1088/0004-637X/782/2/81},
  \eprint{1311.4093}

\bibitem[{{von Steiger} and {Zurbuchen}(2011)}]{vonSteiger2011}
{von Steiger} R, {Zurbuchen} TH (2011) {Polar coronal holes during the past
  solar cycle: Ulysses observations}. Journal of Geophysical Research (Space
  Physics), 116:A01105, \doi{10.1029/2010JA015835}

\bibitem[{{von Steiger} et~al.(1997){von Steiger}, {Geiss}, and
  {Gloeckler}}]{vonSteiger1997}
{von Steiger} R, {Geiss} J, {Gloeckler} G (1997) {Composition of the Solar
  Wind}. In: {Jokipii} JR, {Sonett} CP, {Giampapa} MS (eds) Cosmic Winds and
  the Heliosphere, p 581

\bibitem[{{von Steiger} et~al.(2000){von Steiger}, {Schwadron}, {Fisk},
  {Geiss}, {Gloeckler}, {Hefti}, {Wilken}, {Wimmer-Schweingruber}, and
  {Zurbuchen}}]{vonSteiger2000}
{von Steiger} R, {Schwadron} NA, {Fisk} LA, {Geiss} J, {Gloeckler} G, {Hefti}
  S, {Wilken} B, {Wimmer-Schweingruber} RF, {Zurbuchen} TH (2000) {Composition
  of quasi-stationary solar wind flows from Ulysses/Solar Wind Ion Composition
  Spectrometer}. \jgr, 105:27217--27238, \doi{10.1029/1999JA000358}

\bibitem[{{Wang}(2012)}]{WangYM2012}
{Wang} YM (2012) {Semiempirical Models of the Slow and Fast Solar Wind}. \ssr,
  172(1-4):123--143, \doi{10.1007/s11214-010-9733-0}

\bibitem[{{Wang} and {Sheeley}(1990)}]{WangYM1990}
{Wang} YM, {Sheeley} J N~R (1990) {Solar Wind Speed and Coronal Flux-Tube
  Expansion}. \apj, 355:726, \doi{10.1086/168805}

\bibitem[{{Wang} and {Sheeley}(1992)}]{WangYM1992}
{Wang} YM, {Sheeley} NR Jr (1992) {On potential field models of the solar
  corona}. \apj, 392:310--319, \doi{10.1086/171430}

\bibitem[{{Welsch} and {Fisher}(2016)}]{Welsch2016}
{Welsch} BT, {Fisher} GH (2016) {Deriving Potential Coronal Magnetic Fields
  from Vector Magnetograms}. \solphys, 291:1681--1710,
  \doi{10.1007/s11207-016-0938-6}, \eprint{1503.08754}

\bibitem[{{Wenzel} et~al.(1992){Wenzel}, {Marsden}, {Page}, and
  {Smith}}]{wenzel1992}
{Wenzel} KP, {Marsden} RG, {Page} DE, {Smith} EJ (1992) {The ULYSSES Mission}.
  \aaps, 92:207

\bibitem[{{Wheatland} et~al.(2000){Wheatland}, {Sturrock}, and
  {Roumeliotis}}]{2000ApJ...540.1150W}
{Wheatland} MS, {Sturrock} PA, {Roumeliotis} G (2000) {An Optimization Approach
  to Reconstructing Force-free Fields}. \apj, 540:1150--1155,
  \doi{10.1086/309355}

\bibitem[{{Wiegelmann}(2007)}]{2007SoPh..240..227W}
{Wiegelmann} T (2007) {Computing Nonlinear Force-Free Coronal Magnetic Fields
  in Spherical Geometry}. \solphys, 240:227--239,
  \doi{10.1007/s11207-006-0266-3}, \eprint{astro-ph/0612124}

\bibitem[{{Wiegelmann} and {Sakurai}(2012)}]{2012LRSP....9....5W}
{Wiegelmann} T, {Sakurai} T (2012) {Solar Force-free Magnetic Fields}. Living
  Reviews in Solar Physics, 9(1):5, \doi{10.12942/lrsp-2012-5},
  \eprint{1208.4693}

\bibitem[{{Wiegelmann} and {Sakurai}(2021)}]{wiegelmann2021}
{Wiegelmann} T, {Sakurai} T (2021) {Solar force-free magnetic fields}. Living
  Reviews in Solar Physics, 18(1):1, \doi{10.1007/s41116-020-00027-4},
  \eprint{1208.4693}

\bibitem[{{Wimmer-Schweingruber} et~al.(1997){Wimmer-Schweingruber}, {von
  Steiger}, and {Paerli}}]{wimmer-etal-1997}
{Wimmer-Schweingruber} RF, {von Steiger} R, {Paerli} R (1997) {Solar wind
  stream interfaces in corotating interaction regions: SWICS/Ulysses results}.
  \jgr, 102(A8):17407--17418, \doi{10.1029/97JA00951}

\bibitem[{{Wimmer-Schweingruber} et~al.(1999){Wimmer-Schweingruber}, {von
  Steiger}, and {Paerli}}]{wimmer-etal-1999}
{Wimmer-Schweingruber} RF, {von Steiger} R, {Paerli} R (1999) {Solar wind
  stream interfaces in corotating interaction regions: New SWICS/Ulysses
  results}. \jgr, 104(A5):9933--9946, \doi{10.1029/1999JA900038}

\bibitem[{Wyper et~al.(2017)Wyper, Antiochos, and DeVore}]{Wyper2017}
Wyper PF, Antiochos SK, DeVore CR (2017) A universal model for solar eruptions.
  Nature, 544(7651):452--455,
  \urlprefix\url{http://dx.doi.org/10.1038/nature22050}

\bibitem[{{Yeates} et~al.(2018){Yeates}, {Amari}, {Contopoulos}, {Feng},
  {Mackay}, {Miki{\'c}}, {Wiegelmann}, {Hutton}, {Lowder}, {Morgan}, {Petrie},
  {Rachmeler}, {Upton}, {Canou}, {Chopin}, {Downs}, {Druckm{\"u}ller},
  {Linker}, {Seaton}, and {T{\"o}r{\"o}k}}]{2018SSRv..214...99Y}
{Yeates} AR, {Amari} T, {Contopoulos} I, {Feng} X, {Mackay} DH, {Miki{\'c}} Z,
  {Wiegelmann} T, {Hutton} J, {Lowder} CA, {Morgan} H, {Petrie} G, {Rachmeler}
  LA, {Upton} LA, {Canou} A, {Chopin} P, {Downs} C, {Druckm{\"u}ller} M,
  {Linker} JA, {Seaton} DB, {T{\"o}r{\"o}k} T (2018) {Global Non-Potential
  Magnetic Models of the Solar Corona During the March 2015 Eclipse}. \ssr,
  214(5):99, \doi{10.1007/s11214-018-0534-1}, \eprint{1808.00785}

\bibitem[{{Young} et~al.(2009){Young}, {Watanabe}, {Hara}, and
  {Mariska}}]{young09}
{Young} PR, {Watanabe} T, {Hara} H, {Mariska} JT (2009) {High-precision density
  measurements in the solar corona. I. Analysis methods and results for Fe XII
  and Fe XIII}. \aap, 495(2):587--606, \doi{10.1051/0004-6361:200810143},
  \eprint{0805.0958}

\bibitem[{{Zambrana Prado} and {Buchlin}(2019)}]{ZambranaPrado2019}
{Zambrana Prado} N, {Buchlin} {\'E} (2019) {Measuring relative abundances in
  the solar corona with optimised linear combinations of spectral lines}. \aap,
  632:A20, \doi{10.1051/0004-6361/201834735}, \eprint{1910.02886}

\bibitem[{{Zhao} et~al.(2013{\natexlab{a}}){Zhao}, {Gibson}, and
  {Fisk}}]{ZhaoL2013a}
{Zhao} L, {Gibson} SE, {Fisk} LA (2013{\natexlab{a}}) {Association of solar
  wind proton flux extremes with pseudostreamers}. \jgr, 118:2834--2841,
  \doi{10.1002/jgra.50335}

\bibitem[{{Zhao} et~al.(2013{\natexlab{b}}){Zhao}, {Landi}, and
  {Gibson}}]{ZhaoL2013b}
{Zhao} L, {Landi} E, {Gibson} SE (2013{\natexlab{b}}) {Two Novel Parameters to
  Evaluate the Global Complexity of the Sun's Magnetic Field and Track the
  Solar Cycle}. \apj, 773:157, \doi{10.1088/0004-637X/773/2/157}

\bibitem[{{Zhao} et~al.(2017{\natexlab{a}}){Zhao}, {Landi}, {Lepri}, {Gilbert},
  {Zurbuchen}, {Fisk}, and {Raines}}]{ZhaoL2017a}
{Zhao} L, {Landi} E, {Lepri} ST, {Gilbert} JA, {Zurbuchen} TH, {Fisk} LA,
  {Raines} JM (2017{\natexlab{a}}) {On the Relation between the In Situ
  Properties and the Coronal Sources of the Solar Wind}. \apj, 846(2):135,
  \doi{10.3847/1538-4357/aa850c}

\bibitem[{{Zhao} et~al.(2017{\natexlab{b}}){Zhao}, {Landi}, {Lepri}, {Kocher},
  {Zurbuchen}, {Fisk}, and {Raines}}]{ZhaoL2017b}
{Zhao} L, {Landi} E, {Lepri} ST, {Kocher} M, {Zurbuchen} TH, {Fisk} LA,
  {Raines} JM (2017{\natexlab{b}}) {An Anomalous Composition in Slow Solar Wind
  as a Signature of Magnetic Reconnection in its Source Region}. \apjs, 228:4,
  \doi{10.3847/1538-4365/228/1/4}

\bibitem[{{Zhao} and {Hoeksema}(1993)}]{ZhaoXP1993}
{Zhao} X, {Hoeksema} JT (1993) {Unique Determination of Model Coronal Magnetic
  Fields Using Photospheric Observations}. \solphys, 143(1):41--48,
  \doi{10.1007/BF00619095}

\bibitem[{{Zurbuchen} et~al.(2002){Zurbuchen}, {Fisk}, {Gloeckler}, and {von
  Steiger}}]{Zurbuchen2002}
{Zurbuchen} TH, {Fisk} LA, {Gloeckler} G, {von Steiger} R (2002) {The solar
  wind composition throughout the solar cycle: A continuum of dynamic states}.
  \grl, 29:1352, \doi{10.1029/2001GL013946}

\bibitem[{{Zurbuchen} et~al.(2012){Zurbuchen}, {von Steiger}, {Gruesbeck},
  {Landi}, {Lepri}, {Zhao}, and {Hansteen}}]{zurbuchen12}
{Zurbuchen} TH, {von Steiger} R, {Gruesbeck} J, {Landi} E, {Lepri} ST, {Zhao}
  L, {Hansteen} V (2012) {Sources of Solar Wind at Solar Minimum: Constraints
  from Composition Data}. \ssr, 172(1-4):41--55,
  \doi{10.1007/s11214-012-9881-5}

\end{thebibliography}

\end{document}